\documentclass[12pt,english]{report}
\usepackage[T1]{fontenc}
\usepackage[latin1]{inputenc}
\usepackage{geometry}
\usepackage{amsmath}
\usepackage{graphicx}
\usepackage{amssymb}

\makeatletter

\providecommand{\tabularnewline}{\\}

\usepackage{feynarts}

\let\myTOC\tableofcontents
\renewcommand\tableofcontents{%
\myTOC
\clearpage
\listoffigures
\listoftables
\clearpage
\pagenumbering{arabic}
}

\date{}

\AtBeginDocument{

}

\usepackage{babel}
\makeatother
\begin{document}

\title{\textbf{\Huge Top quark rare decays in a two Higgs doublet model
for the top}}

\author{{\Large Itzhak Baum}}

\maketitle
\begin{titlepage}

\begin{center}
\vspace{4cm}
\par\end{center}

\begin{center}
\textbf{\LARGE Top quark rare decays in a two Higgs doublet model
for the top}
\par\end{center}{\LARGE \par}

\begin{center}
\vspace{2cm}
\par\end{center}

\begin{center}
Research Thesis
\par\end{center}

\begin{center}
\vspace{1cm}
\par\end{center}

\begin{center}
\textit{In partial Fulfillment of the }
\par\end{center}

\begin{center}
\textit{Requirements for the Degree of }
\par\end{center}

\begin{center}
\textit{Master of Science in Physics}
\par\end{center}

\begin{center}
\vspace{1cm}
\par\end{center}

\begin{center}
\textbf{\large Itzhak Baum}
\par\end{center}{\large \par}

\begin{center}
\vspace{2cm}
\par\end{center}

\begin{center}
\textit{Submitted to the Senate of }
\par\end{center}

\begin{center}
\textit{the Technion - Israel Institute of Technology}
\par\end{center}

\begin{center}
\vspace{2cm}
\par\end{center}

\begin{center}
Tishrei, 5767 \ \ \ \ \ \ \ \ \ \ \ \ Haifa\ \ \ \ \ \ \ \ \ \ \ \ September,
2007
\par\end{center}

\end{titlepage}

\begin{titlepage}

I would like to thank my supervisors, Prof. Gad Eilam and Dr. Shaouly
Bar-Shalom, for their long patience, and insights, and help, without
whom this work would not have been possible. May all students have
such guidance.

\end{titlepage}

\tableofcontents{}

\begin{abstract}
The two Higgs doublet model for the top (T2HDM) is a model with two
scalar doublets in which the top quark receives a special status.
The special status of the top is manifest in the Yukawa potential,
by coupling it to the second Higgs doublet, while all other quarks
couple to the first Higgs doublet. The working assumption of the model
is that the vacuum expectation value (VEV) of the second Higgs ($v_{2}$)
is much larger than the first Higgs VEV ($v_{1}$), so that the top
receives a much larger mass than all other quarks in a natural manner,
and $\tan\beta\equiv v_{2}/v_{1}$ is large. In addition, these Yukawa
couplings generate potentially enhanced flavor-changing (FC) interactions,
both in the charged and the neutral sectors. These interactions can
greatly enhance FC decays such as $t\rightarrow ch$ and $h\rightarrow\bar{t}c$.

In this work we explicitly (and independently) derive the Yukawa and
Higgs potential of the T2HDM, obtaining the scalar to quarks and triple
scalar interactions Feynman rules. We calculate the branching ratio
(BR) of the one-loop and tree-level rare FC decays $t\rightarrow ch$
and $h\rightarrow\bar{t}c$ in the T2HDM. We explore the BR within
the parameter space of the T2HDM, focusing on regions in which $BR\left(t\rightarrow ch\right)$
and $BR\left(h\rightarrow\bar{t}c\right)$ in the T2HDM can be enhanced
compared to these BR's in the standard model (SM) and two Higgs doublet
models (2HDM) of types I and II. We find that the BR of the rare decays
$t\rightarrow ch$ and $h\rightarrow\bar{t}c$ can be enhanced by
many orders of magnitude in the T2HDM compared to the BR in the SM
and in the 2HDM-I,II, especially in regions of the parameter space
where the decays are dominated by dynamics of the neutral scalar sector.

The $BR\left(t\rightarrow ch\right)$ can be measured in the upcoming
Large Hadron Collider (LHC), if its value is above $\sim5\times10^{-5}$
which is the discovery threshold of the LHC. We find that the $BR\left(t\rightarrow ch\right)$
can exceed the LHC threshold in certain regions of the parameter space
of the T2HDM, reaching up to $\sim10^{-4}$ . Moreover, we find that
the $BR\left(h\rightarrow\bar{t}c\right)$ in the T2HDM can exceed
$\sim10^{-4}$. Discovering these processes at the LHC will show a
clear indication of new physics beyond the standard model, and will
particularly motivate the special dynamics of the T2HDM setup.
\end{abstract}

\chapter{Introduction}

The standard model (SM) of elementary particles has been highly successful
in describing observed and measured phenomena. It contains, however,
an unexplored sector, namely, the Higgs sector. The SM also has several
problems, one of which is the quark mass hierarchy problem, especially
the top quark having a much larger mass than all other quarks.

In its minimal form the SM Higgs sector is comprised of one Higgs
doublet, but that is not necessarily the case. Non minimal extensions
of the Higgs sector can describe the same observed phenomena, and
predict additional phenomena, which are as yet unobserved, but not
ruled out.

This work will describe one such extension of the SM -- the two-Higgs
doublet model {}``for the top'' (T2HDM). The T2HDM features particular
Yukawa couplings whereby the top quark receives a special status.
This particular Yukawa structure also gives rise to potentially large
flavor-changing (FC) couplings in the up-quark sector.

In this work we will explicitly (and independently) derive the Yukawa
potential of the model, though it has been shown elsewhere \cite{Das,soni 34 best}.

The FC rare decays $t\rightarrow ch$ and $h\rightarrow\bar{t}c$
have a very low branching ratio (BR) in the SM, of $\sim10^{-13}$
\cite{t-ch SM,h-tc SM arhrib}. This low BR makes these decays extremely
sensitive to new physics in the scalar sector. In this work we will
explore the BR of the FC rare decays $t\rightarrow ch$ and $h\rightarrow\bar{t}c$
in the parameter space of the T2HDM, at the 1-loop level (we adhere
in this work to the t'Hooft Feynman gauge) and at the tree-level order.
We will focus on regions of the parameter space in which the $BR\left(t\rightarrow ch\right)$
can exceed the detection limit of the upcoming large hadron collider
(LHC), and also on regions where the $BR\left(t\rightarrow ch\right)$
and $BR\left(h\rightarrow\bar{t}c\right)$ can be enhanced significantly
compared to other two Higgs doublet models (2HDM).

\section{\label{sub:intro 2hdm}The two Higgs doublets model}

The minimal extension of the SM is the two Higgs doublet model (2HDM).
A comprehensive review of the principles of the 2HDM can be found
in \cite{HHG}. Basically, the model is comprised of two Higgs doublets,
$\Phi_{1}$ and $\Phi_{2}$. They usually obey discrete symmetries,
whose aim is to define the Yukawa terms, and which divide them into
several types. The 3 most common types are: type I, where $\Phi_{2}$
couples to all quarks, and $\Phi_{1}$ does not couple to quarks;
type II, where $\Phi_{1}$ couples to down quarks, and $\Phi_{2}$
couples to up quarks; type III, which denotes a general case in which
both $\Phi_{1}$ and $\Phi_{2}$ couple to all quarks.

The type II 2HDM describes the Yukawa structure of the minimal supersymmetric
standard model (MSSM), and is therefore of particular interest in
the literature.

Several properties are common to all types of 2HDM. All feature additional
physical scalars: initially there are two complex doublets, hence
8 $(2\times4)$ degrees of freedom, and the electroweak breaking absorbs
3. We are therefore left with 5 degrees of freedom which are equivalent
to 5 physical scalars, plus the 3 (unphysical) Goldstone bosons which
are present also in the minimal SM, and which are {}``eaten'' by
the gauge bosons. The components of the scalar doublets mix to produce
the mass eigenstates. They are denoted as follows: $h^{0},H^{0}$
-- CP-even neutral scalars, $A^{0}$ -- CP-odd neutral scalar, $G^{0}$
-- neutral (unphysical) Goldstone boson, $H^{\pm}$ -- charged scalars,
$G^{\pm}$ -- charged (unphysical) Goldstone bosons. The mixing conserves
the symmetries of the theory: the mass matrix does not mix scalars
with different charges, and in CP conserving theories there is no
mixing between CP-even and CP-odd scalars.

\section{\label{sub:intro T2HDM}The two Higgs doublets model \char`\"{}for
the top\char`\"{}}

\noindent The 2HDM \char`\"{}for the top\char`\"{} (T2HDM) was first
introduced by Das and Kao \cite{Das} as an effective approach for
providing the top its mass in a natural way. They proposed a 2HDM
in which the second Higgs field couples only to the top, while the
first Higgs field couples to all other quarks.

The choice of the coupling can be expressed also in terms of a discrete
symmetry imposed on the Lagrangian \cite{Das}, under which the fields
transform as follows:

\begin{eqnarray}
 & \Phi_{1}\rightarrow-\Phi_{1}, & \quad d_{R}\rightarrow-d_{R},\quad u_{R}\rightarrow-u_{R}\;(u=u,c),\nonumber \\
 & \Phi_{2}\rightarrow+\Phi_{2}, & \quad Q_{L}\rightarrow+Q_{L},\quad t_{R}\rightarrow+t_{R}\label{eq:discrete symm}\end{eqnarray}

where: $\Phi_{i}$ are the Higgs fields, $\left(u,c,t\right)_{R}$
are the right-handed $SU(2)$ singlet up-type quarks, $d_{R}=\left(d,s,b\right)_{R}$
are the down-type right-handed quarks, and $Q_{L}$ is the left-handed
$SU(2)$ quark doublet. The discrete symmetry \eqref{eq:discrete symm}
produces the Yukawa couplings of the T2HDM, described below. This
discrete symmetry is softly broken by the $\lambda_{5}$ term of \eqref{eq:higgs potential}
as discussed in \cite{georgi soft CP and Z2}.

With these Yukawa couplings, the top gets its mass primarily from
the second Higgs vacuum expectation value (VEV), which we will choose
to be much larger than the first Higgs VEV: \begin{align}
\frac{v_{2}}{v_{1}} & \gg1.\end{align}
\textit{This is the working assumption of the T2HDM.}

This particular Yukawa coupling can also give rise to large FC interactions,
as we shall later show.

Distinct features of the T2HDM are:

\begin{itemize}
\item The $H^{+}\bar{c}b$ vertex is enhanced by the ratio of CKM matrix
elements $V_{tb}/V_{cb}$ compared to other 2HDM's. This property
motivated our work, as well as the analysis in \cite{soni 15 Z-bs,soni 21 most calcs}.
\item There are tree-level FC interactions in the up-quark sector; but there
are no tree-level FC interactions in the down-quark sector, unlike
the case of the 2HDM-III in which the tree-level FC interactions are
both in the up and down-quark sectors.
\item The couplings of the neutral scalars ($H^{0},h^{0},A^{0}$) to all
the quarks except for the top quark, increase with $\tan\beta$. This
property motivated the analysis in \cite{soni 13 3jet}.
\end{itemize}
These points will be further elaborated upon in Sec. \ref{sec:FC sector T2HDM}.

The T2HDM could stand on its own, although the couplings and symmetries
defined above do not seem {}``naturally derived''. However, it could
also be viewed as an effective low energy realization of a more fundamental
theory. Some examples are:

\begin{itemize}
\item An extra-dimensions scenario, Randall-Sundrum like, in which the couplings
are derived from the location of fields in the 5th dimension \cite{soni RS extra dims}.
\item A technicolor scenario with a topcolor condensate scalar having a
large VEV, which by construction couples only to the top quark \cite{topcolor}.
\item A non-minimal supersymmetry scenario, in which $\Phi_{1}$ couples
to down quarks and $\Phi_{2}$ to up quarks, but the couplings for
$u_{R}$, $c_{R}$ are very small, and get most of their value from
loop corrections.
\end{itemize}

\section{\label{sub:intro rare proc}Rare processes and the $t\rightarrow ch$
and $h\rightarrow\bar{t}c$ rare decays}

Rare decays are a sensitive probe for new physics \cite{isidori mele rare}.
Such decays are defined as rare because in the SM they are subject
to a suppression mechanism, which can be either highly effective cancellations,
such as the GIM mechanism, or the conservation of a fundamental symmetry,
such as lepton number.

New physics models may greatly enhance such processes, by working
around the suppression mechanism. For example, the process $\mu\rightarrow e\gamma$
which is forbidden by lepton flavor conservation, can be realized
by relaxing the symmetry in the neutrino sector (see e.g. \cite{seesaw neutrino}).

In this work, we have chosen to explore the BR of the rare decays
$h\rightarrow\bar{t}c$ and $t\rightarrow ch$ in the T2HDM, as a
potential new-physics signal at the LHC. The LHC discovery limit for
the $t\rightarrow ch$ decay process is $BR\geq5.8\cdot10^{-5}$ \cite{t-ch LHC aguilar-saavedra}
for an integrated luminosity of $100fb^{-1}$. As mentioned before,
the SM BR is about $\sim10^{-13}$, and, therefore, unobservable at
the LHC. Previous studies \cite{bejar} have shown that the $BR\left(t\rightarrow ch\right)$,
where $h=H^{0},h^{0},A^{0}$, could reach up to $\sim10^{-4}$ in
the 2HDM type II and in the MSSM, and about $\sim10^{-6}$ in the
2HDM type I.

In the T2HDM the FC decays can be enhanced due to the large $H^{+}cb$
coupling which is proportional to $V_{tb}\times\tan\beta$ instead
of $V_{cb}\times\tan\beta$ in other 2HDM's, as we shall later show.
This large coupling motivated us in calculating the BR of $t\rightarrow ch$.

As an aside, we will briefly recall how the experimental detection
of the process will proceed at the LHC \cite{t-ch LHC aguilar-saavedra}.
At the LHC the top will be mainly produced in $t\bar{t}$ pairs. One
then searches for processes in which the $t$ decays to $ch$, while
the $\bar{t}$ decays in the main $\bar{b}W^{-}$channel. The $h$
is most likely to decay into $b\bar{b}$ pairs when its mass is below
130 GeV, whereas above this mass the $W^{+}W^{-}$ decay channel starts
to dominate \cite{djouadi I}. The full process (in the lower mass
range) will look like: $gg\rightarrow t\,\bar{t}\rightarrow hc\,\bar{b}W^{-}\rightarrow b\bar{b}c\,\bar{b}l\bar{\nu}$.
The main background will come from a similar process in which the
$t\rightarrow bW^{+}\rightarrow bjc$ (where $j$ denotes a quark
jet), and the $\bar{t}$ decays as before. In this case a misidentification
of the jet as $\bar{b}$ will result in an erroneous $t\rightarrow ch$
identification \cite{t-ch LHC aguilar-saavedra}.

The $h\rightarrow\bar{t}c$ decay is the complementary process to
$t\rightarrow ch$ if $m_{h}>m_{t}+m_{c}$. The amplitude of the process
is equal to the amplitude of $t\rightarrow ch$, by applying crossing
symmetry \cite{peskin}, and therefore it is subject to the same enhancements
as the $t\rightarrow ch$ process, compared with other 2HDM's and
the SM.

\section{\label{sub:intro predictions of t2hdm}Predictions and constraints
on the T2HDM}

To date several rare decays and other observables have been calculated
in the T2HDM:

\begin{itemize}
\item The electric dipole moment (EDM) of the electron was calculated in
the T2HDM \cite{Das}, and the neutron EDM in \cite{bramon shabalin 32},
for their dependence on the CP violating mixing in the Higgs sector.
The experimental results constrain this mixing.
\item The process $b\rightarrow s\gamma$ was calculated in the T2HDM in
the leading order, for its contribution to $C_{7,8}$ \cite{soni 25 first short,soni 19 2nd best}.
By adding this result to the SM prediction one can compare the theory
to the experimental result: $BR\left(b\rightarrow s\gamma\right)=\left(3.55\pm0.26\right)\times10^{-4}$
\cite{heavy flavor group}, and derive bounds on the model. A prediction
was also given for the partial rate asymmetry of the decay, which
is very different from the SM prediction, and can be measured in B
factories. Newer measurements in \cite{heavy flavor group} seem to
further restrict the additional CP violating phase in the Yukawa sector
of the T2HDM (see below).
\item The meson mixings $B-\bar{B}$ \cite{lu-xiao 2 B-Bbar}, $K-\bar{K}$
\cite{soni 21 most calcs}, $D-\bar{D}$ \cite{soni 21 most calcs,Das}
were calculated for contributions to the mass splittings $\Delta m_{B,K,D}$,
$\epsilon_{K}$ of $K-\bar{K}$ mixing, the ratio p/q of $D-\bar{D}$
mixing. These results further constrain the parameter space of the
T2HDM, in a manner similar to $b\rightarrow s\gamma$, as was discussed
above, ruling out different regions of the parameter space \cite{Das,soni 21 most calcs,lu-xiao 2 B-Bbar}.
\item The process $b\rightarrow sl_{i}^{+}l_{j}^{-}$ was calculated in
\cite{lu-xiao 1 and 3  - b-sll}, and was found to constrain the T2HDM
weakly, so that those bounds are included within other calculated
bounds.
\item The process $gq\rightarrow qqq$ was considered in \cite{soni 13 3jet},
where $q$ denotes a $b$ or $c$ quark. It was found that the T2HDM
cross section for this process will be detectable at the LHC, while
the MSSM and the 2HDM type II are not expected to have a detectable
signal. Therefore if such a signal is observed at the LHC then it
will stand out as a clear indication in favor of the T2HDM.
\item The process $Z\rightarrow b\bar{s}+\bar{b}s$ was calculated in various
models \cite{soni 15 Z-bs}. Experimentally the BR has a weak upper
bound. It was found that the T2HDM BR for this decay is comparable
in size to the SM predicted value ($\sim10^{-8}$), and to the MSSM
with $\tilde{t}-\tilde{c}$ mixing. In comparison, MSSM with $\tilde{b}-\tilde{s}$
mixing is about two orders of magnitude higher, while the 2HDM type
II is about two orders of magnitude lower.
\item In a recent article \cite{soni 34 best} some of the above calculations
were simultaneously combined for a fit to recent experimental data,
mostly from B-factories. The best-fit values and $1\sigma$ intervals
for all the parameters in the fit were calculated. As this is the
most comprehensive work constraining the T2HDM parameters, these were
the bounds used in the present work. 
\end{itemize}
The rare decays $t\rightarrow ch$ and $h\rightarrow\bar{t}c$ have
not been calculated yet in the T2HDM. This work is aimed at this calculation,
with the intention of giving a prediction which will hopefully be
verifiable at the LHC.

\chapter{\label{sec:Yukawa}Yukawa interactions in the T2HDM}

In this section we give an explicit derivation of the Feynman rules
of scalar-quark-quark interactions in the T2HDM. We will start from
the interaction-basis Lagrangian, which follows from the symmetries
imposed. We will rotate the quark fields and the scalar fields to
their mass basis. Finally, we will write the Yukawa terms in the mass
basis, arranged by interactions, in terms of standard parameters.

The Lagrangian density of the T2HDM Yukawa interactions is of the
following form \cite{Das}:\begin{align}
\mathcal{L}_{Y} & =-\bar{Q}_{Li}\Phi_{1}F_{ij}d_{Rj}-\bar{Q}_{Li}\tilde{\Phi}_{1}G_{ij=1,2}\left(\begin{array}{c}
u\\
c\end{array}\right)_{R}-\bar{Q}_{Li}\tilde{\Phi}_{2}G_{i3}t_{R}+h.c.\mbox{ ,}\end{align}

where: $i,j=1,2,3$ are flavour indices, $L(R)\equiv\left(1-(+)\gamma^{5}\right)/2$
are the chiral left (right) projection operators, $f_{L(R)}=L(R)f$
are left(right)-handed fermion fields, $F,G$ are general $3\times3$
Yukawa matrices, and:\begin{align*}
\Phi & =\left(\begin{array}{c}
\Phi^{+}\\
\frac{v+\Phi^{0}}{\sqrt{2}}\end{array}\right),\quad\tilde{\Phi}=\left(\begin{array}{c}
\frac{v^{*}+\Phi^{0*}}{\sqrt{2}}\\
-\Phi^{-}\end{array}\right)\mbox{ .}\end{align*}

The Higgs potential can be generically written as (assuming CP conservation)
\cite{HHG}:\begin{eqnarray}
\mathcal{L}_{H} & = & \lambda_{1}\left(\Phi_{1}^{+}\Phi_{1}-v_{1}^{2}/2\right)^{2}+\lambda_{2}\left(\Phi_{2}^{+}\Phi_{2}-v_{2}^{2}/2\right)^{2}+\lambda_{3}\left[\left(\Phi_{1}^{+}\Phi_{1}-v_{1}^{2}/2\right)+\left(\Phi_{2}^{+}\Phi_{2}-v_{2}^{2}/2\right)\right]^{2}+\nonumber \\
 &  & +\lambda_{4}\left[\left(\Phi_{1}^{+}\Phi_{1}\right)\left(\Phi_{2}^{+}\Phi_{2}\right)-\left(\Phi_{1}^{+}\Phi_{2}\right)\left(\Phi_{2}^{+}\Phi_{1}\right)\right]+\lambda_{5}\left|\Phi_{1}^{+}\Phi_{2}-v_{1}v_{2}/2\right|^{2}\mbox{ .}\label{eq:higgs potential}\end{eqnarray}

The absence of CP violation implies that the CP-even and CP-odd Higgs
mass-eigenstates do not mix, and that the VEV's can be taken to be
real without affecting the Lagrangian of the theory \cite{HHG}.

Dropping the flavor indices and defining: $\mathbb{I}^{(12)}={\rm diag}(1,1,0)$,
$\mathbb{I}^{(3)}={\rm diag}(0,0,1)$, the Yukawa potential reads:\begin{eqnarray}
\mathcal{L}_{Y} & = & -\bar{Q}_{L}\Phi_{1}Fd_{R}-\bar{Q}_{L}\left(\tilde{\Phi}_{1}G\mathbb{I}^{(12)}+\tilde{\Phi}_{2}G\mathbb{I}^{(3)}\right)u_{R}+h.c.\mbox{ .}\end{eqnarray}

Inserting $\Phi$, $\tilde{\Phi}$ into the Yukawa potential and rearranging
the Yukawa terms:\begin{eqnarray}
\mathcal{L}_{Y} & = & -\bar{d}_{L}\frac{v_{1}}{\sqrt{2}}Fd_{R}-\bar{u}_{Li}\frac{1}{\sqrt{2}}\left(v_{1}G\mathbb{I}^{(12)}+v_{2}G\mathbb{I}^{(3)}\right)u_{R}-\bar{Q}_{L}\Phi_{1}Fd_{R}-\nonumber \\
 &  & -\bar{Q}_{L}\left(\tilde{\Phi}_{1}G\mathbb{I}^{(12)}+\tilde{\Phi}_{2}G\mathbb{I}^{(3)}\right)u_{R}-\frac{v_{2}^{*}}{v_{1}^{*}}\bar{Q}_{Li}\tilde{\Phi}_{1}G_{i3}t_{R}+\frac{v_{2}^{*}}{v_{1}^{*}}\bar{Q}_{Li}\tilde{\Phi}_{1}G_{i3}t_{R}+h.c.=\nonumber \\
 & = & -\bar{d}_{L}\frac{v_{1}}{\sqrt{2}}Fd_{R}-\bar{u}_{L}\frac{1}{\sqrt{2}}\left(v_{1}G\mathbb{I}^{(12)}+v_{2}G\mathbb{I}^{(3)}\right)u_{R}-\bar{Q}_{L}\Phi_{1}Fd_{R}-\nonumber \\
 &  & -\bar{Q}_{L}\left(\tilde{\Phi}_{1}G\mathbb{I}^{(12)}+\frac{v_{2}}{v_{1}}\tilde{\Phi}_{1}G\mathbb{I}^{(3)}\right)u_{R}-\bar{Q}_{Li}\left(\tilde{\Phi}_{2}-\frac{v_{2}}{v_{1}}\tilde{\Phi}_{1}\right)G_{i3}t_{R}+h.c.=\nonumber \\
 & = & -\bar{d}_{L}\frac{v_{1}}{\sqrt{2}}Fd_{R}-\bar{u}_{L}\frac{1}{\sqrt{2}}\left(v_{1}G\mathbb{I}^{(12)}+v_{2}G\mathbb{I}^{(3)}\right)u_{R}-\bar{d}_{L}\frac{1}{\sqrt{2}}\Phi_{1}^{0}Fd_{R}-\bar{u}_{L}\Phi_{1}^{+}Fd_{R}+\nonumber \\
 &  & -\bar{u}_{L}\frac{1}{\sqrt{2}}\Phi_{1}^{0*}G\left(\mathbb{I}^{(12)}+\frac{v_{2}}{v_{1}}\mathbb{I}^{(3)}\right)u_{R}+\bar{d}_{L}\Phi_{1}^{-}G\left(\mathbb{I}^{(12)}+\frac{v_{2}}{v_{1}}\mathbb{I}^{(3)}\right)u_{R}+\nonumber \\
 &  & -\bar{u}_{L}\frac{1}{\sqrt{2}}\left(\Phi_{2}^{0*}-\frac{v_{2}}{v_{1}}\Phi_{1}^{0*}\right)G\mathbb{I}^{(3)}t_{R}+\bar{d}_{L}\left(\Phi_{2}^{-}-\frac{v_{2}}{v_{1}}\Phi_{1}^{-}\right)G\mathbb{I}^{(3)}t_{R}+h.c.\mbox{ .}\end{eqnarray}

If the $G_{ij}$ are of $O(1)$, and $v_{2}$ is much larger than
$v_{1}$, then the eigenvalues of the matrix $\left(v_{1}G\mathbb{I}^{(12)}+v_{2}G\mathbb{I}^{(3)}\right)$
can be expanded as a series of $v_{1}/v_{2}$. After expanding to
the leading order, the mass matrix eigenvalues are: $O(1)\cdot\left[v_{1},v_{1},v_{2}\right]$.
As can be seen, the top quark receives a mass contribution from the
second, and larger, VEV, while the up and charm receive their masses
from the first VEV.

Rotating to the quark mass basis, we define: $d_{L,R}\rightarrow D_{L,R}d_{L,R}$,
$u_{L,R}\rightarrow U_{L,R}u_{L,R}$, such that:\begin{align}
M_{d} & \equiv\frac{v_{1}}{\sqrt{2}}D_{L}^{\dagger}FD_{R}=\mbox{diag}\left(m_{d},m_{s},m_{b}\right),\nonumber \\
\mbox{ }M_{u} & \equiv U_{L}^{\dagger}\frac{1}{\sqrt{2}}\left(v_{1}G\mathbb{I}^{(12)}+v_{2}G\mathbb{I}^{(3)}\right)U_{R}=\mbox{diag}\left(m_{u},m_{c},m_{t}\right).\end{align}

We define the CKM matrix: $V_{CKM}\equiv U_{L}^{\dagger}D_{L}$, $V_{CKM}^{\dagger}\equiv D_{L}^{\dagger}U_{L}$
(we will henceforward drop the subscript CKM when referring to the
CKM matrix), and a new mixing matrix for the up-quarks:\begin{align}
\Sigma & \equiv M_{u}U_{R}^{\dagger}\mathbb{I}^{(3)}U_{R},\end{align}

as was originally defined in \cite{Das}.

The matrix $U_{R}$ can be generally parametrized by multiplying 3
rotation matrices \cite{soni 21 most calcs}. Defining 3 rotation
angles: $\alpha_{12}=\phi$, $\alpha_{23}=\sin^{-1}\left(\epsilon_{ct}\xi\right)$
and $\alpha_{13}=\sin^{-1}\left(\epsilon_{ct}\xi'\right)$, where
$\epsilon_{ct}\equiv\frac{m_{c}}{m_{t}}$, and $\xi,\xi'$ are parameters
naturally of $O(1)$, and $\xi\equiv\left|\xi\right|^{i\varphi_{\xi}}$,
we get:\begin{align}
U_{R} & =\left(\begin{array}{ccc}
\cos\phi & -\sin\phi & 0\\
\sin\phi & \cos\phi & 0\\
0 & 0 & 1\end{array}\right)\left(\begin{array}{ccc}
1 & 0 & 0\\
0 & \sqrt{1-\left|\epsilon_{ct}\xi\right|^{2}} & -\epsilon_{ct}\xi^{*}\\
0 & \epsilon_{ct}\xi & \sqrt{1-\left|\epsilon_{ct}\xi\right|^{2}}\end{array}\right)\left(\begin{array}{ccc}
\sqrt{1-\left|\epsilon_{ct}\xi'\right|^{2}} & 0 & -\epsilon_{ct}\xi'^{*}\\
0 & 1 & 0\\
\epsilon_{ct}\xi' & 0 & \sqrt{1-\left|\epsilon_{ct}\xi'\right|^{2}}\end{array}\right)=\nonumber \\
 & =\left(\begin{array}{ccc}
* & * & *\\
* & * & *\\
\epsilon_{ct}\xi'\sqrt{1-\left|\epsilon_{ct}\xi\right|^{2}} & \quad\epsilon_{ct}\xi\quad & \sqrt{1-\left|\epsilon_{ct}\xi\right|^{2}}\sqrt{1-\left|\epsilon_{ct}\xi'\right|^{2}}\end{array}\right),\label{eq:U_R def}\end{align}

where the asterisks ($*$) denote terms which are not relevant for
our calculations to follow.

Using Eq. \eqref{eq:U_R def} we can now write the $\Sigma$ matrix:{\small \begin{align}
\frac{\Sigma}{m_{t}} & =\left(\begin{array}{ccc}
\frac{m_{u}}{m_{t}}\epsilon_{ct}^{2}\left|\xi'\right|^{2}\left(1-\left|\epsilon_{ct}\xi\right|^{2}\right) & \frac{m_{u}}{m_{t}}\epsilon_{ct}^{2}\xi'^{*}\xi\sqrt{1-\left|\epsilon_{ct}\xi\right|^{2}} & \frac{m_{u}}{m_{t}}\epsilon_{ct}\xi'^{*}\left(1-\left|\epsilon_{ct}\xi\right|^{2}\right)\sqrt{1-\left|\epsilon_{ct}\xi'\right|^{2}}\\
\epsilon_{ct}^{3}\xi^{*}\xi'\sqrt{1-\left|\epsilon_{ct}\xi\right|^{2}} & \epsilon_{ct}^{3}\left|\xi\right|^{2} & \epsilon_{ct}^{2}\xi^{*}\sqrt{1-\left|\epsilon_{ct}\xi\right|^{2}}\sqrt{1-\left|\epsilon_{ct}\xi'\right|^{2}}\\
\epsilon_{ct}\xi'\left(1-\left|\epsilon_{ct}\xi\right|^{2}\right)\sqrt{1-\left|\epsilon_{ct}\xi'\right|^{2}} & \;\epsilon_{ct}\xi\sqrt{1-\left|\epsilon_{ct}\xi\right|^{2}}\sqrt{1-\left|\epsilon_{ct}\xi'\right|^{2}} & \left(1-\left|\epsilon_{ct}\xi\right|^{2}\right)\left(1-\left|\epsilon_{ct}\xi'\right|^{2}\right)\end{array}\right).\label{eq:sigma}\end{align}
}{\small \par}

The special mixing matrix between up and down quarks via the charged
Higgs, neglecting terms of $O\left(\epsilon_{ct}^{2}\right)$ or $O\left(\frac{m_{u}}{m_{t}}\right)$
(recall: $\epsilon_{ct}\equiv\frac{m_{c}}{m_{t}}$), can then be written
approximately as:

\begin{align}
\left(\Sigma^{\dagger}V\right)/m_{t} & =\left(\begin{array}{ccc}
\epsilon_{ct}\xi'^{*}V_{td} & \;\epsilon_{ct}^{3}\xi'^{*}\xi V_{cs}+\epsilon_{ct}\xi'^{*}V_{ts}\; & \epsilon_{ct}\xi'^{*}V_{tb}\\
\epsilon_{ct}\xi^{*}V_{td} & \epsilon_{ct}^{3}\left|\xi\right|^{2}V_{cs}+\epsilon_{ct}\xi^{*}V_{ts} & \epsilon_{ct}\xi^{*}V_{tb}\\
V_{td} & \epsilon_{ct}^{2}\xi V_{cs}+V_{ts} & V_{tb}-V_{tb}\epsilon_{ct}^{2}\left(\left|\xi\right|^{2}+\left|\xi'\right|^{2}\right)\end{array}\right).\label{eq:sigmadag*V}\end{align}

We note that the matrix was approximated for the purpose of illustration.
In all calculations the matrix was used without neglecting anything.

The Yukawa terms in the quark mass basis are: 

\begin{eqnarray}
\mathcal{L}_{Y} & = & -\bar{d}_{L}M_{d}d_{R}-\bar{u}_{L}M_{u}u_{R}-\bar{d}_{L}\Phi_{1}^{0}\frac{M_{d}}{v_{1}}d_{R}-\bar{u}_{L}\Phi_{1}^{0*}\frac{M_{u}}{v_{1}}u_{R}-\nonumber \\
 &  & -\bar{u}_{L}U_{L}^{\dagger}\Phi_{1}^{+}D_{L}D_{L}^{\dagger}FD_{R}d_{R}+\bar{d}_{L}D_{L}^{\dagger}\Phi_{1}^{-}U_{L}U_{L}^{\dagger}\frac{1}{v_{1}}\left(v_{1}G\mathbb{I}^{(12)}+v_{2}G\mathbb{I}^{(3)}\right)U_{R}u_{R}-\\
 &  & -\bar{u}_{L}U_{L}^{\dagger}\frac{1}{\sqrt{2}}\left(\Phi_{2}^{0*}-\frac{v_{2}}{v_{1}}\Phi_{1}^{0*}\right)G\mathbb{I}^{(3)}U_{R}u_{R}+\bar{d}_{L}D_{L}^{\dagger}\left(\Phi_{2}^{-}-\frac{v_{2}}{v_{1}}\Phi_{1}^{-}\right)U_{L}U_{L}^{\dagger}G\mathbb{I}^{(3)}U_{R}u_{R}+h.c.\mbox{ .}\nonumber \end{eqnarray}

We will henceforward drop the mass terms.

Using: $\mathbb{I}^{(12)}\cdot\mathbb{I}^{(3)}=\left[0\right]^{3\times3}$,
and: $U_{L}^{\dagger}G\mathbb{I}^{(3)}U_{R}=U_{L}^{\dagger}\frac{1}{v_{2}}\left(v_{1}G\mathbb{I}^{(12)}+v_{2}G\mathbb{I}^{(3)}\right)U_{R}U_{R}^{\dagger}\mathbb{I}^{(3)}U_{R}=\frac{\sqrt{2}}{v_{2}}\Sigma$,
we get:\begin{eqnarray}
\mathcal{L}_{Y} & = & -\bar{d}_{L}\Phi_{1}^{0}\frac{M_{d}}{v_{1}}d_{R}-\bar{u}_{L}\Phi_{1}^{0*}\frac{M_{u}}{v_{1}}u_{R}-\Phi_{1}^{+}\bar{u}_{L}V\frac{\sqrt{2}M_{d}}{v_{1}}d_{R}+\Phi_{1}^{-}\bar{d}_{L}V_{CKM}^{\dagger}\frac{\sqrt{2}M_{u}}{v_{1}}u_{R}-\nonumber \\
 &  & -\left(\Phi_{2}^{0*}-\frac{v_{2}}{v_{1}}\Phi_{1}^{0*}\right)\bar{u}_{L}\frac{1}{v_{2}}\Sigma u_{R}+\left(\Phi_{2}^{-}-\frac{v_{2}}{v_{1}}\Phi_{1}^{-}\right)\bar{d}_{L}V^{\dagger}\frac{\sqrt{2}}{v_{2}}\Sigma u_{R}+h.c.\mbox{ .}\end{eqnarray}

The Higgs fields are not in their mass basis yet. The mass basis is
derived from the Higgs potential in the Lagrangian, see App. \ref{app:Higgs-potential}
(see also \cite{HHG}). 

We note that the real parts of the neutral Higgs fields generally
rotate with the angle $\alpha$, whereas the other Higgs fields rotate
with the angle $\beta$. This is due to their shift by the VEV's.
This point is sometimes overlooked in the literature, where often
the value for $\alpha$ is chosen arbitrarily. This omission of $\alpha$
can perhaps be attributed to the MSSM, where the angle $\alpha$ is
constrained by the Higgs masses \cite{HHG}.

The Higgs fields in the mass basis are defined as follows:\begin{align}
 & \Phi^{0}=\Phi^{0r}+i\Phi^{0i},\qquad\tan\left(\beta\right)\equiv\frac{v_{2}}{v_{1}},\nonumber \\
 & \Phi_{1}^{or}=H^{0}\cos\alpha-h^{0}\sin\alpha,\qquad\Phi_{1}^{oi}=G^{0}\cos\beta-A^{0}\sin\beta,\qquad\Phi_{1}^{+}=G^{+}\cos\beta-H^{+}\sin\beta,\nonumber \\
 & \Phi_{2}^{or}=H^{0}\sin\alpha+h^{0}\cos\alpha,\qquad\Phi_{2}^{oi}=G^{0}\sin\beta+A^{0}\cos\beta,\qquad\Phi_{2}^{+}=G^{+}\sin\beta+H^{+}\cos\beta.\end{align}

The Yukawa terms in the quark and Higgs mass basis, are then:\begin{eqnarray}
\mathcal{L}_{Y} & = & -\bar{d}_{L}\left[\left(H^{0}\cos\alpha-h^{0}\sin\alpha\right)+i\left(G^{0}\cos\beta-A^{0}\sin\beta\right)\right]\frac{M_{d}}{v_{1}}d_{R}-\nonumber \\
 &  & -\bar{u}_{L}\left[\left(H^{0}\cos\alpha-h^{0}\sin\alpha\right)-i\left(G^{0}\cos\beta-A^{0}\sin\beta\right)\right]\frac{M_{u}}{v_{1}}u_{R}-\nonumber \\
 &  & -\left(G^{+}\cos\beta-H^{+}\sin\beta\right)\bar{u}_{L}V\frac{\sqrt{2}M_{d}}{v_{1}}d_{R}+\left(G^{-}\cos\beta-H^{-}\sin\beta\right)\bar{d}_{L}V^{\dagger}\frac{\sqrt{2}M_{u}}{v_{1}}u_{R}-\nonumber \\
 &  & -\left[\left(H^{0}\sin\alpha+\phi_{2}^{0r}\cos\alpha\right)-i\left(G^{0}\sin\beta+A^{0}\cos\beta\right)\right]\bar{u}_{L}\frac{1}{v_{2}}\Sigma u_{R}+\nonumber \\
 &  & +\frac{v_{2}}{v_{1}}\left[\left(H^{0}\cos\alpha-\phi_{2}^{0r*}\sin\alpha\right)-i\left(G^{0}\cos\beta-A^{0}\sin\beta\right)\right]\bar{u}_{L}\frac{1}{v_{2}}\Sigma u_{R}+\nonumber \\
 &  & +\left[\left(G^{-}\sin\beta+H^{-}\cos\beta\right)-\frac{v_{2}}{v_{1}}\left(G^{-}\cos\beta-H^{-}\sin\beta\right)\right]\bar{d}_{L}V^{\dagger}\frac{\sqrt{2}}{v_{2}}\Sigma u_{R}+h.c.\mbox{ .}\end{eqnarray}

Collecting identical interactions:\begin{eqnarray}
\mathcal{L}_{Y} & = & -\bar{d}_{L}\left[\left(H^{0}\cos\alpha-h^{0}\sin\alpha\right)+i\left(G^{0}\cos\beta-A^{0}\sin\beta\right)\right]\frac{M_{d}}{v_{1}}d_{R}+\nonumber \\
 &  & +H^{0}\bar{u}_{L}\left[-\frac{M_{u}}{v_{1}}\cos\alpha-\frac{1}{v_{2}}\Sigma\sin\alpha+\frac{1}{v_{1}}\Sigma\cos\alpha\right]u_{R}+\nonumber \\
 &  & +h^{0}\bar{u}_{L}\left[\frac{M_{u}}{v_{1}^{*}}\sin\alpha-\frac{1}{v_{2}}\Sigma\cos\alpha-\frac{1}{v_{1}}\Sigma\sin\alpha\right]u_{R}+\nonumber \\
 &  & +iG^{0}\bar{u}_{L}\left[\frac{M_{u}}{v_{1}}\cos\beta+\frac{1}{v_{2}}\Sigma\sin\beta-\frac{1}{v_{1}}\Sigma\cos\beta\right]u_{R}+\nonumber \\
 &  & +iA^{0}\bar{u}_{L}\left[-\frac{M_{u}}{v_{1}}\sin\beta+\frac{1}{v_{2}}\Sigma\cos\beta+\frac{1}{v_{1}}\Sigma\sin\beta\right]u_{R}+\nonumber \\
 &  & +\left(-G^{+}\cos\beta+H^{+}\sin\beta\right)\bar{u}_{L}V\frac{\sqrt{2}M_{d}}{v_{1}}d_{R}+\nonumber \\
 &  & +G^{-}\bar{d}_{L}V^{\dagger}\left[\cos\beta\frac{\sqrt{2}M_{u}}{v_{1}}+\sin\beta\frac{\sqrt{2}}{v_{2}}\Sigma-\cos\beta\frac{\sqrt{2}}{v_{1}}\Sigma\right]u_{R}+\nonumber \\
 &  & +H^{-}\bar{d}_{L}V^{\dagger}\left[-\sin\beta\frac{\sqrt{2}M_{u}}{v_{1}}+\cos\beta\frac{\sqrt{2}}{v_{2}}\Sigma+\sin\beta\frac{\sqrt{2}}{v_{1}}\Sigma\right]u_{R}+h.c.\mbox{ .}\end{eqnarray}

Using: $v_{1}=v\cos\beta$, $v_{2}=v\sin\beta$, $v=\sqrt{v_{1}^{2}+v_{2}^{2}}=\frac{2m_{W}}{g}$,
and adding the h.c., we get the final Yukawa terms, in the physical
mass basis, arranged by interactions:\begin{align}
\mathcal{L}_{Y}= & H^{0}\bar{d}\left[-\frac{gM_{d}}{2m_{W}}\frac{\cos\alpha}{\cos\beta}\right]d+h^{0}\bar{d}\left[\frac{gM_{d}}{2m_{W}}\frac{\sin\alpha}{\cos\beta}\right]d+\nonumber \\
 & +A^{0}\bar{d}\left[i\frac{gM_{d}}{2m_{W}}\tan\beta\left(R-L\right)\right]d+G^{0}\bar{d}\left[-i\frac{gM_{d}}{2m_{W}}\left(R-L\right)\right]d+\nonumber \\
 & +H^{0}\bar{u}\left[\frac{g}{2m_{W}}\left(-M_{u}\frac{\cos\alpha}{\cos\beta}+\Sigma\left(-\frac{\sin\alpha}{\sin\beta}+\frac{\cos\alpha}{\cos\beta}\right)\right)R+\left(h.c.\right)L\right]u+\nonumber \\
 & +h^{0}\bar{u}\left[\frac{g}{2m_{W}}\left(M_{u}\frac{\sin\alpha}{\cos\beta}-\Sigma\left(\frac{\cos\alpha}{\sin\beta}+\frac{\sin\alpha}{\cos\beta}\right)\right)R+\left(h.c.\right)L\right]u+\nonumber \\
 & +A^{0}\bar{u}\left[i\frac{g}{2m_{W}}\left(-M_{u}\tan\beta+\Sigma\left(\tan\beta+\cot\beta\right)\right)R+\left(h.c.\right)L\right]u+G^{0}\bar{u}\left[i\frac{gM_{u}}{2m_{W}}\left(R-L\right)\right]u+\nonumber \\
 & +G^{+}\bar{u}\frac{g}{\sqrt{2}m_{W}}\left[-VM_{d}R+M_{u}VL\right]d+h.c.+\nonumber \\
 & +H^{+}\bar{u}\frac{g}{\sqrt{2}m_{W}}\left[\tan\beta VM_{d}R+\left(-M_{u}\tan\beta+\Sigma^{\dagger}\left(\tan\beta+\cot\beta\right)\right)VL\right]d+h.c.\mbox{ .}\end{align}

For completeness, in App. \ref{app:Feynman-rules} we give the complete
list of Feynman rules for the T2HDM model, as derived above.

\newpage

\chapter{\label{sec:FC sector T2HDM}The flavor-changing sector of the T2HDM}

The T2HDM features unique flavor-changing (FC) couplings of both charged
Higgs and neutral Higgs, where the neutral Higgs FC couplings are
in the up-quark sector only. These couplings can enhance FC processes,
in particular the $t\rightarrow ch$ decay.

\section{\label{sec:FC charged}Charged-Higgs FC Yukawa interactions}

In the SM the leading order diagrams of the $t\rightarrow ch$ decay
are at the 1-loop level, through the mediation of $W^{+}$ gauge bosons
and b-quarks in the loop, being $\propto V_{cb}$. In general, in
any 2HDM there is a corresponding vertex of $H^{+}\bar{c}b$, and
the process $t\rightarrow ch$ proceeds also via similar diagrams
with $H^{+}$ and $b$ in the loop. In the T2HDM the $H^{+}\bar{c}b$
vertex receives a particular value distinct from other 2HDM's:\begin{align}
L & \supset\frac{g}{\sqrt{2}m_{W}}H^{+}\bar{c}\left[\tan\beta VM_{d}R+\left(-M_{u}\tan\beta+\Sigma^{\dagger}\left(\tan\beta+\cot\beta\right)\right)VL\right]_{cb}b\sim\nonumber \\
 & \sim\frac{g}{\sqrt{2}m_{W}}H^{+}\bar{c}\left[\tan\beta V_{cb}m_{b}R+m_{c}\left(-\tan\beta V_{cb}+\xi^{*}\left(\tan\beta+\cot\beta\right)V_{tb}\right)L\right]b\mbox{ .}\end{align}

As can be seen, the $H^{+}\bar{c}b$ vertex has a term proportional
to $V_{cb}\times\tan\beta$ which is common to other 2HDM's, but has
an additional term proportional to $\left(\tan\beta+\cot\beta\right)\times\left(\Sigma^{\dagger}V\right)_{cb}\sim m_{c}\xi^{*}V_{tb}$,
as shown above in \eqref{eq:sigmadag*V}. The main contribution for
1-loop diagrams with internal $H^{+}$ and $b$, in the T2HDM, will
therefore come from $m_{c}\xi^{*}V_{tb}$ terms, and is thus not CKM
suppressed.

For our analysis we also consider the $H^{+}\bar{t}b$ vertex in the
T2HDM:\begin{align}
L & \supset\frac{g}{\sqrt{2}m_{W}}H^{+}\bar{t}\left\{ \tan\beta V_{tb}m_{b}R+\right.\nonumber \\
 & \left.\qquad+\left[-m_{t}V_{tb}\tan\beta+m_{t}\left(V_{tb}-V_{tb}\epsilon_{ct}^{2}\left(\left|\xi\right|^{2}+\left|\xi'\right|^{2}\right)\right)\left(\tan\beta+\cot\beta\right)\right]L\right\} b=\nonumber \\
 & =\frac{g}{\sqrt{2}m_{W}}H^{+}\bar{t}\left\{ \tan\beta V_{tb}m_{b}R+\right.\nonumber \\
 & \left.\qquad+\left[m_{t}V_{tb}\cot\beta-m_{t}V_{tb}\epsilon_{ct}^{2}\left(\left|\xi\right|^{2}+\left|\xi'\right|^{2}\right)\left(\tan\beta+\cot\beta\right)\right]L\right\} b\label{eq:H+tb}\end{align}

We can see that by taking $\xi,\xi'\rightarrow0$ the $H^{+}\bar{t}b$
interaction in \eqref{eq:H+tb} becomes equivalent to that of a 2HDM
type I or II.

\section{\label{sec:FC neutral}Neutral-Higgs FC Yukawa interactions}

\textit{A priori} there is no distinction between $h^{0}$ and $H^{0}$
other than the rotation angle $\alpha$. In this work we therefore
adopt $\alpha=\beta$, and in the following discussion we explore
the consequences of this choice, in particular the elimination of
the tree-level $H^{0}\bar{t}c$ vertex.

The $t\rightarrow cH^{0}$ decay in the T2HDM can proceed at tree
level, from the following interaction term (see Sec. \ref{sec:Yukawa}):\begin{align}
L & \supset H^{0}\bar{t}\left[\frac{g}{2m_{W}}\left(-M_{u}\frac{\cos\alpha}{\cos\beta}+\Sigma\left(-\frac{\sin\alpha}{\sin\beta}+\frac{\cos\alpha}{\cos\beta}\right)\right)R+\left(h.c.\right)L\right]_{tc}c\sim\nonumber \\
 & \sim H^{0}\bar{t}\left[\frac{g}{2m_{W}}\left(-\frac{\sin\alpha}{\sin\beta}+\frac{\cos\alpha}{\cos\beta}\right)\left(m_{c}\xi R+m_{c}\epsilon_{ct}\xi L\right)\right]c,\end{align}

where we used the off-diagonal terms of $\Sigma$ from Eq. \ref{eq:sigma},
neglecting terms of order $\epsilon_{ct}^{2}$ (recall $\epsilon_{ct}=m_{c}/m_{t}$),
$\Sigma_{tc}\sim m_{c}\xi$ and $\left(\Sigma^{\dagger}\right)_{tc}\sim m_{c}\epsilon_{ct}\xi$.

For arbitrary $\alpha$ and $\beta$, this will lead to $t\rightarrow cH^{0}$
decay at tree level. On the other hand, the $H^{0}\bar{t}c$ vertex
can vanish if:

1)$\xi=0$, 

2)$-\frac{\sin\alpha}{\sin\beta}+\frac{\cos\alpha}{\cos\beta}=0\quad\Leftrightarrow\quad\alpha=\beta+n\pi\;.$

In this work we wish to examine the case of a vanishing $H^{0}\bar{t}c$
tree-level interaction, therefore adopting $\alpha=\beta$, since:

\begin{enumerate}
\item $\xi=0$ is strongly disfavoured by the bounds in \cite{soni 34 best},
as we will discuss in Sec. \ref{sub:calcs vals&bounds}.
\item $\xi=0$ cancels the potentially enhanced term of the $H^{+}\bar{c}b$
coupling, as shown above.
\item The limit $\alpha=\beta$ is a natural result of the MSSM, when the
mass of the CP-odd neutral Higgs, $A^{0}$, is large (see e.g. \cite{HHG}
in the limit $m_{A^{0}}\rightarrow\infty$). Note that $\alpha=\beta$
is widely used in the literature, partly for this reason. Even though
the T2HDM setup is not natural within the MSSM, this will help us
compare our results with other existing results in different types
of 2HDM's.
\item The limit $\alpha=\beta$ sets the scalar $H^{0}$ to be SM-like.
As such, it will have SM-Higgs Yukawa couplings:\begin{align}
\mathcal{L}_{Y}^{\alpha=\beta} & \supset H^{0}\bar{d}\left[-\frac{gM_{d}}{2m_{W}}\right]d+H^{0}\bar{u}\left[-\frac{gM_{u}}{2m_{W}}\right]u\mbox{ .}\end{align}
In that case, the direct mass bounds on the SM Higgs may roughly apply
to $H^{0}$.
\end{enumerate}
We turn now to the $h^{0}$-up quarks Yukawa interactions. The 1-loop
$t\rightarrow cH^{0}$ can proceed also through the mediation of $h^{0}$
scalars and top quarks in the loop. In these diagrams the important
interactions are the $h^{0}\bar{t}c$ and the $h^{0}\bar{t}t$ vertices,
which get special values in the T2HDM.

The $h^{0}\bar{t}c$ interaction reads:\begin{align}
L & \supset h^{0}\bar{t}\left[\frac{g}{2m_{W}}\left(M_{u}\frac{\sin\alpha}{\cos\beta}-\Sigma\left(\frac{\cos\alpha}{\sin\beta}+\frac{\sin\alpha}{\cos\beta}\right)\right)R+\left(h.c.\right)L\right]_{tc}c\sim\nonumber \\
 & \sim h^{0}\bar{t}\left[-\frac{g}{2m_{W}}\left(\frac{\cos\alpha}{\sin\beta}+\frac{\sin\alpha}{\cos\beta}\right)\left(m_{c}\xi R+m_{c}\epsilon_{ct}\xi L\right)\right]c,\end{align}
where setting $\alpha=\beta$ gives:\begin{align}
\mathcal{L}^{\alpha=\beta} & \supset h^{0}\bar{t}\left[-\frac{g}{2m_{W}}\left(\tan\beta+\cot\beta\right)\left(m_{c}\xi R+m_{c}\epsilon_{ct}\xi L\right)\right]c.\end{align}

One can see that the $h^{0}\bar{t}c$ interaction can be enhanced
by $\tan\beta$. 

The $h^{0}\bar{t}t$ interaction, with $\alpha=\beta$, reads:\begin{align}
L & \supset h^{0}\bar{t}\left\{ \frac{g}{2m_{W}}\left[m_{t}\tan\beta-\Sigma_{tt}\left(\tan\beta+\cot\beta\right)\right]R+\left(h.c.\right)L\right\} t\sim\nonumber \\
 & \sim h^{0}\bar{t}\left\{ \frac{g}{2m_{W}}\left[-m_{t}\cot\beta+m_{c}\epsilon_{ct}\left(\left|\xi\right|^{2}+\left|\xi'\right|^{2}\right)\left(\tan\beta+\cot\beta\right)\right]R+\left(h.c.\right)L\right\} t,\end{align}
where we use $\Sigma_{tt}=\left(\Sigma^{\dagger}\right)_{tt}\sim m_{t}-m_{c}\epsilon_{ct}\left(\left|\xi\right|^{2}+\left|\xi'\right|^{2}\right)$.
As in the case of $H^{+}\bar{t}b$, the leading term $\propto m_{t}\tan\beta$
cancels, leaving terms that are suppressed either by $\epsilon_{ct}^{2}$
or $\cot^{2}\beta$.

\chapter{\label{sec:Calculations}Calculations}

\section{\label{sub:calcs 1L}One-loop amplitude}

\begin{figure}
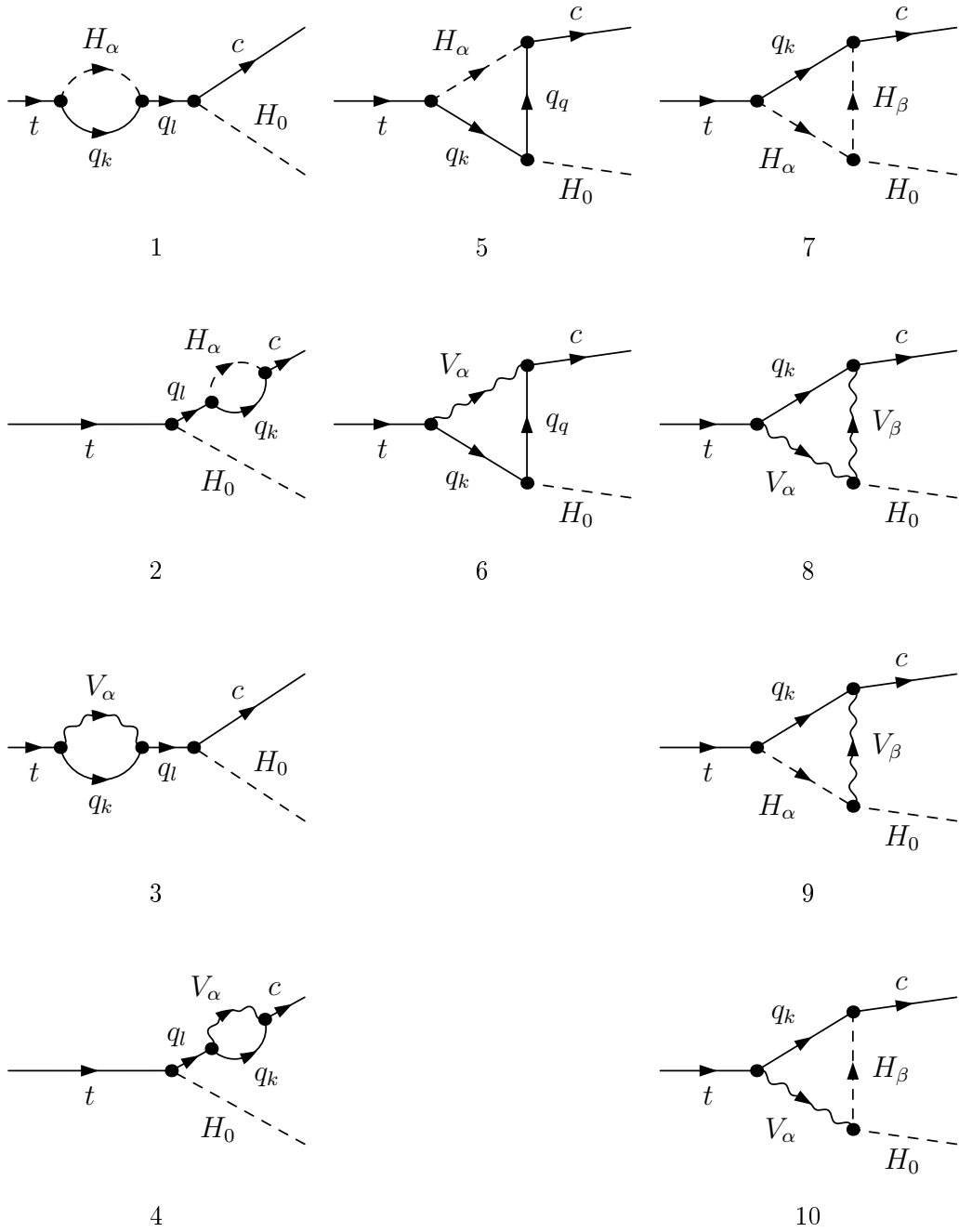

\begin{feynartspicture}(432,525)(3,4)
\FALabel(33.,95.)[]{\large $t\,\to\,c\,h\;diagrams$}

\FADiagram{1}
\FAProp(0.,10.)(3.5,10.)(0.,){/Straight}{1}
\FALabel(1.75,8.93)[t]{$t$}
\FAProp(20.,15.)(12.5,10.)(0.,){/Straight}{-1}
\FALabel(15.8702,13.3097)[br]{$c$}
\FAProp(20.,5.)(12.5,10.)(0.,){/ScalarDash}{0}
\FALabel(16.4911,8.10167)[bl]{$H_0$}
\FAProp(12.5,10.)(9.,10.)(0.,){/Straight}{-1}
\FALabel(10.75,8.93)[t]{$q_l$}
\FAProp(3.5,10.)(9.,10.)(0.8,){/Straight}{1}
\FALabel(6.25,6.73)[t]{$q_k$}
\FAProp(3.5,10.)(9.,10.)(-0.8,){/ScalarDash}{1}
\FALabel(6.25,13.27)[b]{$H_\alpha$}
\FAVert(3.5,10.){0}
\FAVert(12.5,10.){0}
\FAVert(9.,10.){0}

\FADiagram{5}
\FAProp(0.,10.)(6.5,10.)(0.,){/Straight}{1}
\FALabel(3.25,8.93)[t]{$t$}
\FAProp(20.,15.)(13.,14.)(0.,){/Straight}{-1}
\FALabel(16.2808,15.5544)[b]{$c$}
\FAProp(20.,5.)(13.,6.)(0.,){/ScalarDash}{0}
\FALabel(16.3162,4.69307)[t]{$H_0$}
\FAProp(6.5,10.)(13.,14.)(0.,){/ScalarDash}{1}
\FALabel(9.20801,13.1807)[br]{$H_\alpha$}
\FAProp(6.5,10.)(13.,6.)(0.,){/Straight}{1}
\FALabel(9.20801,6.81927)[tr]{$q_k$}
\FAProp(13.,14.)(13.,6.)(0.,){/Straight}{-1}
\FALabel(14.274,10.)[l]{$q_q$}
\FAVert(6.5,10.){0}
\FAVert(13.,14.){0}
\FAVert(13.,6.){0}

\FADiagram{7}
\FAProp(0.,10.)(6.5,10.)(0.,){/Straight}{1}
\FALabel(3.25,8.93)[t]{$t$}
\FAProp(20.,15.)(13.,14.)(0.,){/Straight}{-1}
\FALabel(16.2808,15.5544)[b]{$c$}
\FAProp(20.,5.)(13.,6.)(0.,){/ScalarDash}{0}
\FALabel(16.3162,4.69307)[t]{$H_0$}
\FAProp(6.5,10.)(13.,14.)(0.,){/Straight}{1}
\FALabel(9.20801,13.1807)[br]{$q_k$}
\FAProp(6.5,10.)(13.,6.)(0.,){/ScalarDash}{1}
\FALabel(9.20801,6.81927)[tr]{$H_\alpha$}
\FAProp(13.,14.)(13.,6.)(0.,){/ScalarDash}{-1}
\FALabel(14.274,10.)[l]{$H_\beta$}
\FAVert(6.5,10.){0}
\FAVert(13.,14.){0}
\FAVert(13.,6.){0}

\FADiagram{2}
\FAProp(0.,10.)(11.,10.)(0.,){/Straight}{1}
\FALabel(5.5,8.93)[t]{$t$}
\FAProp(20.,15.)(17.3,13.5)(0.,){/Straight}{-1}
\FALabel(18.3773,15.1249)[br]{$c$}
\FAProp(20.,5.)(11.,10.)(0.,){/ScalarDash}{0}
\FALabel(15.3487,6.8436)[tr]{$H_0$}
\FAProp(11.,10.)(13.7,11.5)(0.,){/Straight}{1}
\FALabel(12.0773,11.6249)[br]{$q_l$}
\FAProp(17.3,13.5)(13.7,11.5)(-0.8,){/Straight}{-1}
\FALabel(16.5727,10.1851)[tl]{$q_k$}
\FAProp(17.3,13.5)(13.7,11.5)(0.8,){/ScalarDash}{-1}
\FALabel(14.4273,14.8149)[br]{$H_\alpha$}
\FAVert(11.,10.){0}
\FAVert(17.3,13.5){0}
\FAVert(13.7,11.5){0}

\FADiagram{6}
\FAProp(0.,10.)(6.5,10.)(0.,){/Straight}{1}
\FALabel(3.25,8.93)[t]{$t$}
\FAProp(20.,15.)(13.,14.)(0.,){/Straight}{-1}
\FALabel(16.2808,15.5544)[b]{$c$}
\FAProp(20.,5.)(13.,6.)(0.,){/ScalarDash}{0}
\FALabel(16.3162,4.69307)[t]{$H_0$}
\FAProp(6.5,10.)(13.,14.)(0.,){/Sine}{1}
\FALabel(9.20801,13.1807)[br]{$V_\alpha$}
\FAProp(6.5,10.)(13.,6.)(0.,){/Straight}{1}
\FALabel(9.20801,6.81927)[tr]{$q_k$}
\FAProp(13.,14.)(13.,6.)(0.,){/Straight}{-1}
\FALabel(14.274,10.)[l]{$q_q$}
\FAVert(6.5,10.){0}
\FAVert(13.,14.){0}
\FAVert(13.,6.){0}

\FADiagram{8}
\FAProp(0.,10.)(6.5,10.)(0.,){/Straight}{1}
\FALabel(3.25,8.93)[t]{$t$}
\FAProp(20.,15.)(13.,14.)(0.,){/Straight}{-1}
\FALabel(16.2808,15.5544)[b]{$c$}
\FAProp(20.,5.)(13.,6.)(0.,){/ScalarDash}{0}
\FALabel(16.3162,4.69307)[t]{$H_0$}
\FAProp(6.5,10.)(13.,14.)(0.,){/Straight}{1}
\FALabel(9.20801,13.1807)[br]{$q_k$}
\FAProp(6.5,10.)(13.,6.)(0.,){/Sine}{1}
\FALabel(9.20801,6.81927)[tr]{$V_\alpha$}
\FAProp(13.,14.)(13.,6.)(0.,){/Sine}{-1}
\FALabel(14.274,10.)[l]{$V_\beta$}
\FAVert(6.5,10.){0}
\FAVert(13.,14.){0}
\FAVert(13.,6.){0}

\FADiagram{3}
\FAProp(0.,10.)(3.5,10.)(0.,){/Straight}{1}
\FALabel(1.75,8.93)[t]{$t$}
\FAProp(20.,15.)(12.5,10.)(0.,){/Straight}{-1}
\FALabel(15.8702,13.3097)[br]{$c$}
\FAProp(20.,5.)(12.5,10.)(0.,){/ScalarDash}{0}
\FALabel(16.4911,8.10167)[bl]{$H_0$}
\FAProp(12.5,10.)(9.,10.)(0.,){/Straight}{-1}
\FALabel(10.75,8.93)[t]{$q_l$}
\FAProp(3.5,10.)(9.,10.)(0.8,){/Straight}{1}
\FALabel(6.25,6.73)[t]{$q_k$}
\FAProp(3.5,10.)(9.,10.)(-0.8,){/Sine}{1}
\FALabel(6.25,13.27)[b]{$V_\alpha$}
\FAVert(3.5,10.){0}
\FAVert(12.5,10.){0}
\FAVert(9.,10.){0}

\FADiagram{}

\FADiagram{9}
\FAProp(0.,10.)(6.5,10.)(0.,){/Straight}{1}
\FALabel(3.25,8.93)[t]{$t$}
\FAProp(20.,15.)(13.,14.)(0.,){/Straight}{-1}
\FALabel(16.2808,15.5544)[b]{$c$}
\FAProp(20.,5.)(13.,6.)(0.,){/ScalarDash}{0}
\FALabel(16.3162,4.69307)[t]{$H_0$}
\FAProp(6.5,10.)(13.,14.)(0.,){/Straight}{1}
\FALabel(9.20801,13.1807)[br]{$q_k$}
\FAProp(6.5,10.)(13.,6.)(0.,){/ScalarDash}{1}
\FALabel(9.20801,6.81927)[tr]{$H_\alpha$}
\FAProp(13.,14.)(13.,6.)(0.,){/Sine}{-1}
\FALabel(14.274,10.)[l]{$V_\beta$}
\FAVert(6.5,10.){0}
\FAVert(13.,14.){0}
\FAVert(13.,6.){0}

\FADiagram{4}
\FAProp(0.,10.)(11.,10.)(0.,){/Straight}{1}
\FALabel(5.5,8.93)[t]{$t$}
\FAProp(20.,15.)(17.3,13.5)(0.,){/Straight}{-1}
\FALabel(18.3773,15.1249)[br]{$c$}
\FAProp(20.,5.)(11.,10.)(0.,){/ScalarDash}{0}
\FALabel(15.3487,6.8436)[tr]{$H_0$}
\FAProp(11.,10.)(13.7,11.5)(0.,){/Straight}{1}
\FALabel(12.0773,11.6249)[br]{$q_l$}
\FAProp(17.3,13.5)(13.7,11.5)(-0.8,){/Straight}{-1}
\FALabel(16.5727,10.1851)[tl]{$q_k$}
\FAProp(17.3,13.5)(13.7,11.5)(0.8,){/Sine}{-1}
\FALabel(14.4273,14.8149)[br]{$V_\alpha$}
\FAVert(11.,10.){0}
\FAVert(17.3,13.5){0}
\FAVert(13.7,11.5){0}

\FADiagram{}

\FADiagram{10}
\FAProp(0.,10.)(6.5,10.)(0.,){/Straight}{1}
\FALabel(3.25,8.93)[t]{$t$}
\FAProp(20.,15.)(13.,14.)(0.,){/Straight}{-1}
\FALabel(16.2808,15.5544)[b]{$c$}
\FAProp(20.,5.)(13.,6.)(0.,){/ScalarDash}{0}
\FALabel(16.3162,4.69307)[t]{$H_0$}
\FAProp(6.5,10.)(13.,14.)(0.,){/Straight}{1}
\FALabel(9.20801,13.1807)[br]{$q_k$}
\FAProp(6.5,10.)(13.,6.)(0.,){/Sine}{1}
\FALabel(9.20801,6.81927)[tr]{$V_\alpha$}
\FAProp(13.,14.)(13.,6.)(0.,){/ScalarDash}{-1}
\FALabel(14.274,10.)[l]{$H_\beta$}
\FAVert(6.5,10.){0}
\FAVert(13.,14.){0}
\FAVert(13.,6.){0}

\end{feynartspicture}

\caption{\label{fig: 1-loop diags}1-loop Feynman diagrams for $t\rightarrow cH^{0}$}
\end{figure}

The 1-loop decay amplitude is composed of 10 Feynman diagrams, shown
in Fig. \ref{fig: 1-loop diags}. Their explicit calculation is given
in App. \ref{app:1L diags}. 

The calculation was aimed to be model-independent. Thus, the Feynman
diagrams in Fig. \ref{fig: 1-loop diags} were drawn by assuming general
vertices, which are defined in Fig. \ref{fig:vertices definition}
in App. \ref{app:1L diags}, and by assuming general fields:

$q_{i}$-- denotes a quark (up or down type)

$V_{\alpha}$-- denotes vector (gauge) fields

$H_{\alpha}$-- denotes scalar fields

In this way it was possible to calculate the same process in different
models, by inserting the appropriate vertices and fields.

The 1-loop integrals were calculated numerically with FORTRAN using
the FF package \cite{ff vanold}.

The calculations were done using the Passarino-Veltman reduction scheme,
which expresses the integrals in terms of basic scalar n-point functions.
All other (vector, tensor) integrals can be computed using combinations
of the scalar functions (for explicit furmulae see e.g. \cite{bejar}
App. A). In App. \ref{app:dijcij definition} we give the definitions
of the reduced functions that we used to calculate the 1-loop integrals.

\section{\label{sub:calcs vals&bounds}Bounds on the parameter space of the
T2HDM}

As was mentioned in the introduction, bounds on the T2HDM charged-sector
parameter space were simultaneously calculated in \cite{soni 34 best}
to give a best fit to various experimental results. The processes
that were selected were the ones most sensitive to the T2HDM.

We describe here the processes for which the parameter fit was done
in \cite{soni 34 best}, and summarize in Eq. \eqref{eq:bounds} the
allowed values of the parameters of the T2HDM at $1\sigma$.

\begin{itemize}
\item The BR of $B\rightarrow X_{s}\gamma$ \cite{soni 19 2nd best} was
estimated from the shifts in the Wilson coefficients $C_{7,8}$ caused
by the T2HDM. The BR constrains mainly the parameters $m_{H^{+}}$,
$\tan\beta$ and $\xi$.
\item The BR of $B^{+}\rightarrow\tau^{+}\nu_{\tau}$ was calculated in
the T2HDM \cite{soni 34 best}. This process receives a tree-level
contribution from charged-Higgs exchange, and has a large impact in
constraining the parameter space, especially $m_{H^{+}}$ and $\xi'$.
\item The CP-violating parameter $\varepsilon_{K}$ was calculated in the
T2HDM \cite{soni 21 most calcs}, and was found in \cite{soni 34 best}
to severely constrain the parameters $m_{H^{+}}$, $\tan\beta$ and
$\xi$.
\item The time dependent amplitude of the CP asymmetry $a_{\Psi K}=A\left(\bar{B}^{0}\rightarrow J/\psi K_{s}\right)$,
was found to constrain the parameters $m_{H^{+}}$ and $\tan\beta$.
\item The neutron EDM constrains mainly $m_{H^{+}}$, $\tan\beta$ and $\xi'$.
\item The $\Delta m_{D}$ mass difference from $D-\bar{D}$ mixing can be
completely dominated by NP effects, and so it was required in the
analysis that the T2HDM value would not exceed the experimental value.
This requirement constrains $m_{H^{+}}$, $\tan\beta$, $\xi$ and
$\xi'$.
\item The ratio $\Delta m_{B_{s}}/\Delta m_{B_{d}}$ was included in the
fit, although it gave weaker constraints than $\varepsilon_{K}$.
\end{itemize}
In the analysis of \cite{soni 34 best}, a $\chi^{2}$ function was
defined, that featured as variables the T2HDM charged Higgs sector
parameters -- $\tan\beta$, $m_{H}^{+}$, $\xi$ and $\xi'$ and the
SM CKM-matrix parameters -- $\rho$, $\eta$ and $\left[\alpha,\beta,\gamma\right]$
(unitarity triangle). These parameters were simultaneously fit to
the processes described above. As this is the most comprehensive work
constraining the T2HDM parameters, these were the bounds used in the
present work. Since they are directly relevant to the present work,
we list the final results of \cite{soni 34 best} below (recall that
$\xi=\left|\xi\right|e^{i\varphi_{\xi}}$):

\begin{eqnarray}
m_{H^{\pm}} & = & \left(660_{-280}^{+390}\right)\mbox{ GeV},\nonumber \\
\tan\beta & = & 28_{-8}^{+44},\nonumber \\
\left|\xi\right| & \sim & 0.8,\quad0.5<\left|\xi\right|<1,\nonumber \\
\varphi_{\xi} & = & \left(110_{-65}^{+30}\right)^{\circ},\nonumber \\
\left|\xi'\right| & \sim & 0.21,\nonumber \\
\varphi_{\xi'} & \sim & 250^{\circ}.\label{eq:bounds}\end{eqnarray}

We note that these values are also allowed for the 2HDM-II as given
in \cite{arhrib}. We further note that the $\Delta m_{D}$ mass difference
from $D-\bar{D}$ mixing receives a contribution from the neutral
sector that can constrain the neutral sector parameters, but that
contribution is suppressed by a factor of $\left(\frac{1}{\tan\beta}\frac{m_{c}}{m_{t}}\frac{m_{H^{+}}}{m_{h^{0}}}\right)^{2}$
compared to the charged Higgs contribution, and therefore does not
impose significant constraints on the parameter space of the neutral
sector \cite{atwood reina soni bound neutral,soni 21 most calcs}.

Direct constraints from experiments impose weak bounds \cite{PDG}:
for the SM Higgs, the direct search bound is $m_{H^{0}}>114\mbox{ GeV}$.
For supersymmetry, bounds for neutral scalars are $m_{h}>90\mbox{ GeV}$,
while for charged scalars current bounds are $m_{H^{+}}>80\mbox{ GeV}$.

\section{\label{sub:calcs cross-sec}From amplitude to BR}

From the amplitude we get the width of the decay using \cite{PDG}:

\begin{equation}
\Gamma=4\pi\cdot\lambda^{\frac{1}{2}}\left(1,\frac{m_{c}^{2}}{m_{t}^{2}},\frac{m_{h}^{2}}{m_{t}^{2}}\right)\cdot\frac{\underset{pol}{\sum}\overline{\left|\mathcal{M}\right|^{2}}}{64\pi^{2}m_{t}}\mbox{ ,}\end{equation}

where $\underset{pol}{\sum}\overline{\left|\mathcal{M}\right|^{2}}$
is the squared amplitude averaged over initial polarizations and summed
over final polarizations, and $\lambda\left(x,y,z\right)=x^{2}+y^{2}+z^{2}-2xy-2xz-2yz$.
In the case of an incoming fermion, we have: $\underset{pol}{\sum}\overline{\left|\mathcal{M}\right|^{2}}=\frac{1}{2}\underset{pol}{\sum}\left|\mathcal{M}\right|^{2}$.

Calculating the squared amplitude, we get:

\begin{eqnarray}
\mathcal{M} & = & \frac{i\bar{u}_{c}}{16\pi^{2}}\left(M_{L}L+M_{R}R\right)u_{t},\nonumber \\
\underset{pol}{\sum}\left|\mathcal{M}\right|^{2} & = & \frac{1}{256\pi^{4}}tr\left[\bar{u}_{c}\left(M_{L}L+M_{R}R\right)u_{t}\bar{u}_{t}\left(M_{R}^{*}L+M_{L}^{*}R\right)u_{c}\right]=\\
 & = & \frac{1}{256\pi^{4}}\left[2m_{c}m_{t}\left(M_{L}M_{R}^{*}+M_{R}M_{L}^{*}\right)+\left(m_{c}^{2}+m_{t}^{2}-m_{h}^{2}\right)\left(M_{L}M_{L}^{*}+M_{R}M_{R}^{*}\right)\right].\nonumber \end{eqnarray}

The BR is then: \begin{equation}
BR\left(t\rightarrow ch\right)=\frac{\Gamma\left(t\rightarrow ch\right)}{\underset{x}{\sum}\Gamma\left(t\rightarrow x\right)}\mbox{ .}\end{equation}

Where usually only $x=W^{+}b$ is taken, since its BR is very close
to 1 \cite{PDG}. We take the tree-level value of $\Gamma\left(t\rightarrow W^{+}b\right)=1.55\mbox{ GeV}$.
If the mass of the $H^{+}$ is smaller than the mass of the top, the
process $t\rightarrow H^{+}b$ is possible and significant, and must
be taken into account in the sum.

\section{\label{sub:Higgs-decay-BR}Higgs decay BR}

The 1-loop amplitude of the opposite process, $h\rightarrow\bar{t}c+\bar{c}t$,
is identical to the 1-loop amplitude of $t\rightarrow ch$, by applying
crossing symmetry \cite{peskin}. The width of the process is given
by:

\begin{align}
\Gamma\left(h\rightarrow\bar{t}c+\bar{c}t\right)=2\times\Gamma\left(h\rightarrow\bar{t}c\right) & =2N_{c}\lambda^{\frac{1}{2}}\left(1,\frac{m_{c}^{2}}{m_{h}^{2}},\frac{m_{t}^{2}}{m_{h}^{2}}\right)\cdot\frac{\underset{pol}{\sum}\overline{\left|\mathcal{M}\right|^{2}}}{16\pi m_{h}},\end{align}
where $N_{c}=3$ is the color factor, and $\underset{pol}{\sum}\overline{\left|\mathcal{M}\right|^{2}}=\underset{pol}{\sum}\left|\mathcal{M}\right|^{2}$
in this case (an incoming scalar).

In order to calculate the BR for $h\rightarrow\bar{t}c$, one has
to calculate the total width of the scalars. In this work we have
included leading-order contributions to the Higgs width from $h\rightarrow\bar{q}q$,
$h\rightarrow VV$, $h\rightarrow\mbox{2 scalars}$ and $h\rightarrow\mbox{vector+scalar}$
\cite{djouadi II}. The last can be important in some regions of the
parameter space such as low $\tan\beta$. The formulae used for the
calculation of the total width are given in App. \ref{app:higgs width formulas},
along with a plot of the SM Higgs width in the leading order approximation
compared with higher order predictions, in Fig. \ref{fig:higgs width SM}.

\section{\label{sub:calcs checks}Checks of the calculations}

\begin{itemize}
\item We have successfully reproduced the results for $BR(t\rightarrow ch)$
obtained in \cite{t-ch SM} in the SM and in \cite{bejar} in the
2HDM-II. 
\item We have successfully reproduced the results for $BR\left(h\rightarrow\bar{t}c\right)$
obtained in \cite{h-tc SM arhrib} in the SM and in \cite{arhrib}
in the 2HDM-II. We note that Arhrib's results matched for $\alpha_{EW}\sim1/128.9$,
which is different from the one reportedly used, $\alpha_{EW}\sim1/137$,
as also confirmed by him in a private communication. However, we were
not able to reproduce the $BR\left(h\rightarrow\bar{t}c\right)$ values
of \cite{bejar}, as also stated in \cite{arhrib}.
\item Some amplitudes have a divergent part. Since the process is calculated
at leading order, no renormalization is needed to cancel the divergent
terms, and they should cancel among themselves. This cancellation
is demonstrated in App. \ref{app:cancel divergences} analytically.
It was also verified numerically in the FORTRAN code. Checking that
the results do not diverge is also a test of the self-consistency
of the 1-loop amplitude calculations.
\end{itemize}

\section{\label{sub:calcs Tree}Tree-level amplitude}

As stated in the introduction, there are FC tree-level interactions
of $h^{0}$ in the T2HDM, when $\alpha=\beta$, as opposed to $H^{0}$
which has no FC tree-level interactions when $\alpha=\beta$. Therefore,
the decays $t\rightarrow ch^{0}$ and $h^{0}\rightarrow\bar{t}c$
at tree-level are possible when allowed kinematically. The tree-level
coupling $h^{0}\bar{t}c$ was given in Sec. \ref{sec:FC sector T2HDM}.
Here we derive the leading order tree-level BR values of these two
decays. We will neglect throughout the derivation terms of order:
$m_{c}^{2}/m_{t}^{2},$ $m_{b}^{2}/m_{t}^{2},$ $m_{c}^{2}/m_{h^{0}}^{2}$
and $\cot^{2}\beta$. The last term is neglected in accordance with
the working assumption of the T2HDM, which is a large $\tan\beta$.
We also set $\alpha=\beta$. The last requirement is not imperative,
but it renders simpler formulae and makes the following derivation
consistent with the 1-loop calculations of $t\rightarrow cH^{0}$
and $H^{0}\rightarrow\bar{t}c$ decays in this work.

The tree-level amplitude for the process $t\rightarrow ch^{0}$ is:

\begin{align}
\mathcal{M}\left(t\rightarrow ch^{0}\right) & =\bar{u}_{c}\left[\frac{g}{2m_{W}}\left(\left(M_{u}\right)_{ct}\frac{\sin\alpha}{\cos\beta}-\Sigma_{ct}\left(\frac{\cos\alpha}{\sin\beta}+\frac{\sin\alpha}{\cos\beta}\right)\right)R+\left(h.c.\right)L\right]u_{t},\end{align}
where from \eqref{eq:sigma} we have:\begin{align}
\Sigma_{ct} & =m_{t}\epsilon_{ct}^{2}\xi^{*}\sqrt{1-\left|\epsilon_{ct}\xi\right|^{2}}\sqrt{1-\left|\epsilon_{ct}\xi'\right|^{2}}\sim m_{c}\epsilon_{ct}\xi^{*},\nonumber \\
\left(\Sigma^{\dagger}\right)_{ct} & =m_{t}\epsilon_{ct}\xi^{*}\sqrt{1-\left|\epsilon_{ct}\xi\right|^{2}}\sqrt{1-\left|\epsilon_{ct}\xi'\right|^{2}}\sim m_{c}\xi^{*},\end{align}
which we insert in the amplitude with $\alpha=\beta$ to get:\begin{align}
\mathcal{M}\left(t\rightarrow ch^{0}\right) & =\bar{u}_{c}\frac{-g}{2m_{W}}\left(\cot\beta+\tan\beta\right)m_{c}\xi^{*}\left[\epsilon_{ct}R+L\right]_{ct}u_{t}\equiv\nonumber \\
 & \equiv\bar{u}_{c}\left[M_{R}R+M_{L}L\right]u_{t}.\end{align}

The squared amplitude summed over external spinors is:

\begin{align}
\underset{pol}{\sum}\left|\mathcal{M}\right|^{2} & =2m_{c}m_{t}\left(M_{L}M_{R}^{*}+M_{R}M_{L}^{*}\right)+\left(m_{t}^{2}+m_{c}^{2}-m_{h}^{2}\right)\left(M_{L}M_{L}^{*}+M_{R}M_{R}^{*}\right)=\nonumber \\
 & =\left[\frac{g^{2}m_{c}^{2}}{4m_{W}^{2}}\left(\cot\beta+\tan\beta\right)^{2}\left|\xi\right|^{2}\right]\left[2m_{c}m_{t}\cdot2\epsilon_{ct}+\left(m_{t}^{2}+m_{c}^{2}-m_{h^{0}}^{2}\right)\left(1+\epsilon_{ct}^{2}\right)\right]\sim\nonumber \\
 & \sim\frac{g^{2}m_{c}^{2}}{4m_{W}^{2}}\tan^{2}\beta\left|\xi\right|^{2}\left[m_{t}^{2}-m_{h^{0}}^{2}\right].\end{align}

From $\underset{pol}{\sum}\overline{\left|\mathcal{M}\right|^{2}}=\frac{1}{2}\underset{pol}{\sum}\left|\mathcal{M}\right|^{2}$
we can calculate the BR's of the processes $t\rightarrow ch^{0}$
and $h^{0}\rightarrow\bar{t}c$.

The width of $t\rightarrow ch^{0}$ reads:

\begin{align}
\Gamma\left(t\rightarrow ch^{0}\right) & =4\pi\cdot\lambda^{\frac{1}{2}}\left(1,\frac{m_{c}^{2}}{m_{t}^{2}},\frac{m_{h^{0}}^{2}}{m_{t}^{2}}\right)\cdot\frac{\underset{pol}{\sum}\overline{\left|\mathcal{M}\right|^{2}}}{64\pi^{2}m_{t}}\sim\nonumber \\
 & \sim\frac{g^{2}\left|\xi\right|^{2}m_{t}m_{c}^{2}}{128\pi m_{W}^{2}}\tan^{2}\beta\left(1-\frac{m_{h^{0}}^{2}}{m_{t}^{2}}\right)^{2}.\end{align}

The width for $t\rightarrow bW^{+}$ (at tree-level and neglecting
terms of order $m_{b}^{2}/m_{t}^{2}$ and $\alpha_{s}$) is \cite{PDG}:\begin{align}
\Gamma\left(t\rightarrow bW^{+}\right) & \sim\frac{g^{2}m_{t}}{64\pi}\left(1-\frac{m_{W}^{2}}{m_{t}^{2}}\right)\left(1-2\frac{m_{W}^{2}}{m_{t}^{2}}+\frac{m_{t}^{2}}{m_{W}^{2}}\right),\end{align}

from which we get the leading order $BR\left(t\rightarrow ch^{0}\right)$
(for large $\tan\beta$):\begin{align}
BR\left(t\rightarrow ch^{0}\right) & \sim\frac{\left|\xi\right|^{2}m_{c}^{2}}{2m_{W}^{2}}\tan^{2}\beta\left(1-\frac{m_{h^{0}}^{2}}{m_{t}^{2}}\right)^{2}\left(1-\frac{m_{W}^{2}}{m_{t}^{2}}\right)^{-1}\left(1-2\frac{m_{W}^{2}}{m_{t}^{2}}+\frac{m_{t}^{2}}{m_{W}^{2}}\right)^{-1}.\end{align}

For instance, for the best-fit parameters of Eq. \eqref{eq:bounds},
$\tan\beta=28$, $\left|\xi\right|=0.8$, and for $\alpha=\beta$,
and $m_{h^{0}}=91\mbox{ GeV}$, we get $BR\left(t\rightarrow ch^{0}\right)=0.0077$.

By applying crossing symmetry on $\left|\mathcal{M}\right|^{2}$,
the tree-level squared amplitude of the opposite process, $h^{0}\rightarrow\bar{t}c$,
is:

\begin{align}
\underset{pol}{\sum}\overline{\left|\mathcal{M}\right|^{2}} & \sim\frac{g^{2}m_{c}^{2}}{4m_{W}^{2}}\tan^{2}\beta\left|\xi\right|^{2}\left[m_{h^{0}}^{2}-m_{t}^{2}\right],\end{align}

from which we get:\begin{align}
\Gamma\left(h\rightarrow\bar{t}c+\bar{c}t\right)=2\times\Gamma\left(h^{0}\rightarrow\bar{t}c\right) & =2N_{c}\lambda^{\frac{1}{2}}\left(1,\frac{m_{c}^{2}}{m_{h^{0}}^{2}},\frac{m_{t}^{2}}{m_{h^{0}}^{2}}\right)\cdot\frac{\underset{pol}{\sum}\overline{\left|\mathcal{M}\right|^{2}}}{16\pi m_{h^{0}}}\sim\nonumber \\
 & \sim\frac{N_{c}\left|\xi\right|^{2}g^{2}m_{h^{0}}m_{c}^{2}}{32\pi m_{W}^{2}}\tan^{2}\beta\left(1-\frac{m_{t}^{2}}{m_{h^{0}}^{2}}\right)^{2}.\end{align}

For $\alpha=\beta$, (assuming also that $m_{h^{0}}<2m_{A^{0}},2m_{H^{+}}$)
the total width of $h^{0}$ is mainly comprised of fermion decays,
since the couplings $W^{+}W^{-}h^{0}$, $Z^{0}Z^{0}h^{0}$ and $H^{0}H^{0}h^{0}$
are all $\propto\sin\left(\beta-\alpha\right)$ (see table \ref{tab:vvh feyn rules}
and App. \ref{app:higgs width formulas}). Below the $t\bar{t}$ threshold
(at about $340\mbox{ GeV}$) the $b\bar{b}$ decays dominate. The
width of $h^{0}\rightarrow\bar{b}b$ is then (see App. \ref{app:higgs width formulas}):

\begin{align}
\Gamma\left(h^{0}\rightarrow\bar{b}b\right) & \sim\frac{N_{c}g^{2}m_{b}^{2}m_{h^{0}}}{32\pi m_{W}^{2}}\tan^{2}\beta.\end{align}

In this case, the BR of $h^{0}\rightarrow\bar{t}c$ is: \begin{align}
BR\left(h^{0}\rightarrow\bar{t}c+\bar{c}t\right) & \sim\left|\xi\right|^{2}\frac{m_{c}^{2}}{m_{b}^{2}}\left(1-\frac{m_{t}^{2}}{m_{h^{0}}^{2}}\right)^{2}.\end{align}

For $\left|\xi\right|=0.8$ and $m_{h^{0}}=300\mbox{ GeV}$, we get
$BR\left(h^{0}\rightarrow\bar{t}c+\bar{c}t\right)=0.023$ .

\chapter{\label{sec:Results}Results}

In this section we give our results and discussion for the 1-loop
decays $BR\left(t\rightarrow cH^{0}\right)$ and $BR\left(H^{0}\rightarrow\bar{t}c\right)$
in the T2HDM. All the masses are in units of GeV. We set $\alpha=\beta$
for the reasons explained above. Other parameters are set to their
best-fit value of \eqref{eq:bounds} unless stated otherwise. Our
calculations were done in the t'Hooft Feynman gauge. The SM Higgs
has a best-fit mass to EW precision data of $91_{-32}^{+45}\mbox{ GeV}$
\cite{PDG}. In our setup in which $H^{0}$ has couplings identical
to the SM Higgs, we expect these bounds to be roughly applicable,
and therefore we set $m_{H^{0}}=91\mbox{ GeV}$ in the process $t\rightarrow cH^{0}$.
We take the total $t$-quark width $\Gamma\left(t\rightarrow W^{+}b\right)=1.55\mbox{ GeV}$.
For the process $H^{0}\rightarrow\bar{t}c$ we arbitrarily choose
$m_{H^{0}}=300\mbox{ GeV}$.

For definiteness, other values we used for the calculations were \cite{PDG}:
$m_{t}=172.5\mbox{ GeV}$ (pole mass), $m_{c}=\overline{m}_{c}\left(\overline{m}_{c}\right)=1.24\mbox{ GeV}$,
$m_{b}=\overline{m}_{b}\left(\overline{m}_{b}\right)=4.20\mbox{ GeV}$
($m_{c}$ and $m_{b}$ are in the $\overline{MS}$ renormalization
scheme), $m_{W}=80.40\mbox{ GeV}$, $m_{Z}=91.188\mbox{ GeV}$, $\cos\theta_{W}=m_{W}/m_{Z}$,
$\alpha_{EW}\left(m_{z}\right)=1/127.9$. The values were used as
given here, without running in energy scale. The BR results were found
to be sensitive to $m_{c}$ and $m_{b}$: for example, for the BR
value $5.99\times10^{-5}$ quoted in table \ref{tab:t-ch results compare}
in the upper row for the T2HDM with $m_{b}=4.2$ and $m_{c}=1.25$,
setting $m_{b}(m_{Z})\sim3$ and $m_{c}(m_{Z})\sim0.7$ \cite{running masses 2000}
yields $1.28\times10^{-5}$. We note that our results were not sensitive
to $m_{s}$.

We used $m_{A^{0}}=1000\mbox{ GeV}$ to enhance the triple-scalar
coupling which is roughly $\propto m_{A^{0}}^{2}$, as can be seen
in App. \ref{app:Higgs-potential} and \ref{app:Feynman-rules}. The
$\xi$ and $\xi'$ parameters are set to their best-fit value of Eq.
\eqref{eq:bounds}: $\left|\xi\right|=0.8$, $\varphi_{\xi}=110^{\circ}$,
$\left|\xi'\right|=0.21$, and $\varphi_{\xi'}=250^{\circ}$.

\section{\label{sub:res t-ch}Results for the 1-loop top rare decay $t\rightarrow cH^{0}$}

\begin{figure}
\begin{tabular}{cc}
\includegraphics[width=11cm,keepaspectratio]{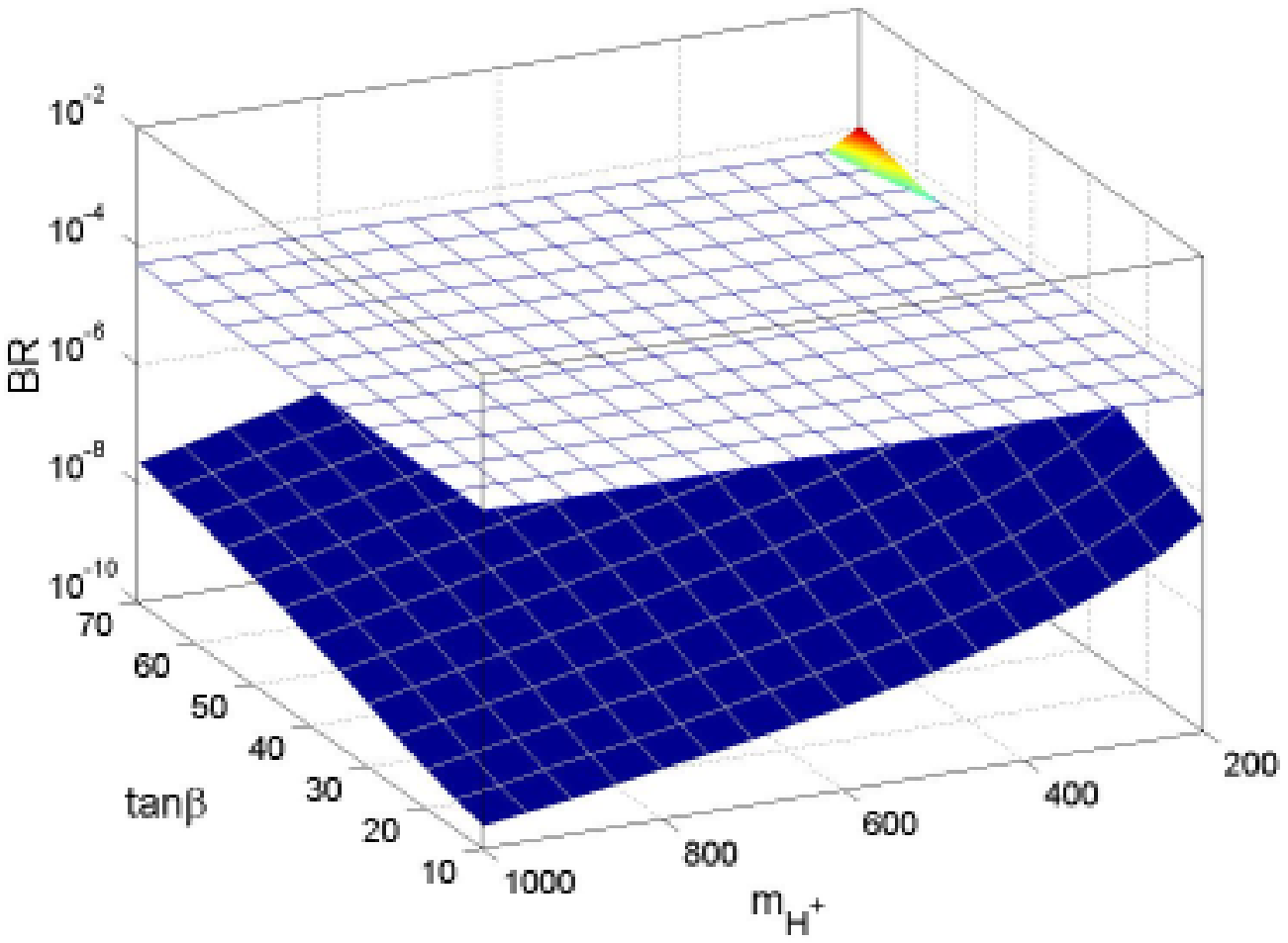}&
\begin{tabular}{c}
\vspace{-10cm}
\tabularnewline
\begin{feynartspicture}(150,150)(1,1)
\FADiagram{}
\FAProp(0.,10.)(6.5,10.)(0.,){/Straight}{1}
\FALabel(3.25,8.93)[t]{$t$}
\FAProp(20.,15.)(13.,14.)(0.,){/Straight}{-1}
\FALabel(16.2808,15.5544)[b]{$c$}
\FAProp(20.,5.)(13.,6.)(0.,){/ScalarDash}{0}
\FALabel(16.3162,4.69307)[t]{$H^0$}
\FAProp(6.5,10.)(13.,14.)(0.,){/Straight}{1}
\FALabel(9.20801,13.1807)[br]{$b$}
\FAProp(6.5,10.)(13.,6.)(0.,){/ScalarDash}{1}
\FALabel(9.20801,6.81927)[tr]{$H^+$}
\FAProp(13.,14.)(13.,6.)(0.,){/ScalarDash}{-1}
\FALabel(14.274,10.)[l]{$H^+$}
\FAVert(6.5,10.){0}
\FAVert(13.,14.){0}
\FAVert(13.,6.){0}
\end{feynartspicture}\tabularnewline
\end{tabular}\tabularnewline
(a)&
(b)\tabularnewline
\end{tabular}

\caption{\label{fig:t-ch 3D mH+:tanb}(a) 3D plot of $BR\left(t\rightarrow cH^{0}\right)$
in the $m_{H^{+}}-\tan\beta$ plane in the T2HDM, and (b) the dominant
diagram. We set $m_{h^{0}}=1000\mbox{ GeV}$ and $m_{A^{0}}=1200\mbox{ GeV}$.
The color scale represents the BR: the blue represents the lowest
BR and red the highest.}
\end{figure}

\begin{figure}
\begin{tabular}{cc}
\includegraphics[width=11cm]{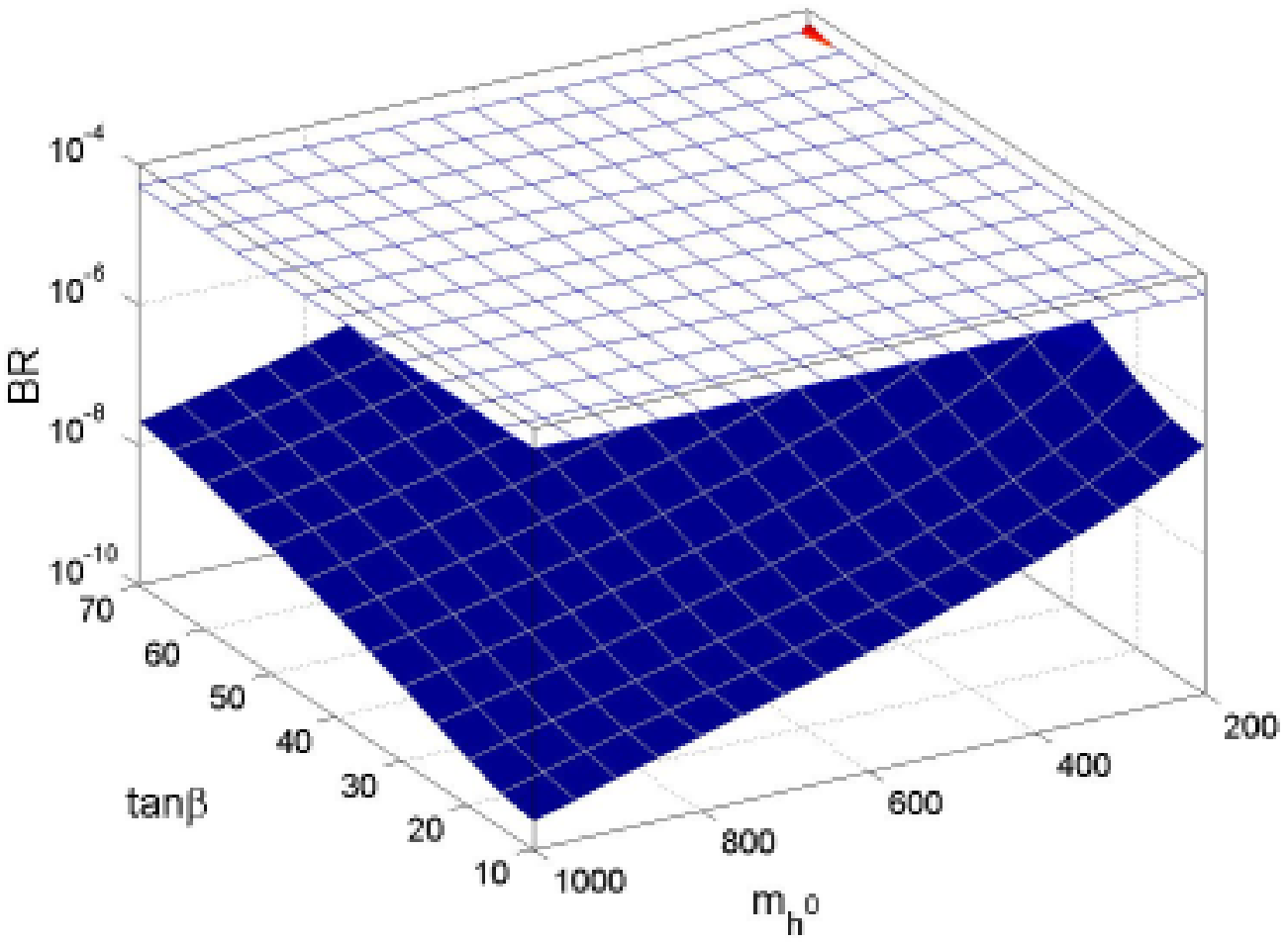}&
\begin{tabular}{c}
\vspace{-10cm}
\tabularnewline
\begin{feynartspicture}(150,150)(1,1)
\FADiagram{}
\FAProp(0.,10.)(6.5,10.)(0.,){/Straight}{1}
\FALabel(3.25,8.93)[t]{$t$}
\FAProp(20.,15.)(13.,14.)(0.,){/Straight}{-1}
\FALabel(16.2808,15.5544)[b]{$c$}
\FAProp(20.,5.)(13.,6.)(0.,){/ScalarDash}{0}
\FALabel(16.3162,4.69307)[t]{$H^0$}
\FAProp(6.5,10.)(13.,14.)(0.,){/Straight}{1}
\FALabel(9.20801,13.1807)[br]{$t$}
\FAProp(6.5,10.)(13.,6.)(0.,){/ScalarDash}{1}
\FALabel(9.20801,6.81927)[tr]{$h^0$}
\FAProp(13.,14.)(13.,6.)(0.,){/ScalarDash}{-1}
\FALabel(14.274,10.)[l]{$h^0$}
\FAVert(6.5,10.){0}
\FAVert(13.,14.){0}
\FAVert(13.,6.){0}
\end{feynartspicture}\tabularnewline
\end{tabular}\tabularnewline
(a)&
(b)\tabularnewline
\end{tabular}

\caption{\label{fig:t-ch 3D tanb:mh0}(a) 3D plot of $BR\left(t\rightarrow cH^{0}\right)$
in the $m_{h^{0}}-\tan\beta$ plane in the T2HDM, and (b) the dominant
diagram. We set $m_{H^{+}}=1000\mbox{ GeV}$ and $m_{A^{0}}=1200\mbox{ GeV}$.}
\end{figure}

In Fig. (\ref{fig:t-ch 3D mH+:tanb} a) we show a 3D plot of $BR\left(t\rightarrow cH^{0}\right)$
in the $m_{H^{+}}-\tan\beta$ plane in the T2HDM. The flat grid in
Figs. \ref{fig:t-ch 3D mH+:tanb} and \ref{fig:t-ch 3D tanb:mh0}
is the LHC detection limit of $BR>5\cdot10^{-5}$, so that the colored
surface above the grid is the region in the parameter space which
has (in the T2HDM) a BR high enough to be detected at the LHC. 

The choice $m_{h^{0}}=1000\mbox{ GeV}$ in Fig. \ref{fig:t-ch 3D mH+:tanb}
suppresses the diagrams in which the neutral $h^{0}$ Higgs runs in
the loop and, thus, better explores the charged Higgs sector properties.
As expected the BR rises with $\tan\beta$ and is highest when $m_{H^{+}}$
is lowest. The dominant Feynman diagram in this case is the one which
has two $H^{+}$ scalars and a $b$ quark in the loop, and is shown
in Fig. (\ref{fig:t-ch 3D mH+:tanb} b). This diagram receives an
enhancement from the 3-scalar vertex, as discussed above.

In Fig. \ref{fig:t-ch 3D tanb:mh0} we show the BR in the $m_{h^{0}}-\tan\beta$
plane in the T2HDM.

We took $m_{H^{+}}=1000\mbox{ GeV}$ in Fig. \ref{fig:t-ch 3D tanb:mh0}
so that the diagrams in which the charged $H^{+}$ Higgs runs in the
loop will be suppressed, to better explore the neutral Higgs sector
properties. As we can see, the BR is highest when $m_{h^{0}}$ is
lowest, and rises with $\tan\beta$. The dominant diagrams in this
case are the ones which have two $h^{0}$ or two $H^{+}$ scalars
in the loop: The diagram with two $h^{0}$ dominates in the low $\tan\beta$
-- low $m_{h^{0}}$ region, while the diagram with two $H^{+}$ dominates
in the high $\tan\beta$ -- high $m_{h^{0}}$ region, and is responsible
for the rise of the BR with $\tan\beta$. Both of these diagrams receive
an enhancement from the 3-scalar vertex with large $m_{A^{0}}$, as
mentioned above.

The plots are similar, yet the BR are higher when the $H^{+}$ runs
in the loop. That is a distinctive property of the T2HDM: the charged
Higgs coupling $H^{+}\bar{b}c$ can be enhanced by as much as $V_{tb}/V_{cb}$
compared with any other 2HDM.

\begin{figure}
\begin{centering}
\includegraphics[width=11cm]{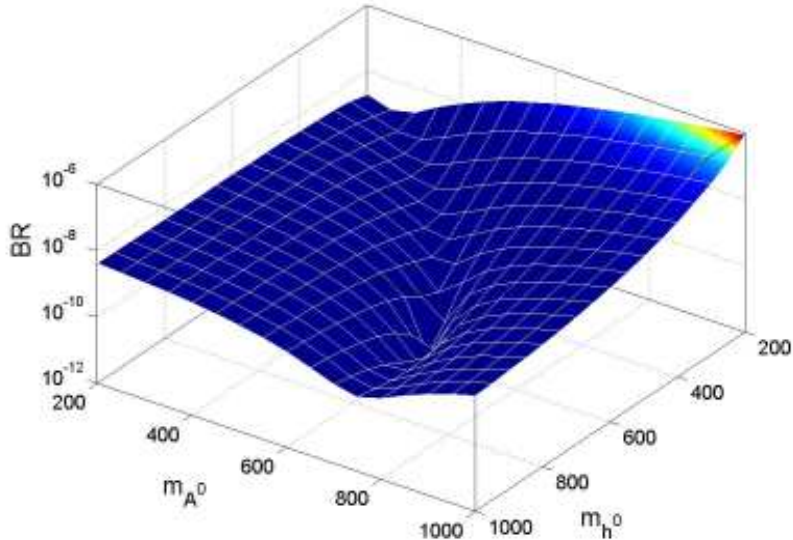}
\par\end{centering}

\caption{\label{fig:t-ch 3D mA0:mh0}3D plot of $BR\left(t\rightarrow cH^{0}\right)$
in the $m_{A^{0}}-m_{h^{0}}$ plane in the T2HDM. We set $m_{H^{+}}=660\mbox{ GeV}$
and $\tan\beta=28$.}
\end{figure}

In Fig. \ref{fig:t-ch 3D mA0:mh0} we show the BR in the $m_{A^{0}}-m_{h^{0}}$
plane in the T2HDM, at the best fit of the charged sector parameters.
The graph shows that the BR rises when $m_{A^{0}}$ is highest and
$m_{h^{0}}$ lowest, since then the diagram with two $m_{h^{0}}$
starts to dominate. The dip in the middle of the surface is due to
cancellation in the $H^{0}H^{+}H^{+}$ vertex.

\begin{figure}
\begin{centering}
\includegraphics[keepaspectratio]{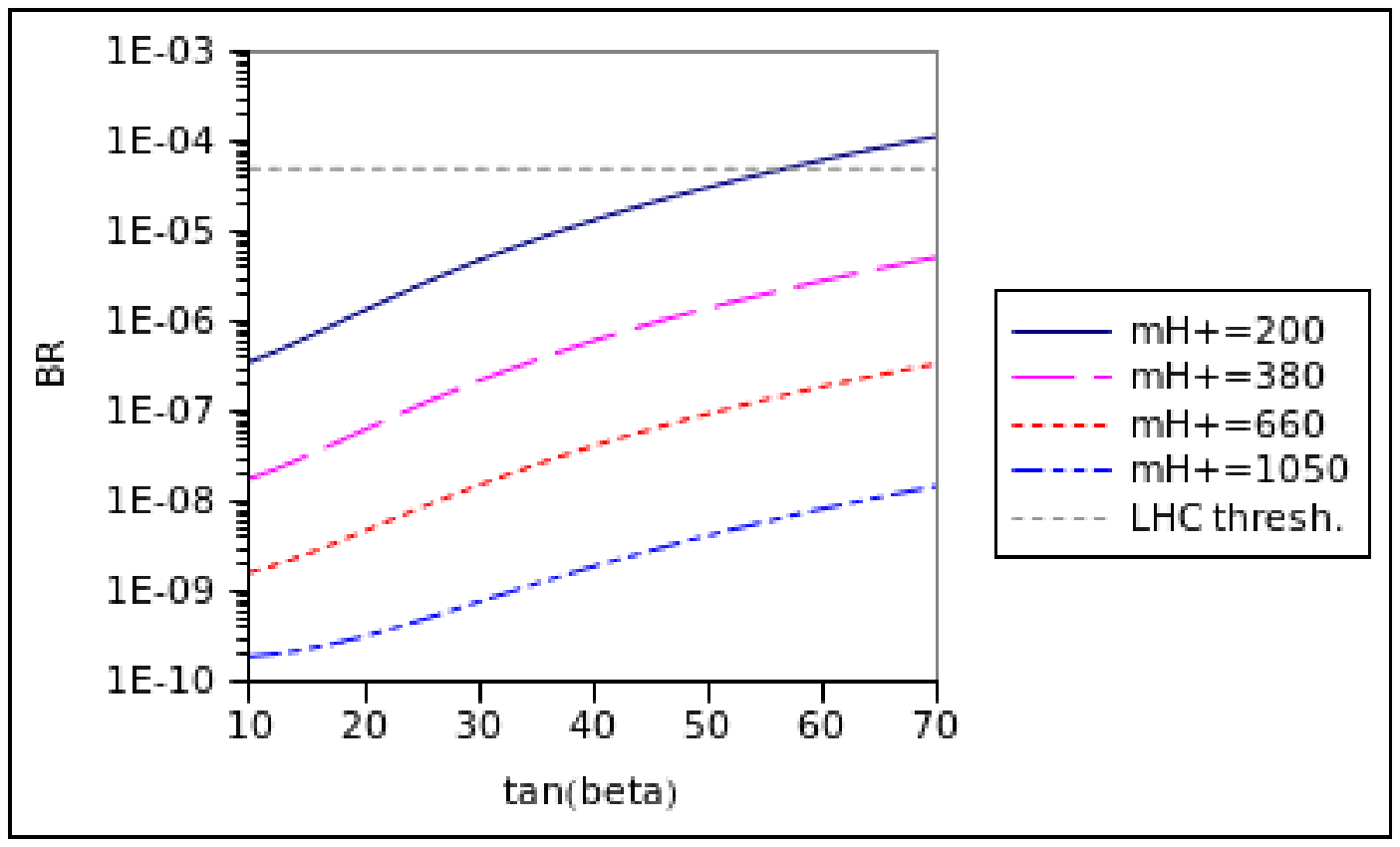}
\par\end{centering}

\caption{\label{fig:t-ch tanb}The $BR\left(t\rightarrow cH^{0}\right)$ as
a function of $\tan\beta$ at various $m_{H^{+}}$ in the T2HDM. We
set $m_{h^{0}}=1000\mbox{ GeV}$ and $m_{A^{0}}=1200\mbox{ GeV}$.
{}``LHC thresh.'' stands for the limit of the LHC sensitivity at
100 $fb^{-1}$.}
\end{figure}

\begin{figure}
\begin{centering}
\includegraphics[keepaspectratio]{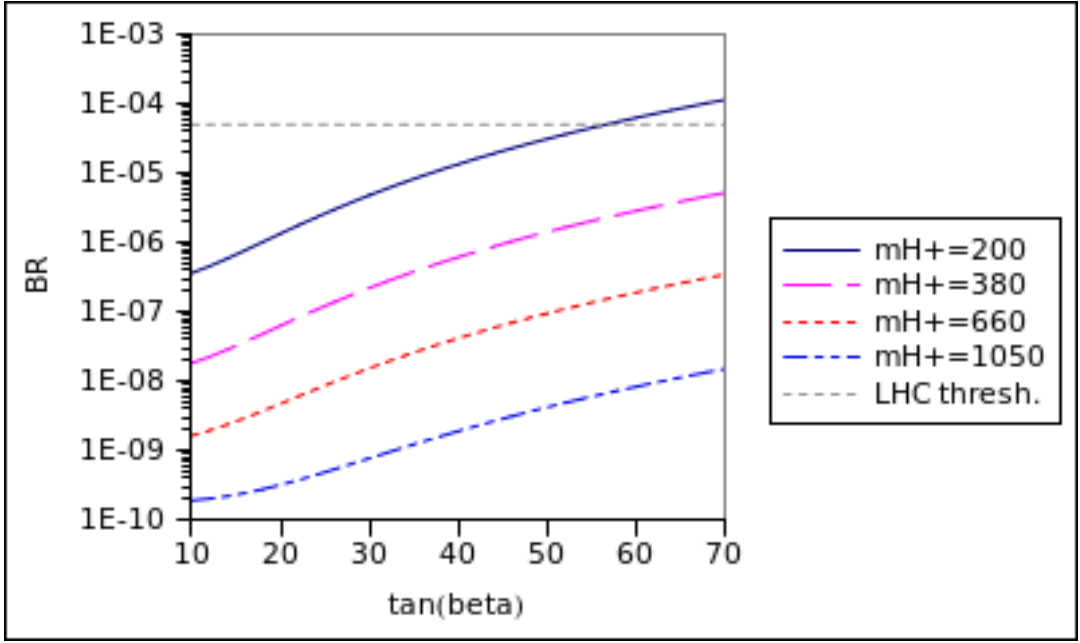}
\par\end{centering}

\caption{\label{fig:t-ch mH+}The $BR\left(t\rightarrow cH^{0}\right)$ as
a function of $m_{H^{+}}$ at various $\tan\beta$ in the T2HDM. We
set $m_{h^{0}}=1000\mbox{ GeV}$ and $m_{A^{0}}=1200\mbox{ GeV}$.
{}``LHC thresh.'' stands for the limit of the LHC sensitivity at
100 $fb^{-1}$.}
\end{figure}

In Figs. \ref{fig:t-ch tanb} and \ref{fig:t-ch mH+} we give 2D plots
of the BR as a function of $\tan\beta$ and $m_{H^{+}}$ respectively,
with the same parameters as in Fig. \ref{fig:t-ch 3D mH+:tanb}. We
can now see the dependence of the BR on $\tan\beta$ and $m_{H^{+}}$
more clearly: the BR rises with $\tan\beta$ but increases with smaller
$m_{H^{+}}$.

Finally, we wish to illustrate more clearly the difference between
the T2HDM, the 2HDM-II, and the SM. For that purpose we give in table
\ref{tab:t-ch results compare} the $BR\left(t\rightarrow cH^{0}\right)$
values within these 3 different models, for several points in the
relevant parameter space. We recall that the 2HDM-II has a Yukawa
potential similar to the MSSM, and has no tree-level FC interactions.

\begin{table}
\begin{tabular}{|c|c|c|c|}
\hline 
parameters&
SM&
2HDM-II&
T2HDM\tabularnewline
\hline
\hline 
$m_{h^{0}}=800$, $m_{A^{0}}=1000$, $\tan\beta=72$, $m_{H^{+}}=200$&
$\vphantom{\begin{array}{c}
a\\
b\end{array}}6.03\times10^{-14}$&
$4.25\times10^{-5}$&
$5.99\times10^{-5}$\tabularnewline
\hline 
$m_{h^{0}}=800$, $m_{A^{0}}=1000$, $\tan\beta=72$, $m_{H^{+}}=380$&
$\vphantom{\begin{array}{c}
a\\
b\end{array}}6.03\times10^{-14}$&
$1.79\times10^{-6}$&
$2.57\times10^{-6}$\tabularnewline
\hline 
$m_{h^{0}}=200$, $m_{A^{0}}=4000$, $\tan\beta=20$, $m_{H^{+}}=1050$&
$\vphantom{\begin{array}{c}
a\\
b\end{array}}6.03\times10^{-14}$&
$5.15\times10^{-8}$&
$9.39\times10^{-5}$\tabularnewline
\hline 
$m_{h^{0}}=200$, $m_{A^{0}}=1000$, $\tan\beta=20$, $m_{H^{+}}=1050$&
$\vphantom{\begin{array}{c}
a\\
b\end{array}}6.03\times10^{-14}$&
$3.34\times10^{-12}$&
$3.14\times10^{-7}$\tabularnewline
\hline
\end{tabular}

\caption{\label{tab:t-ch results compare}Comparison of the $BR\left(t\rightarrow cH^{0}\right)$
within the T2HDM, the 2HDM-II, and the SM. Masses are in units of
GeV.}
\end{table}

The first two rows illustrate the impact of the charged sector, by
setting a high $m_{h^{0}}$. The BR is a bit higher in the T2HDM than
in the 2HDM-II. We note that the value $m_{H^{+}}=200\mbox{ GeV}$
is outside the $1\sigma$ bounds. We can see that, for a high $m_{h^{0}}$,
the $BR\left(t\rightarrow cH^{0}\right)$ in the T2HDM is not as enhanced
as expected relative to the 2HDM-II. We expected that the diagram
with two charged scalar and a $b$ quark will be enhanced in the T2HDM,
since the $H^{+}\bar{c}b$ interaction is enhanced. We recall the
Feynman rule for this interaction:

\begin{align}
 & \frac{g}{\sqrt{2}m_{W}}\left[\tan\beta V_{cb}m_{b}R+m_{c}\left(-\tan\beta V_{cb}+\xi^{*}\left(\tan\beta+\cot\beta\right)V_{tb}\right)L\right],\end{align}

and the amplitude of this diagram:

\begin{align}
M_{7} & =\frac{-i\bar{u}_{c}}{16\pi^{2}}g_{H^{+}H^{+}h}^{3h}\left[m_{b}C_{0}\left(A_{cb}^{H^{+}}B_{tb}^{H^{+}*}L+B_{cb}^{H^{+}}A_{tb}^{H^{+}*}R\right)-m_{c}C_{12}\left(B_{cb}^{H^{+}}B_{tb}^{H^{+}*}L+A_{cb}^{H^{+}}A_{tb}^{H^{+}*}R\right)+\right.\nonumber \\
 & \left.\qquad\qquad+m_{t}\left(-C_{11}+C_{12}\right)\left(A_{cb}^{H^{+}}A_{tb}^{H^{+}*}L+B_{cb}^{H^{+}}B_{tb}^{H^{+}*}R\right)\right]u_{t},\label{eq:amp7 H+H+b}\end{align}
where $C_{ij}$ are the Passarino-Veltman scalar functions (see App.
\ref{app:dijcij definition}). 

The term multiplied by the left projection operator is enhanced. In
our notations it is denoted as $A_{cb}^{H^{+}}$ in \eqref{eq:amp7 H+H+b}.
The leading term in the amplitude in the 2HDM-II, for $\tan\beta\gtrsim10$,
is $\propto m_{t}B_{cb}^{H^{+}}B_{tb}^{H^{+}*}\propto m_{t}m_{b}^{2}\tan^{2}\beta V_{cb}V_{tb}^{*}$
(see Sec. \ref{sec:FC sector T2HDM}). In the T2HDM this term does
not receive a significant enhancement. However, the term $\propto m_{b}A_{cb}^{H^{+}}B_{tb}^{H^{+}*}$,
which in the 2HDM-II is a sub-leading term, is in the T2HDM $\propto\xi^{*}m_{c}m_{b}^{2}\tan^{2}\beta V_{tb}V_{tb}^{*}$,
and is of the same order of magnitude as the leading term of the 2HDM-II
(together with the $C_{ij}$ loop functions). Thus we can summarize
that what would have been a sub-leading term in the 2HDM-II, becomes
in the T2HDM of the same order of magnitude of the leading term, and
therefore the enhancement is not as significant as expected.

In the last two rows of the table we see the impact of the neutral
Higgs sector, by setting a high $m_{H^{+}}$. The results are much
higher in the T2HDM than in the 2HDM-II, which is to be expected,
since the 2HDM-II does not have any tree-level FC interactions. However,
the overall BR's in the small $m_{h^{0}}$ regime are also small due
to the cancellation of the leading term in the $h^{0}\bar{t}t$ vertex.

To complete the picture, we note that the SM value is only dependent
upon the neutral Higgs mass. Setting $m_{H^{0}}=91$ GeV, we get:
$BR_{SM}\left(t\rightarrow cH^{0}\right)=6.03\times10^{-14}$, for
$m_{H^{0}}=100$ GeV we get $BR_{SM}\left(t\rightarrow cH^{0}\right)=4.63\times10^{-14}$,
and for $m_{H^{0}}=150$ GeV we get $BR_{SM}\left(t\rightarrow cH^{0}\right)=5.26\times10^{-15}$.

\section{\label{sub:res h-tc}Results for the 1-loop Higgs rare decay $H^{0}\rightarrow\bar{t}c$}

We recall the values that we use in the following plots, which were
also given above: $m_{H^{0}}=300\mbox{ GeV}$ (chosen arbitrarily),
$m_{t}=172.5\mbox{ GeV}$,  $m_{c}=1.24\mbox{ GeV}$, $m_{b}=4.20\mbox{ GeV}$,
$m_{W}=80.40\mbox{ GeV}$, $m_{Z}=91.188\mbox{ GeV}$, $\alpha_{EW}\left(m_{z}\right)=1/127.9$,
and $\alpha=\beta$. The $\xi$ and $\xi'$ parameters are set to
their best-fit value of Eq. \eqref{eq:bounds}: $\left|\xi\right|=0.8$,
$\varphi_{\xi}=110^{\circ}$, $\left|\xi'\right|=0.21$, and $\varphi_{\xi'}=250^{\circ}$.
The total Higgs width is calculated (see Sec. \ref{sub:Higgs-decay-BR})
from the decays $H^{0}\rightarrow\bar{q}q$, $H^{0}\rightarrow VV$,
$H^{0}\rightarrow h_{i}h_{j}$ and $H^{0}\rightarrow V_{i}h_{j}$,
as defined in App. \ref{app:higgs width formulas}.

\begin{figure}
\begin{centering}
\includegraphics{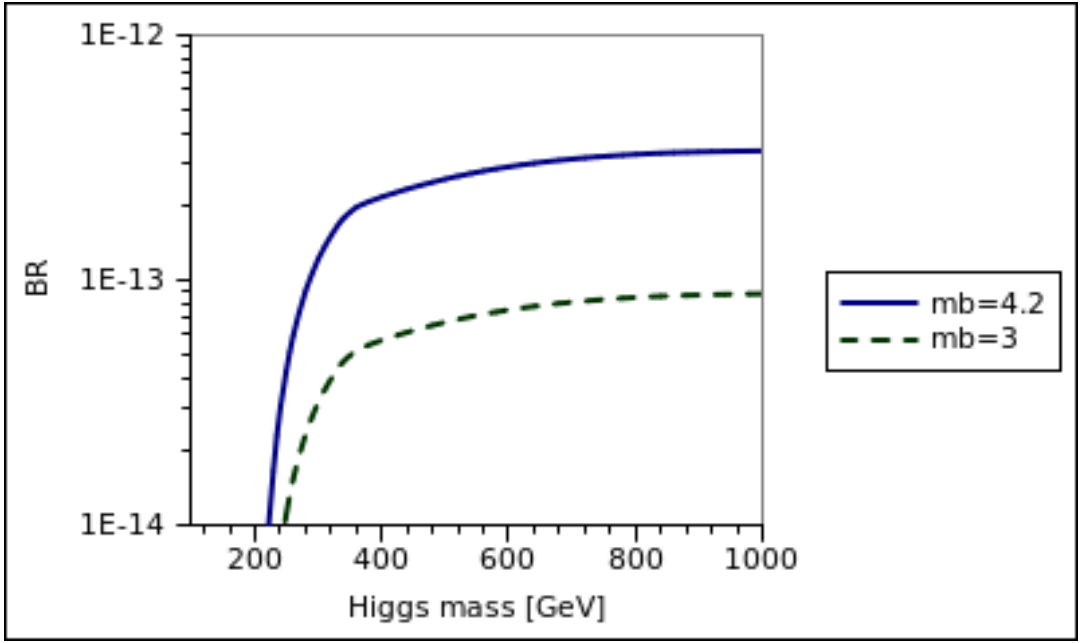}
\par\end{centering}

\caption{\label{fig:h-tc SM} The SM value for the $BR\left(H^{0}\rightarrow\bar{t}c+\bar{c}t\right)$
as a function of the Higgs mass, for $\overline{m_{b}}(\overline{m_{b}})=4.2\mbox{ GeV}$
and for $\overline{m_{b}}(\overline{m_{Z}})=3\mbox{ GeV}$ \cite{PDG}.
The BR is not sensitive to $m_{c}$.}
\end{figure}

In Fig. \ref{fig:h-tc SM} we present the SM value for the $BR\left(H^{0}\rightarrow\bar{t}c+\bar{c}t\right)$.
Our results agree with \cite{h-tc SM arhrib}.

\begin{figure}
\begin{tabular}{cc}
\includegraphics[width=11cm]{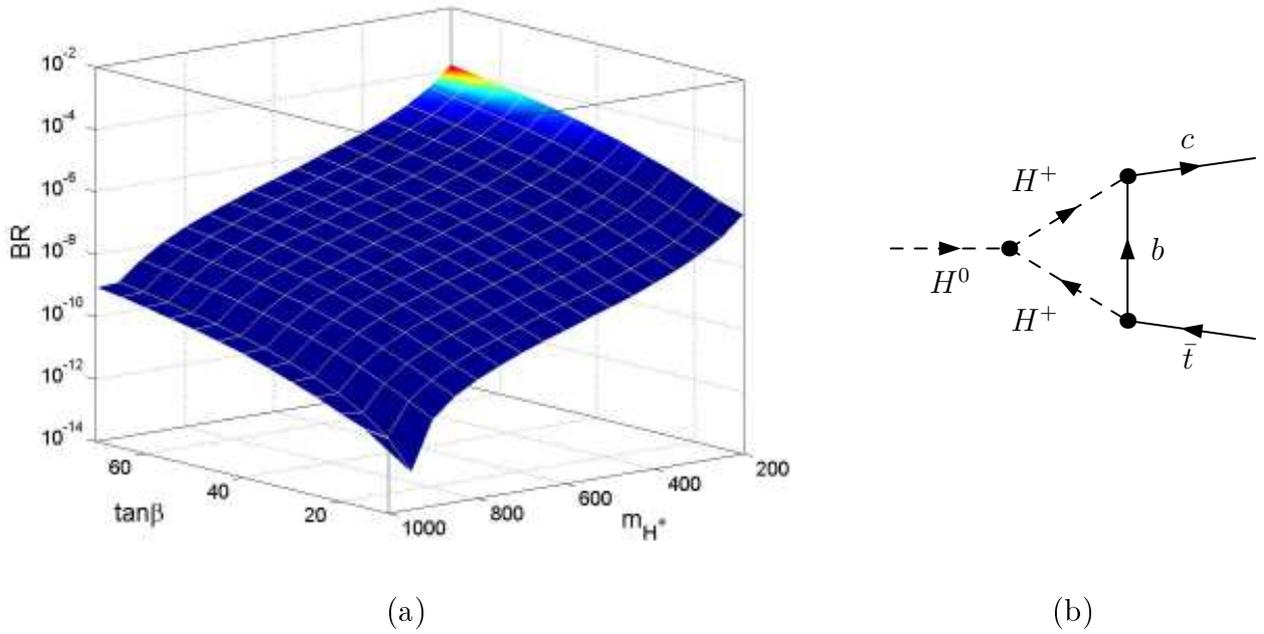}&
\begin{tabular}{c}
\vspace{-9cm}
\tabularnewline
\begin{feynartspicture}(150,150)(1,1)
\FADiagram{}
\FAProp(0.,10.)(6.5,10.)(0.,){/ScalarDash}{1}
\FALabel(3.25,8.93)[t]{$H^0$}
\FAProp(20.,15.)(13.,14.)(0.,){/Straight}{-1}
\FALabel(16.2808,15.5544)[b]{$c$}
\FAProp(20.,5.)(13.,6.)(0.,){/Straight}{1}
\FALabel(16.3162,4.69307)[t]{$\bar t$}
\FAProp(6.5,10.)(13.,14.)(0.,){/ScalarDash}{1}
\FALabel(9.20801,13.1807)[br]{$H^+$}
\FAProp(6.5,10.)(13.,6.)(0.,){/ScalarDash}{-1}
\FALabel(9.20801,6.81927)[tr]{$H^+$}
\FAProp(13.,14.)(13.,6.)(0.,){/Straight}{-1}
\FALabel(14.274,10.)[l]{$b$}
\FAVert(6.5,10.){0}
\FAVert(13.,14.){0}
\FAVert(13.,6.){0}
\end{feynartspicture}\tabularnewline
\end{tabular}\tabularnewline
(a)&
(b)\tabularnewline
\end{tabular}

\caption{\label{fig:h-tc 3D tanb:mhp}(a) 3D plot of $BR\left(H^{0}\rightarrow\bar{t}c+\bar{c}t\right)$
in the $m_{H^{+}}-\mbox{tan}\beta$ plane in the T2HDM, and (b) the
dominant diagram. We set $m_{h^{0}}=1000\mbox{ GeV}$ and $m_{A^{0}}=1000\mbox{ GeV}$.}
\end{figure}

Next we turn to results in the T2HDM. In Fig. \ref{fig:h-tc 3D tanb:mhp}
we give a 3D plot of $BR\left(H^{0}\rightarrow\bar{t}c+\bar{c}t\right)$
in the $m_{H^{+}}-\mbox{tan}\beta$ plane in the T2HDM. We see the
same tendency as in the decay $t\rightarrow cH^{0}$: The BR rises
with $\tan\beta$ and rises with lower $m_{H^{+}}$.

\begin{figure}
\begin{centering}
\includegraphics[keepaspectratio]{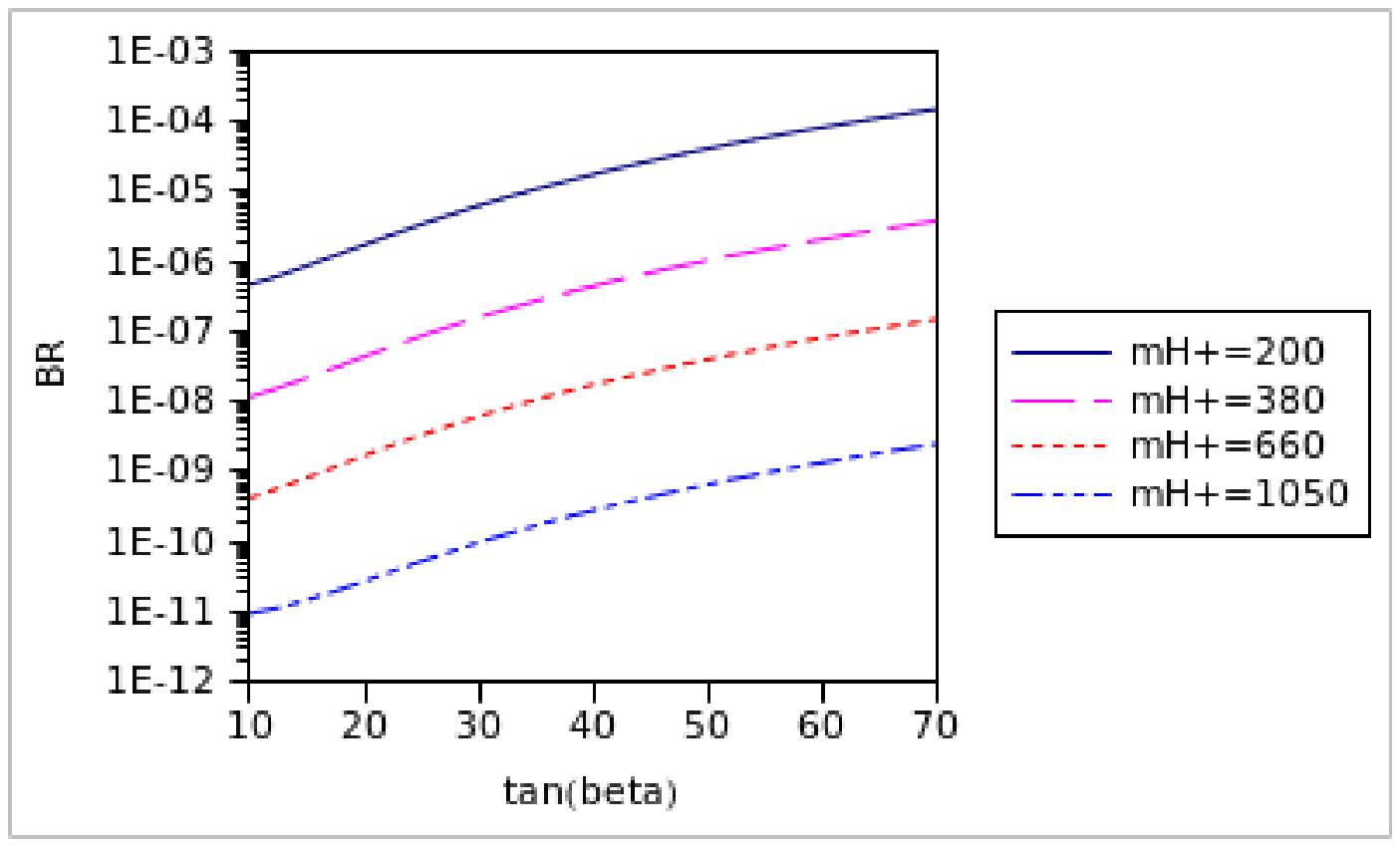}
\par\end{centering}

\caption{\label{fig:h-tc tanb var-mhp}The $BR\left(H^{0}\rightarrow\bar{t}c+\bar{c}t\right)$
as a function of $\tan\beta$ at different $m_{H^{+}}$ in the T2HDM.
We set $m_{h^{0}}=1000\mbox{ GeV}$ and $m_{A^{0}}=1000\mbox{ GeV}$.}
\end{figure}

\begin{figure}
\begin{centering}
\includegraphics[keepaspectratio]{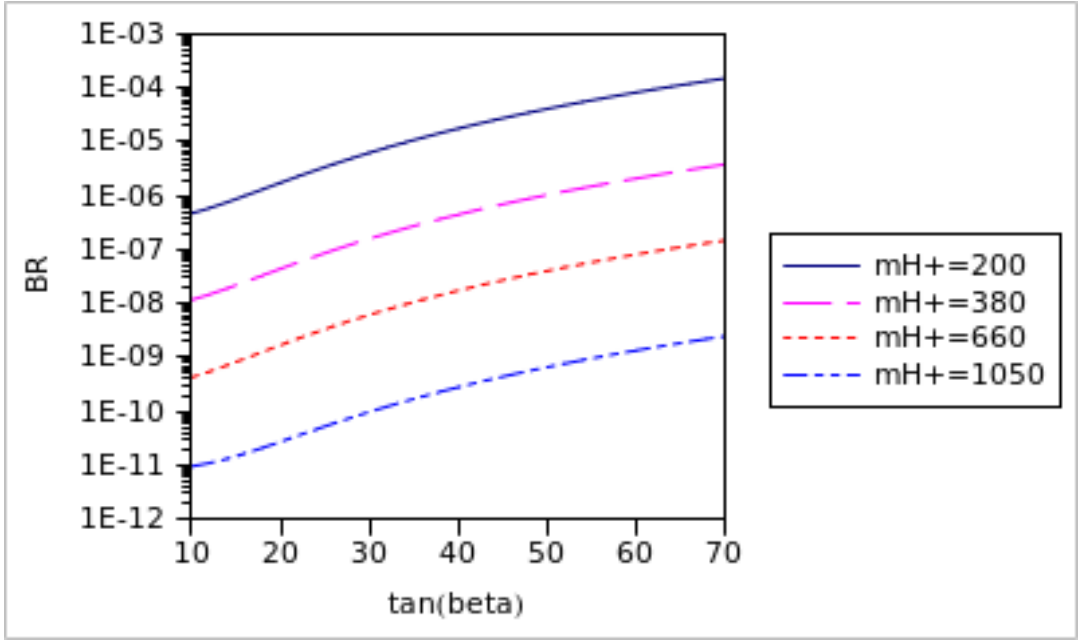}
\par\end{centering}

\caption{\label{fig:h-tc mhp var-tanb}The $BR\left(H^{0}\rightarrow\bar{t}c+\bar{c}t\right)$
as a function of $m_{H^{+}}$ at different $\tan\beta$ in the T2HDM.
We set $m_{h^{0}}=1000\mbox{ GeV}$ and $m_{A^{0}}=1000\mbox{ GeV}$.}
\end{figure}

In Figs. \ref{fig:h-tc tanb var-mhp},\ref{fig:h-tc mhp var-tanb}
we give 2D plots of the BR as a function of $\tan\beta$ and $m_{H^{+}}$
respectively, with the same parameters as in Fig. \ref{fig:h-tc 3D tanb:mhp}.
We see now more clearly the behavior described above.

\begin{figure}
\begin{tabular}{cc}
\includegraphics[width=11cm]{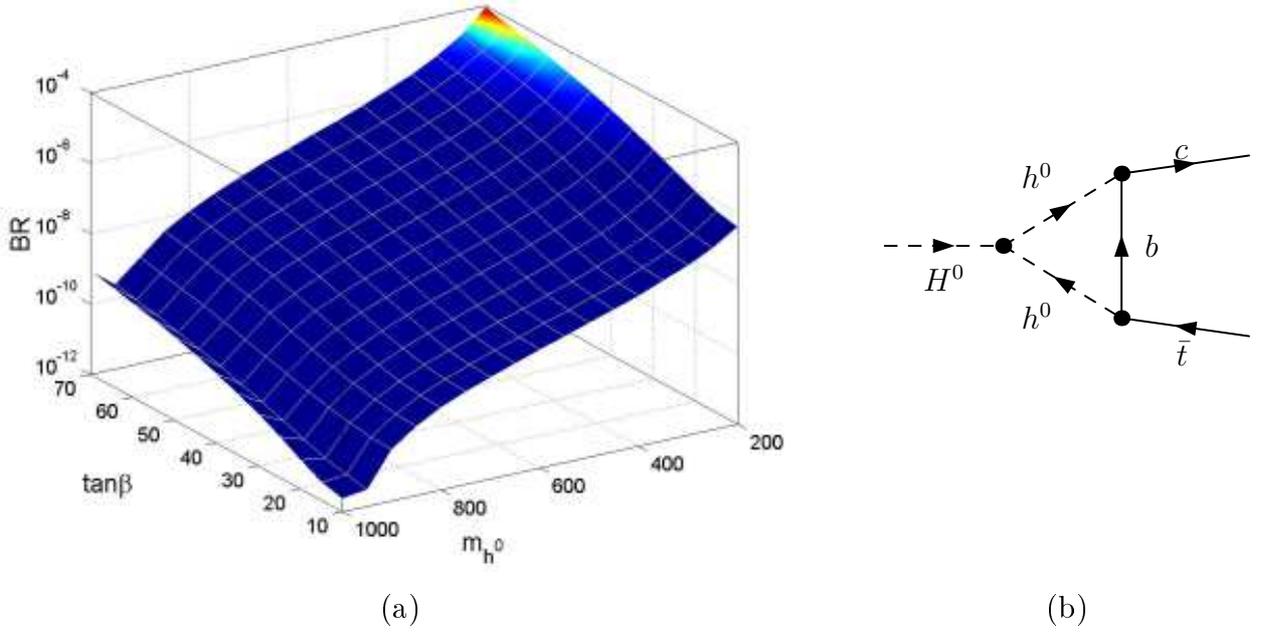}&
\begin{tabular}{c}
\vspace{-9cm}
\tabularnewline
\begin{feynartspicture}(150,150)(1,1)
\FADiagram{}
\FAProp(0.,10.)(6.5,10.)(0.,){/ScalarDash}{1}
\FALabel(3.25,8.93)[t]{$H^0$}
\FAProp(20.,15.)(13.,14.)(0.,){/Straight}{-1}
\FALabel(16.2808,15.5544)[t]{$c$}
\FAProp(20.,5.)(13.,6.)(0.,){/Straight}{1}
\FALabel(16.3162,4.69307)[t]{$\bar t$}
\FAProp(6.5,10.)(13.,14.)(0.,){/ScalarDash}{1}
\FALabel(9.20801,13.1807)[br]{$h^0$}
\FAProp(6.5,10.)(13.,6.)(0.,){/ScalarDash}{-1}
\FALabel(9.20801,6.81927)[tr]{$h^0$}
\FAProp(13.,14.)(13.,6.)(0.,){/Straight}{-1}
\FALabel(14.274,10.)[l]{$b$}
\FAVert(6.5,10.){0}
\FAVert(13.,14.){0}
\FAVert(13.,6.){0}
\end{feynartspicture}\tabularnewline
\end{tabular}\tabularnewline
(a)&
(b)\tabularnewline
\end{tabular}

\caption{\label{fig:h-tc 3D tanb:mh0}(a) 3D plot of $BR\left(H^{0}\rightarrow\bar{t}c+\bar{c}t\right)$
in the $m_{h^{0}}-\mbox{tan}\beta$ plane in the T2HDM, and (b) the
dominant diagram. We set $m_{H^{+}}=1000\mbox{ GeV}$ and $m_{A^{0}}=1000\mbox{ GeV}$.}
\end{figure}

\begin{figure}
\begin{centering}
\includegraphics[keepaspectratio]{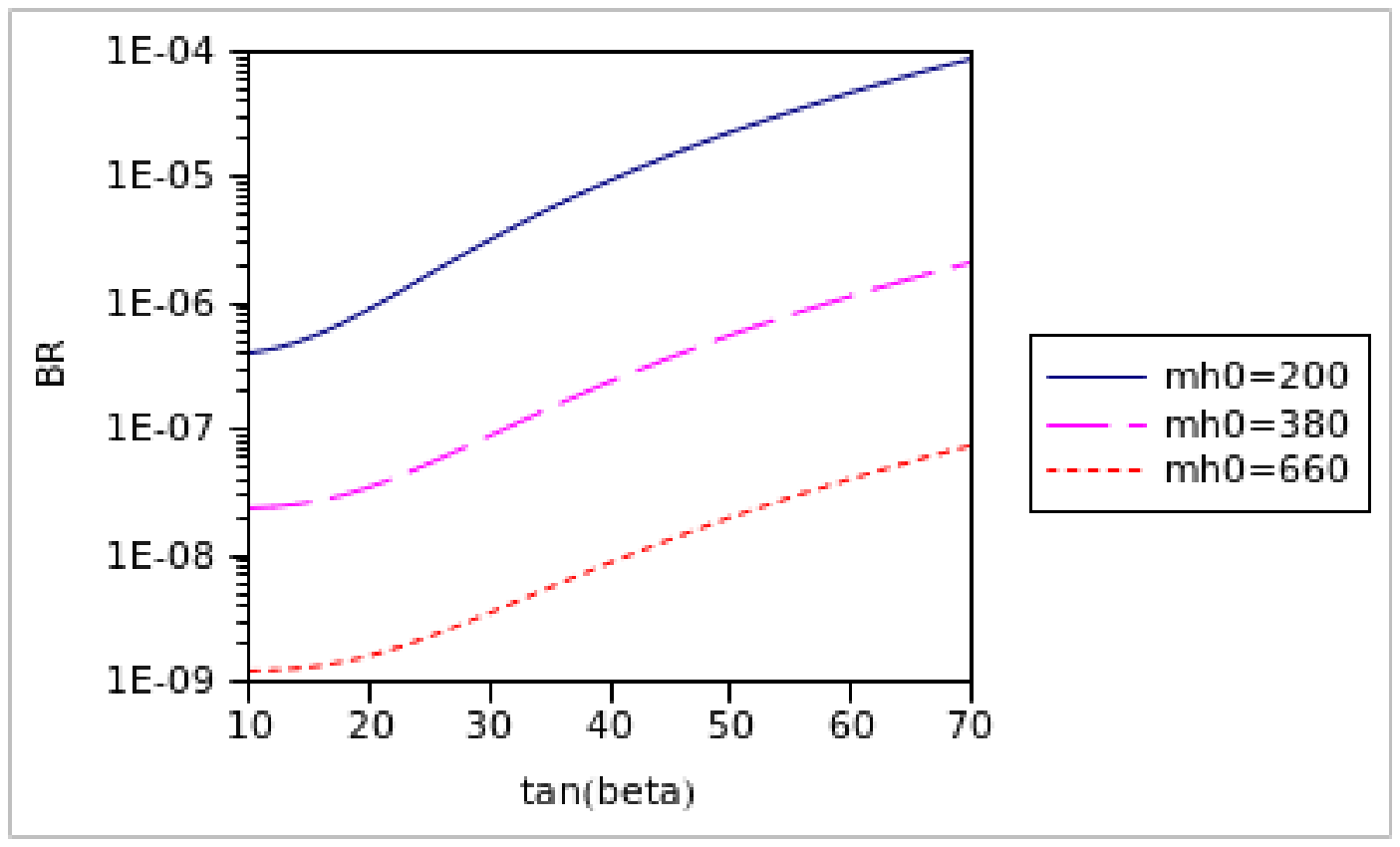}
\par\end{centering}

\caption{\label{fig:h-tc tanb var-mh0}The $BR\left(H^{0}\rightarrow\bar{t}c+\bar{c}t\right)$
as a function of $\tan\beta$ at different $m_{h^{0}}$ in the T2HDM.
We set $m_{H^{+}}=1000\mbox{ GeV}$ and $m_{A^{0}}=1000\mbox{ GeV}$.}
\end{figure}

In Fig. \ref{fig:h-tc 3D tanb:mh0} we give a 3D plot of $BR\left(H^{0}\rightarrow\bar{t}c+\bar{c}t\right)$
in the $m_{h^{0}}-\mbox{tan}\beta$ plane in the T2HDM, and in Fig.
\ref{fig:h-tc tanb var-mh0} we give a 2D plot of the BR as a function
of $\tan\beta$ with the same parameters as Fig. \ref{fig:h-tc 3D tanb:mh0}
at several values of $m_{h^{0}}$. We can see that the BR rises with
lower $m_{h^{0}}$, but has a weak dependence on $\tan\beta$.

In table \ref{tab:h-tc results compare} we give the $BR\left(H^{0}\rightarrow\bar{t}c+\bar{c}t\right)$
values in the different models, for a few points in the parameter
space. As can be seen, the behavior is similar to the $t\rightarrow cH^{0}$
process, although generally the BR values are higher.

\begin{table}
\begin{tabular}{|c|c|c|c|}
\hline 
parameters&
SM&
2HDM-II&
T2HDM\tabularnewline
\hline
\hline 
$m_{h^{0}}=800$, $m_{A^{0}}=1000$, $\tan\beta=72$, $m_{H^{+}}=200$&
$\vphantom{\begin{array}{c}
a\\
b\end{array}}1.23\times10^{-13}$&
$1.26\times10^{-4}$&
$1.70\times10^{-4}$\tabularnewline
\hline 
$m_{h^{0}}=800$, $m_{A^{0}}=1000$, $\tan\beta=72$, $m_{H^{+}}=380$&
$\vphantom{\begin{array}{c}
a\\
b\end{array}}1.23\times10^{-13}$&
$3.09\times10^{-6}$&
$4.45\times10^{-6}$\tabularnewline
\hline 
$m_{h^{0}}=200$, $m_{A^{0}}=4000$, $\tan\beta=20$, $m_{H^{+}}=1050$&
$\vphantom{\begin{array}{c}
a\\
b\end{array}}1.23\times10^{-13}$&
$8.69\times10^{-8}$&
$2.90\times10^{-4}$\tabularnewline
\hline 
$m_{h^{0}}=200$, $m_{A^{0}}=1000$, $\tan\beta=20$, $m_{H^{+}}=1050$&
$\vphantom{\begin{array}{c}
a\\
b\end{array}}1.23\times10^{-13}$&
$8.99\times10^{-12}$&
$9.11\times10^{-7}$\tabularnewline
\hline
\end{tabular}

\caption{\label{tab:h-tc results compare}Comparison of $BR\left(H^{0}\rightarrow\bar{t}c+\bar{c}t\right)$
between the T2HDM, the 2HDM-II, and the SM. Masses are in units of
GeV. We set $m_{H^{0}}=300$, $\alpha=\beta$, and other parameters
to their best-fit value of \eqref{eq:bounds}.}
\end{table}

\chapter{Summary}

The T2HDM is a distinct type of a 2HDM in which the top quark receives
a special status. In this model, the top is coupled to the second
Higgs doublet, while all other quarks are coupled to the first Higgs
doublet. Assuming that the second Higgs VEV ($v_{2}$) is much larger
than the first Higgs VEV ($v_{1}$), the top quark receives a much
larger mass than all other quarks in a natural manner. Therefore the
working assumption of the T2HDM is that $\tan\beta\equiv v_{2}/v_{1}\gg1$.
In addition, these Yukawa couplings generate potentially enhanced
flavor-changing (FC) interactions, both in the charged and the neutral
sectors. These interactions can greatly enhance FC decays such as
$t\rightarrow ch$ and $h\rightarrow\bar{t}c$.

The Yukawa sector of the model was explicitly (and independently)
derived, as well as scalar self interactions. For example, the $H^{+}\bar{b}c$
vertex is enhanced by a factor of $V_{tb}/V_{cb}$ compared to the
corresponding 2HDM-II vertex. This enhancement motivated the present
work, since it is expected to influence the 1-loop $t\rightarrow cH^{0}$
and $H^{0}\rightarrow\bar{t}c$ decays via diagrams involving $H^{+}$
scalars and $b$ quarks inside the loop.

In order to separate the 1-loop decays from the tree-level decays,
we chose $\alpha=\beta$. This choice eliminates the $t\rightarrow cH^{0}$
and $H^{0}\rightarrow\bar{t}c$ tree-level decays, so that these decays
proceed at 1-loop. On the other hand, the decays $t\rightarrow ch^{0}$
and $h^{0}\rightarrow\bar{t}c$ occur at the tree-level for $\alpha=\beta$.
For these tree-level decays we gave explicit formulae. The $h^{0}\bar{t}c$
neutral FC interaction can also enhance the 1-loop $BR\left(t\rightarrow cH^{0}\right)$
and $BR\left(H^{0}\rightarrow\bar{t}c\right)$, via diagrams involving
$h^{0}$ scalars and $t$ quarks inside the loop.

The parameter space of the T2HDM was explored for the resulting 1-loop
$BR\left(t\rightarrow cH^{0}\right)$ and $BR\left(H^{0}\rightarrow\bar{t}c\right)$.
We focused on those regions of the parameter space in which these
BR's can be much higher than in the SM and 2HDM-I,II. We found the
dynamics of the two processes to be similar, which was expected since
their amplitudes and rates are related by crossing symmetry.

The 1-loop $BR\left(t\rightarrow cH^{0}\right)$ can reach $\sim10^{-4}$
in the T2HDM. This is above the LHC detection threshold of $5\times10^{-5}$,
and above the SM, and 2HDM-I,II predictions. The 1-loop $BR\left(H^{0}\rightarrow\bar{t}c\right)$
can reach above $\sim10^{-4}$ in the T2HDM, higher than the SM and
2HDM-I,II predictions.

We conclude that if such decays are indeed identified at the LHC,
then the dynamics of the T2HDM type will be especially motivated.

\appendix

\chapter{\label{app:Higgs-potential}Higgs potential in two Higgs doublet
models}

In the following we introduce the Higgs potential of a general 2HDM,
which corresponds to the T2HDM as well as to 2HDM's of types I, II
and III. 

We assume a CP conserving Higgs potential of the form \cite{HHG}:

\begin{eqnarray}
\mathcal{L}_{H} & = & \lambda_{1}\left(\Phi_{1}^{\dagger}\Phi_{1}-v_{1}^{2}/2\right)^{2}+\lambda_{2}\left(\Phi_{2}^{\dagger}\Phi_{2}-v_{2}^{2}/2\right)^{2}+\lambda_{3}\left[\left(\Phi_{1}^{\dagger}\Phi_{1}-v_{1}^{2}/2\right)+\left(\Phi_{2}^{\dagger}\Phi_{2}-v_{2}^{2}/2\right)\right]^{2}+\nonumber \\
 &  & +\lambda_{4}\left[\left(\Phi_{1}^{\dagger}\Phi_{1}\right)\left(\Phi_{2}^{\dagger}\Phi_{2}\right)-\left(\Phi_{1}^{\dagger}\Phi_{2}\right)\left(\Phi_{2}^{\dagger}\Phi_{1}\right)\right]+\lambda_{5}\left|\Phi_{1}^{\dagger}\Phi_{2}-v_{1}v_{2}/2\right|^{2},\label{eq:higgs pot. app.}\end{eqnarray}

already introduced in Sec. \ref{sec:Yukawa}, and:

\begin{align}
\Phi_{1,2} & =\left(\begin{array}{c}
\Phi_{1,2}^{+}\\
\frac{v_{1,2}+\Phi_{1,2}^{0r}+i\Phi_{1,2}^{0i}}{\sqrt{2}}\end{array}\right).\end{align}
This potential has five couplings $\lambda_{i}$ plus two VEV's $v_{1}$
and $v_{2}$, seven degrees of freedom in total. These will be later
expressed in terms of 5 masses of the physical scalars, plus two angles.
We will then extract the Feynman rules of the 3-scalar interactions,
expressing them in terms of the physical masses and angles.

We assume a potential which conserves CP. The absence of CP violation
implies that the CP-even $h^{0},H^{0}$ and the CP-odd $A^{0},G^{0}$
mix separately, as we will later see. CP violation in the scalar potential
would mix CP-even and odd scalars ($H^{0},h^{0},A^{0}$), as discussed
in \cite{HHG}.

The wearying part is deriving the Feynman rules in terms of the masses
and angles instead of $\lambda_{i}$ ,$v_{1}$ and $v_{2}$. The explicit
derivation is straightforward and we will not follow it completely
here. We will, however, introduce the important formulae, following
the notation of \cite{HHG}.

The fields can always be redefined so that their VEV's are real, without
affecting the potential. The VEV's as defined can be easily seen to
minimize the potential.

The mass terms of the neutral real (CP-even) scalars can be combined
into a symmetric bilinear mass term:

\begin{align}
\mathcal{L}_{m-CPE} & =\left(\Phi_{1}^{0r},\Phi_{2}^{0r}\right)\frac{1}{2}\left[\frac{1}{2}\left(\begin{array}{cc}
4v_{1}^{2}\left(\lambda_{1}+\lambda_{3}\right)+v_{2}^{2}\lambda_{5} & \left(4\lambda_{3}+\lambda_{5}\right)v_{1}v_{2}\\
\left(4\lambda_{3}+\lambda_{5}\right)v_{1}v_{2} & 4v_{2}^{2}\left(\lambda_{2}+\lambda_{3}\right)+v_{1}^{2}\lambda_{5}\end{array}\right)\right]\left(\begin{array}{c}
\Phi_{1}^{0r}\\
\Phi_{2}^{0r}\end{array}\right)\equiv\nonumber \\
 & \equiv\frac{1}{2}\left(\Phi^{0r}\right)^{T}\left[M\right]\left(\Phi^{0r}\right).\end{align}

The mass-squared matrix, $M$, can be diagonalized using the rotation
matrix $\left[\begin{array}{cc}
\cos\alpha & -\sin\alpha\\
\sin\alpha & \cos\alpha\end{array}\right]$, with an angle $\alpha$ such that:

\begin{eqnarray}
\sin2\alpha=\frac{2M_{12}}{\sqrt{\left(M_{11}-M_{22}\right)^{2}+4M_{12}^{2}}} & ,\qquad & \cos2\alpha=\frac{M_{11}-M_{22}}{\sqrt{\left(M_{11}-M_{22}\right)^{2}+4M_{12}^{2}}},\label{eq:sin2alpha}\end{eqnarray}

which defines the CP-even neutral scalars in the mass basis:

\begin{align}
H^{0} & =\Phi_{1}^{or}\cos\alpha+\Phi_{2}^{or}\sin\alpha,\nonumber \\
h^{0} & =-\Phi_{1}^{or}\sin\alpha+\Phi_{2}^{or}\cos\alpha,\end{align}

with masses:

\begin{align}
m_{H^{0},h^{0}}^{2} & =\frac{1}{2}\left(M_{11}+M_{22}\right)\pm\frac{1}{2}\sqrt{\left(M_{11}-M_{22}\right)^{2}+4M_{12}^{2}}\mbox{ .}\label{eq:mh0 mH0}\end{align}

The CP-odd neutral scalars have a simpler mass$^{2}$ matrix: $\lambda_{5}\left[\frac{1}{2}\left(\begin{array}{cc}
v_{2}^{2} & -v_{1}v_{2}\\
-v_{1}v_{2} & v_{1}^{2}\end{array}\right)\right]$ diagonalized using the rotation matrix with the angle $\beta$ such
that:\begin{align}
\tan\beta & =\frac{v_{2}}{v_{1}}\mbox{ ,}\end{align}

which defines the CP-odd neutral scalars and unphysical Goldstone
boson ($G^{0}$), in the mass basis:

\begin{align}
G^{0} & =\Phi_{1}^{oi}\cos\beta+\Phi_{2}^{oi}\sin\beta,\nonumber \\
A^{0} & =-\Phi_{1}^{oi}\sin\beta+\Phi_{2}^{oi}\cos\beta,\end{align}

with mass: $m_{A^{0}}^{2}=\lambda_{5}v^{2}/2$ (recall that: $v=\sqrt{v_{1}^{2}+v_{2}^{2}}=\frac{2m_{W}}{g}$).

We can see that the CP-even $h^{0},H^{0}$ and the CP-odd $A^{0},G^{0}$
mix separately, as promised.

The charged scalars have a similar mass$^{2}$ matrix: $\lambda_{4}\left[\frac{1}{2}\left(\begin{array}{cc}
v_{2}^{2} & -v_{1}v_{2}\\
-v_{1}v_{2} & v_{1}^{2}\end{array}\right)\right]$, diagonalized with the same angle $\beta$, defining the charged
physical scalars and unphysical Goldstone bosons ($G^{\pm}$), in
the mass basis:

\begin{align}
G^{\pm} & =\Phi_{1}^{\pm}\cos\beta+\Phi_{2}^{\pm}\sin\beta,\nonumber \\
H^{\pm} & =-\Phi_{1}^{\pm}\sin\beta+\Phi_{2}^{\pm}\cos\beta,\end{align}

with mass: $m_{H^{\pm}}^{2}=\lambda_{4}v^{2}/2$.

As we stated above, we want to express the triple-scalar couplings
in terms of masses and angles instead of $\lambda_{i},v_{j}$. We
will first connect between $M_{ij}$ and $\lambda_{i},v_{j}$:

\begin{align}
\left(\begin{array}{c}
M_{11}\\
M_{22}\\
M_{12}\end{array}\right) & =\left[\begin{array}{ccc}
4v_{1}^{2} & 0 & 4v_{1}^{2}\\
0 & 4v_{2}^{2} & 4v_{2}^{2}\\
0 & 0 & 4v_{1}v_{2}\end{array}\right]\left(\begin{array}{c}
\lambda_{1}\\
\lambda_{2}\\
\lambda_{3}\end{array}\right)+\left(\begin{array}{c}
v_{2}^{2}\lambda_{5}\\
v_{1}^{2}\lambda_{5}\\
v_{1}v_{2}\lambda_{5}\end{array}\right).\end{align}

We can invert the equation, and insert $\lambda_{5}=\frac{m_{A^{0}}^{2}}{v^{2}}$,
to write:

\begin{eqnarray}
\lambda_{1}+\lambda_{3} & = & \frac{1}{4}\left(\frac{M_{11}}{v_{1}^{2}}-\tan^{2}\beta\frac{m_{A^{0}}^{2}}{v^{2}}\right),\nonumber \\
\lambda_{2}+\lambda_{3} & = & \frac{1}{4}\left(\frac{M_{22}}{v_{2}^{2}}-\cot^{2}\beta\frac{m_{A^{0}}^{2}}{v^{2}}\right),\nonumber \\
\lambda_{3}+\frac{1}{2}\lambda_{5} & = & \frac{1}{4}\left(\frac{M_{12}}{v_{1}v_{2}}+\frac{m_{A^{0}}^{2}}{v^{2}}\right).\label{eq:lam(M_ij)}\end{eqnarray}

The neutral CP-even couplings can be collected from \eqref{eq:higgs pot. app.}:

\begin{eqnarray}
\mathcal{L}_{3h}^{CP-even} & = & 4v_{1}\left(\Phi_{1}^{0r}\right)^{3}\left(\lambda_{1}+\lambda_{3}\right)+4v_{2}\left(\Phi_{2}^{0r}\right)^{3}\left(\lambda_{2}+\lambda_{3}\right)+\nonumber \\
 &  & +4v_{1}\Phi_{1}^{0r}\left(\Phi_{2}^{0r}\right)^{2}\left(\lambda_{3}+\frac{1}{2}\lambda_{5}\right)+4v_{2}\Phi_{2}^{0r}\left(\Phi_{1}^{0r}\right)^{2}\left(\lambda_{3}+\frac{1}{2}\lambda_{5}\right)=\nonumber \\
 & = & v_{1}\left(H_{0}\cos\alpha-h_{0}\sin\alpha\right)^{3}\left(\frac{M_{11}}{v_{1}^{2}}-\tan^{2}\beta\frac{m_{A^{0}}^{2}}{v^{2}}\right)+\nonumber \\
 &  & +v_{2}\left(H_{0}\sin\alpha+h_{0}\cos\alpha\right)^{3}\left(\frac{M_{22}}{v_{2}^{2}}-\cot^{2}\beta\frac{m_{A^{0}}^{2}}{v^{2}}\right)+\nonumber \\
 &  & +v_{1}\left(H_{0}\cos\alpha-h_{0}\sin\alpha\right)\left(H_{0}\sin\alpha+h_{0}\cos\alpha\right)^{2}\left(\frac{M_{12}}{v_{1}v_{2}}+\frac{m_{A^{0}}^{2}}{v^{2}}\right)+\nonumber \\
 &  & +v_{2}\left(H_{0}\sin\alpha+h_{0}\cos\alpha\right)\left(H_{0}\cos\alpha-h_{0}\sin\alpha\right)^{2}\left(\frac{M_{12}}{v_{1}v_{2}}+\frac{m_{A^{0}}^{2}}{v^{2}}\right).\end{eqnarray}

For example, focusing on the $h^{0}h^{0}H^{0}$ term, and using $v_{1}=v\cos\beta$,
$v_{2}=v\sin\beta$, we get: 

\begin{eqnarray}
\mathcal{L}_{h^{0}h^{0}H^{0}} & = & \frac{1}{v}\left\{ 3\cos\beta\cos\alpha\sin^{2}\alpha\left(\frac{M_{11}}{\cos^{2}\beta}-\tan^{2}\beta m_{A^{0}}^{2}\right)+3\sin\beta\sin\alpha\cos^{2}\alpha\left(\frac{M_{22}}{\sin^{2}\beta}-\cot^{2}\beta m_{A^{0}}^{2}\right)+\right.\nonumber \\
 &  & \quad\left.+\left[\cos\left(\alpha+\beta\right)\cos2\alpha-\sin\left(\alpha+\beta\right)\sin\alpha\cos\alpha\right]\left(\frac{M_{12}}{\cos\beta\sin\beta}+m_{A^{0}}^{2}\right)\right\} .\end{eqnarray}

In order to arrive to the final form, more algebraic work is needed.
Using \eqref{eq:sin2alpha}, from which follows also: $2M_{12}\cos2\alpha=\left(M_{11}-M_{22}\right)\sin2\alpha$,
and \eqref{eq:mh0 mH0}, from which follows also $M_{11}+M_{22}=m_{h^{0}}^{2}+m_{H^{0}}^{2}$,
and other trigonometric identities, we arrive at:

\begin{eqnarray}
\mathcal{L}_{h^{0}h^{0}H^{0}} & = & \frac{g\cos\left(\beta-\alpha\right)}{2m_{W}\sin2\beta}\left[\sin2\alpha\left(2m_{h^{0}}^{2}+m_{H^{0}}^{2}\right)-m_{A^{0}}^{2}\left(3\sin2\alpha-\sin2\beta\right)\right],\end{eqnarray}

which agrees with \cite{bejar}, and which is expressed in terms of
the physical masses and mixing angles.

All other 3-scalar interactions are derived similarly. In App. \eqref{app:Feynman-rules}
we listed all the vertices relevant to the present work.

We note that we did not find the $h^{0}H^{0}H^{0}$ coupling for the
higgs potential of \eqref{eq:higgs pot. app.} in the literature and
it was therefore derived by us.

\chapter{\label{app:Feynman-rules}Feynman rules for two Higgs doublet models}

In this section we list the Feynman rules for the 2HDM's that were
used in this work, in the t'Hooft Feynman gauge. The Feynman rules
are presented in Fig. \ref{fig:vertices definition}.

\begin{figure}
\begin{tabular}{cc}
\multicolumn{1}{c}{\begin{tabular}{cc}
\multicolumn{1}{c}{\begin{feynartspicture}(125,125)(1,1)
\FADiagram{}
\FAProp(0.,10.)(11.,10.)(0.,){/ScalarDash}{1}
\FALabel(5.5,8.93)[t]{$h_\delta$}
\FAProp(20.,15.)(11.,10.)(0.,){/Straight}{-1}
\FALabel(15.2273,13.3749)[br]{$q_a$}
\FAProp(20.,5.)(11.,10.)(0.,){/Straight}{1}
\FALabel(15.2273,6.62506)[tr]{$q_b$}
\FAVert(11.,10.){0}
\end{feynartspicture}}&
\begin{tabular}{c}
\vspace{-5cm}
\tabularnewline
$=i\left(A_{ab}^{\delta}L+B_{ab}^{\delta}R\right)$\tabularnewline
\end{tabular}\tabularnewline
\begin{feynartspicture}(125,125)(1,1)
\FADiagram{}
\FAProp(0.,10.)(11.,10.)(0.,){/Sine}{1}
\FALabel(5.5,8.93)[t]{$V_\delta$}
\FAProp(20.,15.)(11.,10.)(0.,){/Straight}{-1}
\FALabel(15.2273,13.3749)[br]{$q_a$}
\FAProp(20.,5.)(11.,10.)(0.,){/Straight}{1}
\FALabel(15.2273,6.62506)[tr]{$q_b$}
\FAVert(11.,10.){0}
\end{feynartspicture}&
\begin{tabular}{c}
\vspace{-5cm}
\tabularnewline
$=i\left(a_{ab}^{\delta}L+b_{ab}^{\delta}R\right)$\tabularnewline
\end{tabular}\tabularnewline
\end{tabular}}&
\begin{tabular}{cc}
\begin{feynartspicture}(125,125)(1,1)
\FADiagram{}
\FAProp(0.,10.)(11.,10.)(0.,){/ScalarDash}{0}
\FALabel(5.5,8.93)[t]{$h_a$}
\FAProp(20.,15.)(11.,10.)(0.,){/ScalarDash}{0}
\FALabel(15.2273,13.3749)[br]{$h_b$}
\FAProp(20.,5.)(11.,10.)(0.,){/ScalarDash}{0}
\FALabel(15.2273,6.62506)[tr]{$h_c$}
\FAVert(11.,10.){0}
\end{feynartspicture}&
\begin{tabular}{c}
\vspace{-5cm}
\tabularnewline
$=ig_{abc}^{3h}$\tabularnewline
\end{tabular}\tabularnewline
\begin{feynartspicture}(125,125)(1,1)
\FADiagram{}
\FAProp(0.,10.)(11.,10.)(0.,){/Sine}{1}
\FALabel(5.5,8.93)[t]{$V^\mu_a$}
\FAProp(20.,15.)(11.,10.)(0.,){/ScalarDash}{-1}
\FALabel(15.2273,13.3749)[br]{$h_b$}
\FAProp(20.,5.)(11.,10.)(0.,){/ScalarDash}{1}
\FALabel(15.2273,6.62506)[tr]{$h_c$}
\FAVert(11.,10.){0}
\end{feynartspicture}&
\begin{tabular}{c}
\vspace{-5cm}
\tabularnewline
$=ig_{abc}^{vhh}\left(P_{b}+P_{c}\right)^{\mu}$\tabularnewline
\end{tabular}\tabularnewline
\begin{feynartspicture}(125,125)(1,1)
\FADiagram{}
\FAProp(0.,15.)(9.,10.)(0.,){/Sine}{0}
\FALabel(4.22725,11.6251)[tr]{$V^\mu_a$}
\FAProp(0.,5.)(9.,10.)(0.,){/Sine}{0}
\FALabel(4.77275,6.62506)[tl]{$V^\nu_b$}
\FAProp(20.,10.)(9.,10.)(0.,){/ScalarDash}{0}
\FALabel(14.5,10.82)[b]{$h_c$}
\FAVert(9.,10.){0}
\end{feynartspicture}&
\begin{tabular}{c}
\vspace{-5cm}
\tabularnewline
$=ig_{abc}^{vvh}g^{\mu\nu}$\tabularnewline
\end{tabular}\tabularnewline
\end{tabular}\tabularnewline
\end{tabular}

\caption{\label{fig:vertices definition}Feynman rules.}
\end{figure}

In table \ref{tab:yukawa feyn rules} we give the Yukawa Feynman rules
of the T2HDM and of the 2HDM type II. The other Feynman rules are
common to all 2HDM's.

\begin{table}
\begin{tabular}{|c|c|c|}
\hline 
&
T2HDM&
2HDM-II \cite{HHG}\tabularnewline
\hline
\hline 
$H^{0}\bar{u}_{j}u_{i}$&
$\frac{g}{2m_{W}}\left(-M_{u}\frac{\cos\alpha}{\cos\beta}+\Sigma\left(-\frac{\sin\alpha}{\sin\beta}+\frac{\cos\alpha}{\cos\beta}\right)\right)R+\left(h.c.\right)L$&
$-\frac{gM_{u}}{2m_{W}}\frac{\sin\alpha}{\sin\beta}$\tabularnewline
\hline 
$h^{0}\bar{u}u$&
$\frac{g}{2m_{W}}\left(M_{u}\frac{\sin\alpha}{\cos\beta}-\Sigma\left(\frac{\cos\alpha}{\sin\beta}+\frac{\sin\alpha}{\cos\beta}\right)\right)R+\left(h.c.\right)L$&
$-\frac{gM_{u}}{2m_{W}}\frac{\cos\alpha}{\sin\beta}$\tabularnewline
\hline 
$A^{0}\bar{u}u$&
$i\frac{g}{2m_{W}}\left(-M_{u}\tan\beta+\Sigma\left(\tan\beta+\cot\beta\right)\right)R+\left(h.c.\right)L$&
$i\frac{gM_{u}}{2m_{W}}\cot\beta\left(R-L\right)$\tabularnewline
\hline 
$G^{0}\bar{u}u$&
$i\frac{gM_{u}}{2m_{W}}\left(R-L\right)$&
$i\frac{gM_{u}}{2m_{W}}\left(R-L\right)$\tabularnewline
\hline 
$H^{0}\bar{d}d$&
$-\frac{gM_{d}}{2m_{W}}\frac{\cos\alpha}{\cos\beta}$&
$-\frac{gM_{d}}{2m_{W}}\frac{\cos\alpha}{\cos\beta}$\tabularnewline
\hline 
$h^{0}\bar{d}d$&
$\frac{gM_{d}}{2m_{W}}\frac{\sin\alpha}{\cos\beta}$&
$\frac{gM_{d}}{2m_{W}}\frac{\sin\alpha}{\cos\beta}$\tabularnewline
\hline 
$A^{0}\bar{d}d$&
$i\frac{gM_{d}}{2m_{W}}\tan\beta\left(R-L\right)$&
$i\frac{gM_{d}}{2m_{W}}\tan\beta\left(R-L\right)$\tabularnewline
\hline 
$G^{0}\bar{d}d$&
$-i\frac{gM_{d}}{2m_{W}}\left(R-L\right)$&
$-i\frac{gM_{d}}{2m_{W}}\left(R-L\right)$\tabularnewline
\hline 
$H^{+}\bar{u}d$&
\begin{tabular}{l}
$\frac{g}{\sqrt{2}m_{W}}\left[\tan\beta V_{CKM}M_{d}R\right.$\tabularnewline
$\left.+\left(-M_{u}\tan\beta+\Sigma\left(\tan\beta+\cot\beta\right)\right)V_{CKM}L\right]$\tabularnewline
\end{tabular}&
\begin{tabular}{l}
$\frac{g}{\sqrt{2}m_{W}}\left[\tan\beta V_{CKM}M_{d}R\right.$\tabularnewline
$\left.\qquad+\cot\beta M_{u}V_{CKM}L\right]$\tabularnewline
\end{tabular}\tabularnewline
\hline 
$G^{+}\bar{u}d$&
$\frac{g}{\sqrt{2}m_{W}}\left(M_{u}V_{CKM}L-V_{CKM}M_{d}R\right)$&
$\frac{g}{\sqrt{2}m_{W}}\left(M_{u}V_{CKM}L-V_{CKM}M_{d}R\right)$\tabularnewline
\hline
\end{tabular}

\caption{\label{tab:yukawa feyn rules}Feynman rules for Yukawa interactions
in the T2HDM and in the 2HDM-II.}
\end{table}

In table \ref{tab:vvh feyn rules} we give the vector-vector-scalar
couplings, common to all 2HDM's, from \cite{HHG}.

\begin{table}
\begin{tabular}{|c|c|}
\hline 
$W^{+}W^{-}H^{0}$&
$igm_{W}\cos\left(\beta-\alpha\right)g^{\mu\nu}$\tabularnewline
\hline 
$W^{+}W^{-}h^{0}$&
$igm_{W}\sin\left(\beta-\alpha\right)g^{\mu\nu}$\tabularnewline
\hline 
$Z^{0}Z^{0}H^{0}$&
$\frac{igm_{Z}}{\cos\theta_{W}}\cos\left(\beta-\alpha\right)g^{\mu\nu}$\tabularnewline
\hline 
$Z^{0}Z^{0}h^{0}$&
$\frac{igm_{Z}}{\cos\theta_{W}}\sin\left(\beta-\alpha\right)g^{\mu\nu}$\tabularnewline
\hline
\end{tabular}

\caption{\label{tab:vvh feyn rules}Feynman rules for vector-vector-scalar
interactions \cite{HHG}.}
\end{table}

In table \ref{tab:vhh feyn rules} we give the vector-scalar-scalar
couplings, we define the vertices as in \cite{HHG}, where the second
particle is outgoing.

\begin{table}
\begin{tabular}{|c|c|}
\hline 
\begin{tabular}{c}
\vspace{-1.5cm}
\tabularnewline
\begin{feynartspicture}(125,125)(1,1)
\FADiagram{}
\FAProp(0.,10.)(11.,10.)(0.,){/Sine}{1}
\FALabel(5.5,8.93)[t]{$V^\mu_a$}
\FAProp(20.,15.)(11.,10.)(0.,){/ScalarDash}{-1}
\FALabel(15.2273,13.3749)[br]{$h_b$}
\FAProp(20.,5.)(11.,10.)(0.,){/ScalarDash}{1}
\FALabel(15.2273,6.62506)[tr]{$h_c$}
\FAVert(11.,10.){0}
\end{feynartspicture}\tabularnewline
\vspace{-1.5cm}
\tabularnewline
\end{tabular}&
\begin{tabular}{c}
\vspace{-1cm}
\tabularnewline
$=ig_{abc}^{vhh}\left(P_{b}+P_{c}\right)^{\mu}$\tabularnewline
\end{tabular}\tabularnewline
\hline
\hline 
$W^{+}H^{+}H^{0}$&
$i\frac{g}{2}\sin\left(\beta-\alpha\right)\left(P_{H^{+}}+P_{H^{0}}\right)^{\mu}$\tabularnewline
\hline 
$W^{+}H^{+}h^{0}$&
$-i\frac{g}{2}\cos\left(\beta-\alpha\right)\left(P_{H^{+}}+P_{h^{0}}\right)^{\mu}$\tabularnewline
\hline 
$W^{+}G^{+}H^{0}$&
-$i\frac{g}{2}\cos\left(\beta-\alpha\right)\left(P_{G^{+}}+P_{H^{0}}\right)^{\mu}$\tabularnewline
\hline 
$W^{+}G^{+}h^{0}$&
$-i\frac{g}{2}\sin\left(\beta-\alpha\right)\left(P_{G^{+}}+P_{h^{0}}\right)^{\mu}$\tabularnewline
\hline 
$Z^{0}A^{0}H^{0}$&
$-\frac{g\sin\left(\beta-\alpha\right)}{2\cos\theta_{W}}\left(P_{A^{0}}+P_{H^{0}}\right)^{\mu}$\tabularnewline
\hline 
$Z^{0}A^{0}h^{0}$&
$\frac{g\cos\left(\beta-\alpha\right)}{2\cos\theta_{W}}\left(P_{A^{0}}+P_{h^{0}}\right)^{\mu}$\tabularnewline
\hline
\end{tabular}

\caption{\label{tab:vhh feyn rules}Feynman rules for vector-scalar-scalar
interactions \cite{HHG}.}
\end{table}

The vertices $Z^{0}G^{0}H^{0}$, $Z^{0}G^{0}h^{0}$ do not participate
in the calculations since the corresponding Yukawa vertex $\bar{q}qG^{0}$
does not generate FC interactions.

Now we turn to the 3-scalar interactions. We were not able to find
all vertices in the literature, and therefore we derived one vertex
($h^{0}H^{0}H^{0}$), while the rest of the scalar self interactions
can be found in \cite{bejar}. We give in table \ref{tab:3h feyn rules}
the complete list of the 3-scalar interactions that were used in this
work.

\begin{table}
\begin{tabular}{|c|c|}
\hline 
$H^{+}H^{-}H^{0}$&
$-\frac{g}{m_{W}}\left[\left(m_{H^{+}}^{2}-m_{A^{0}}^{2}+\frac{1}{2}m_{H^{0}}^{2}\right)\cos\left(\beta-\alpha\right)+\left(m_{A^{0}}^{2}-m_{H^{0}}^{2}\right)\cot2\beta\sin\left(\beta-\alpha\right)\right]$\tabularnewline
\hline 
$H^{+}H^{-}h^{0}$&
$-\frac{g}{m_{W}}\left[\left(m_{H^{+}}^{2}-m_{A^{0}}^{2}+\frac{1}{2}m_{h^{0}}^{2}\right)\sin\left(\beta-\alpha\right)+\left(m_{h^{0}}^{2}-m_{A^{0}}^{2}\right)\cot2\beta\cos\left(\beta-\alpha\right)\right]$\tabularnewline
\hline 
$h^{0}h^{0}H^{0}$&
$-\frac{g\cos\left(\beta-\alpha\right)}{2m_{W}\sin2\beta}\left[\left(2m_{h^{0}}^{2}+m_{H^{0}}^{2}\right)\sin2\alpha-m_{A^{0}}^{2}\left(3\sin2\alpha-\sin2\beta\right)\right]$\tabularnewline
\hline 
$h^{0}H^{0}H^{0}$&
$-\frac{g\sin\left(\beta-\alpha\right)}{2m_{W}\sin2\beta}\left[\left(2m_{H^{0}}^{2}+m_{h^{0}}^{2}\right)\sin2\alpha-m_{A^{0}}^{2}\left(3\sin2\alpha+\sin2\beta\right)\right]$\tabularnewline
\hline 
$A^{0}A^{0}H^{0}$&
$-\frac{g}{2m_{W}}\left[m_{H^{0}}^{2}\cos\left(\beta-\alpha\right)+2\left(m_{H^{0}}^{2}-m_{A^{0}}^{2}\right)\cot2\beta\sin\left(\beta-\alpha\right)\right]$\tabularnewline
\hline 
$A^{0}A^{0}h^{0}$&
$-\frac{g}{2m_{W}}\left[m_{h^{0}}^{2}\sin\left(\beta-\alpha\right)+2\left(m_{h^{0}}^{2}-m_{A^{0}}^{2}\right)\cot2\beta\cos\left(\beta-\alpha\right)\right]$\tabularnewline
\hline 
$H^{+}G^{-}H^{0}$&
$-i\frac{g}{2m_{W}}\left(m_{H^{+}}^{2}-m_{H^{0}}^{2}\right)\sin\left(\beta-\alpha\right)$\tabularnewline
\hline 
$H^{+}G^{-}h^{0}$&
$i\frac{g}{2m_{W}}\left(m_{H^{+}}^{2}-m_{h^{0}}^{2}\right)\cos\left(\beta-\alpha\right)$\tabularnewline
\hline 
$G^{+}G^{-}H^{0}$&
-$i\frac{g}{2m_{W}}m_{H^{0}}^{2}\cos\left(\beta-\alpha\right)$\tabularnewline
\hline 
$G^{+}G^{-}h^{0}$&
$-i\frac{g}{2m_{W}}m_{h^{0}}^{2}\sin\left(\beta-\alpha\right)$\tabularnewline
\hline
\end{tabular}

\caption{\label{tab:3h feyn rules}Feynman rules for triple-scalar interactions
\cite{bejar,HHG}.}
\end{table}

\chapter{\label{app:1L diags}1-loop diagrams calculation}

In this appendix we give the 1-loop calculation of the 10 diagrams
shown in Fig. \ref{fig: 1-loop diags}. The calculation was done in
the t'Hooft Feynman gauge.

In the t'Hooft Feynman gauge the vector bosons propagators reduce
to their simplest form: $\Delta=ig^{\mu\nu}\left[p^{2}-m^{2}+i\epsilon\right]^{-1}$,
and the Goldstone bosons mass is set equal to the respective gauge
bosons: $m_{G^{+}}=m_{W^{+}}$, $m_{G^{0}}=m_{Z}$. The t'Hooft Feynman
gauge was chosen because the calculation of each diagram is simpler.

\textbf{\underbar{definitions:}}

$M_{n}$ -- the amplitude corresponding to diagram $n$

$h$ -- the external neutral scalar

$i$ -- ($=t$) when used as index, the incoming fermion - the top

$j$ -- ($=c$) when used as index, the outgoing fermion - the charm

$\alpha,\beta$ -- when used as indices, internal bosons (vectors
or scalars) in the loop

$l,k,q$ -- when used as indices, internal fermions

$L,R$ -- the Left,Right projection operators

$\bar{u}_{j}$ -- ($=\bar{u}(P_{j})$ ) the outgoing spinor of the
charm

$u_{i}$ -- ($=u(P_{i})$ ) the incoming spinor of the top

$B_{0},B_{1},C_{0},C_{ij}$ -- the n-point integral functions, defined
in App. \ref{app:dijcij definition}

$A_{ab}^{\delta},B_{ab}^{\delta}$ -- the left,right -handed parts
of the fermion-fermion-scalar vertex

$a_{ab}^{\delta},b_{ab}^{\delta}$ -- the left,right -handed parts
of the fermion-fermion-vector vertex, for both charged and neutral
gauge bosons

$g_{abc}^{3h,vhh,vvh}$ -- the vertex of 3-scalars, vector-scalar-scalar,
vector-vector-scalar, respectively

$g^{\mu\nu}$ -- the metric, $g^{\mu\nu}=diag(1,-1,-1,-1)$

\newpage

\begin{align}
M_{1} & =\frac{i\bar{u}_{j}}{16\pi^{2}}\frac{-1}{m_{i}^{2}-m_{l}^{2}}\left[m_{l}m_{k}B_{0}\left(B_{lj}^{h*}A_{lk}^{\alpha}B_{ik}^{\alpha*}L+A_{lj}^{h*}B_{lk}^{\alpha}A_{ik}^{\alpha*}R\right)+\right.\nonumber \\
 & \qquad-m_{l}m_{i}B_{1}\left(B_{lj}^{h*}A_{lk}^{\alpha}A_{ik}^{\alpha*}L+A_{lj}^{h*}B_{lk}^{\alpha}B_{ik}^{\alpha*}R\right)+m_{i}m_{k}B_{0}\left(B_{lj}^{h*}B_{lk}^{\alpha}A_{ik}^{\alpha*}L+A_{lj}^{h*}A_{lk}^{\alpha}B_{ik}^{\alpha*}R\right)+\nonumber \\
 & \qquad\left.-m_{i}^{2}B_{1}\left(B_{lj}^{h*}B_{lk}^{\alpha}B_{ik}^{\alpha*}L+A_{lj}^{h*}A_{lk}^{\alpha}A_{ik}^{\alpha*}R\right)\right]u_{i},\end{align}
where $B=B\left(m_{k}^{2},m_{\alpha}^{2},m_{i}^{2}\right).$

\begin{align}
M_{2} & =\frac{i\bar{u}_{j}}{16\pi^{2}}\frac{-1}{m_{j}^{2}-m_{l}^{2}}\left[m_{l}m_{k}B_{0}\left(A_{jk}^{\alpha}B_{lk}^{\alpha*}B_{il}^{h*}L+B_{jk}^{\alpha}A_{lk}^{\alpha*}A_{il}^{h*}R\right)+\right.\nonumber \\
 & \qquad+m_{k}m_{j}B_{0}\left(B_{jk}^{\alpha}A_{lk}^{\alpha*}B_{il}^{h*}L+A_{jk}^{\alpha}B_{lk}^{\alpha*}A_{il}^{h*}R\right)-m_{j}m_{l}B_{1}\left(B_{jk}^{\alpha}B_{lk}^{\alpha*}B_{il}^{h*}L+A_{jk}^{\alpha}A_{lk}^{\alpha*}A_{il}^{h*}R\right)+\nonumber \\
 & \qquad\left.-m_{j}^{2}B_{1}\left(A_{jk}^{\alpha}A_{lk}^{\alpha*}B_{il}^{h*}L+B_{jk}^{\alpha}B_{lk}^{\alpha*}A_{il}^{h*}R\right)\right]u_{i},\end{align}
where $B=B\left(m_{k}^{2},m_{\alpha}^{2},m_{j}^{2}\right).$

\begin{align}
M_{3} & =\frac{i\bar{u}_{j}}{16\pi^{2}}\frac{1}{m_{i}^{2}-m_{l}^{2}}\left[4m_{l}m_{k}B_{0}\left(B_{lj}^{h*}b_{lk}^{\alpha}a_{ik}^{\alpha*}L+A_{lj}^{h*}a_{lk}^{\alpha}b_{ik}^{\alpha*}R\right)+\right.\nonumber \\
 & \qquad+2m_{l}m_{i}B_{1}\left(B_{lj}^{h*}b_{lk}^{\alpha}b_{ik}^{\alpha*}L+A_{lj}^{h*}a_{lk}^{\alpha}a_{ik}^{\alpha*}R\right)+4m_{i}m_{k}B_{0}\left(B_{lj}^{h*}a_{lk}^{\alpha}b_{ik}^{\alpha*}L+A_{lj}^{h*}b_{lk}^{\alpha}a_{ik}^{\alpha*}R\right)+\nonumber \\
 & \qquad\left.+2m_{i}^{2}B_{1}\left(B_{lj}^{h*}a_{lk}^{\alpha}a_{ik}^{\alpha*}L+A_{lj}^{h*}b_{lk}^{\alpha}b_{ik}^{\alpha*}R\right)\right]u_{i},\end{align}
where $B=B\left(m_{k}^{2},m_{\alpha}^{2},m_{i}^{2}\right).$

\begin{align}
M_{4} & =\frac{i\bar{u}_{j}}{16\pi^{2}}\frac{1}{m_{j}^{2}-m_{l}^{2}}\left[4m_{l}m_{k}B_{0}\left(b_{jk}^{\alpha}a_{lk}^{\alpha*}B_{il}^{h*}L+a_{jk}^{\alpha}b_{lk}^{\alpha*}A_{il}^{h*}R\right)+\right.\nonumber \\
 & \qquad+4m_{k}m_{j}B_{0}\left(a_{jk}^{\alpha}b_{lk}^{\alpha*}B_{il}^{h*}L+b_{jk}^{\alpha}a_{lk}^{\alpha*}A_{il}^{h*}R\right)+2m_{j}m_{l}B_{1}\left(a_{jk}^{\alpha}a_{lk}^{\alpha*}B_{il}^{h*}L+b_{jk}^{\alpha}b_{lk}^{\alpha*}A_{il}^{h*}R\right)+\nonumber \\
 & \qquad\left.+2m_{j}^{2}B_{1}\left(b_{jk}^{\alpha}b_{lk}^{\alpha*}B_{il}^{h*}L+a_{jk}^{\alpha}a_{lk}^{\alpha*}A_{il}^{h*}R\right)\right]u_{i},\end{align}
where $B=B\left(m_{k}^{2},m_{\alpha}^{2},m_{j}^{2}\right).$

\begin{align}
M_{5} & =\frac{-i\bar{u}_{j}}{16\pi^{2}}\left(A_{jq}^{\alpha}L+B_{jq}^{\alpha}R\right)\left\{ \left[\tilde{C}_{0}+m_{i}^{2}C_{11}+\left(m_{h}^{2}-m_{i}^{2}\right)C_{12}\right]\left(A_{jq}^{\alpha}A_{kq}^{h*}B_{ik}^{\alpha*}L+B_{jq}^{\alpha}B_{kq}^{h*}A_{ik}^{\alpha*}R\right)+\right.\nonumber \\
 & -m_{q}m_{i}C_{11}\left(A_{jq}^{\alpha}B_{kq}^{h*}A_{ik}^{\alpha*}L+B_{jq}^{\alpha}A_{kq}^{h*}B_{ik}^{\alpha*}R\right)+m_{q}m_{j}C_{12}\left(B_{jq}^{\alpha}A_{kq}^{h*}B_{ik}^{\alpha*}L+A_{jq}^{\alpha}B_{kq}^{h*}A_{ik}^{\alpha*}R\right)+\nonumber \\
 & +m_{i}m_{j}\left(C_{12}-C_{11}\right)\left(B_{jq}^{\alpha}B_{kq}^{h*}A_{ik}^{\alpha*}L+A_{jq}^{\alpha}A_{kq}^{h*}B_{ik}^{\alpha*}R\right)+m_{q}m_{k}C_{0}\left(A_{jq}^{\alpha}B_{kq}^{h*}B_{ik}^{\alpha*}L+B_{jq}^{\alpha}A_{kq}^{h*}A_{ik}^{\alpha*}R\right)+\nonumber \\
 & -m_{i}m_{k}\left(C_{11}+C_{0}\right)\left(A_{jq}^{\alpha}A_{kq}^{h*}A_{ik}^{\alpha*}L+B_{jq}^{\alpha}B_{kq}^{h*}B_{ik}^{\alpha*}R\right)+\nonumber \\
 & \left.+m_{j}m_{k}\left(C_{12}+C_{0}\right)\left(B_{jq}^{\alpha}B_{kq}^{h*}B_{ik}^{\alpha*}L+A_{jq}^{\alpha}A_{kq}^{h*}A_{ik}^{\alpha*}R\right)\right\} u_{i},\end{align}
where $C=C\left(m_{k}^{2},m_{\alpha}^{2},m_{q}^{2},m_{i}^{2},m_{j}^{2},m_{h}^{2}\right).$

\newpage

\begin{align}
M_{6} & =\frac{i\bar{u}_{j}}{16\pi^{2}}\left\{ \left[4\tilde{C}_{0}+2\left(m_{i}^{2}-m_{j}^{2}+m_{h}^{2}\right)C_{11}+2\left(-m_{i}^{2}+m_{j}^{2}+m_{h}^{2}\right)C_{12}\right]\left(b_{jq}^{\alpha}B_{kq}^{h*}a_{ik}^{\alpha*}L+a_{jq}^{\alpha}A_{kq}^{h*}b_{ik}^{\alpha*}R\right)+\right.\nonumber \\
 & +2m_{q}m_{i}C_{11}\left(b_{jq}^{\alpha}A_{kq}^{h*}b_{ik}^{\alpha*}L+a_{jq}^{\alpha}B_{kq}^{h*}a_{ik}^{\alpha*}R\right)-2m_{q}m_{j}C_{12}\left(a_{jq}^{\alpha}B_{kq}^{h*}a_{ik}^{\alpha*}L+b_{jq}^{\alpha}A_{kq}^{h*}b_{ik}^{\alpha*}R\right)+\nonumber \\
 & +4m_{q}m_{k}C_{0}\left(b_{jq}^{\alpha}A_{kq}^{h*}a_{ik}^{\alpha*}L+a_{jq}^{\alpha}B_{kq}^{h*}b_{ik}^{\alpha*}R\right)+2m_{i}m_{k}\left(C_{11}+C_{0}\right)\left(b_{jq}^{\alpha}B_{kq}^{h*}b_{ik}^{\alpha*}L+a_{jq}^{\alpha}A_{kq}^{h*}a_{ik}^{\alpha*}R\right)+\nonumber \\
 & \left.-2m_{j}m_{k}\left(C_{12}+C_{0}\right)\left(a_{jq}^{\alpha}A_{kq}^{h*}a_{ik}^{\alpha*}L+b_{jq}^{\alpha}B_{kq}^{h*}b_{ik}^{\alpha*}R\right)\right\} u_{i},\end{align}
where $C=C\left(m_{k}^{2},m_{\alpha}^{2},m_{q}^{2},m_{i}^{2},m_{j}^{2},m_{h}^{2}\right).$

\begin{align}
M_{7} & =\frac{-i\bar{u}_{j}}{16\pi^{2}}g_{\alpha\beta h}^{3h}\left[m_{k}C_{0}\left(A_{jk}^{\beta}B_{ik}^{\alpha*}L+B_{jk}^{\beta}A_{ik}^{\alpha*}R\right)-m_{j}C_{12}\left(B_{jk}^{\beta}B_{ik}^{\alpha*}L+A_{jk}^{\beta}A_{ik}^{\alpha*}R\right)+\right.\nonumber \\
 & \left.\qquad\qquad+m_{i}\left(-C_{11}+C_{12}\right)\left(A_{jk}^{\beta}A_{ik}^{\alpha*}L+B_{jk}^{\beta}B_{ik}^{\alpha*}R\right)\right]u_{i},\end{align}
where $C=C\left(m_{k}^{2},m_{\alpha}^{2},m_{\beta}^{2},m_{i}^{2},m_{h}^{2},m_{j}^{2}\right).$

\begin{align}
M_{8} & =\frac{-i\bar{u}_{j}}{16\pi^{2}}g_{\alpha\beta h}^{vvh}\left[4m_{k}C_{0}\left(b_{jk}^{\beta}a_{ik}^{\alpha*}L+a_{jk}^{\beta}b_{ik}^{\alpha*}R\right)+2m_{i}\left(C_{11}-C_{12}\right)\left(b_{jk}^{\beta}b_{ik}^{\alpha*}L+a_{jk}^{\beta}a_{ik}^{\alpha*}R\right)+\right.\nonumber \\
 & \left.\qquad\qquad+2m_{j}C_{12}\left(a_{jk}^{\beta}a_{ik}^{\alpha*}L+b_{jk}^{\beta}b_{ik}^{\alpha*}R\right)\right]u_{i},\end{align}
where $C=C\left(m_{k}^{2},m_{\alpha}^{2},m_{\beta}^{2},m_{i}^{2},m_{h}^{2},m_{j}^{2}\right).$

\begin{align}
M_{9} & =\frac{i\bar{u}_{j}}{16\pi^{2}}g_{\beta\alpha h}^{vhh}\left[\left(\tilde{C}_{0}+2m_{i}^{2}C_{11}+m_{j}^{2}C_{12}-2m_{h}^{2}C_{12}\right)\left(b_{jk}^{\beta}B_{ik}^{\alpha*}L+a_{jk}^{\beta}A_{ik}^{\alpha*}R\right)+\right.\nonumber \\
 & \qquad-m_{i}m_{j}\left(C_{12}+C_{11}\right)\left(a_{jk}^{\beta}A_{ik}^{\alpha*}L+b_{jk}^{\beta}B_{ik}^{\alpha*}R\right)+m_{j}m_{k}\left(C_{0}-C_{12}\right)\left(a_{jk}^{\beta}B_{ik}^{\alpha*}L+b_{jk}^{\beta}A_{ik}^{\alpha*}R\right)+\nonumber \\
 & \left.\qquad+m_{i}m_{k}\left(C_{12}-C_{11}-2C_{0}\right)\left(b_{jk}^{\beta}A_{ik}^{\alpha*}L+a_{jk}^{\beta}B_{ik}^{\alpha*}R\right)\right]u_{i},\end{align}
where $C=C\left(m_{k}^{2},m_{\alpha}^{2},m_{\beta}^{2},m_{i}^{2},m_{h}^{2},m_{j}^{2}\right).$

\begin{align}
M_{10} & =\frac{i\bar{u}_{j}}{16\pi^{2}}g_{\alpha\beta h}^{vhh}\left[\left(-\tilde{C}_{0}+m_{i}^{2}\left(C_{12}-C_{11}\right)-2m_{j}^{2}C_{11}-2m_{h}^{2}\left(C_{12}-C_{11}\right)\right)\left(A_{jk}^{\beta}a_{ik}^{\alpha*}L+B_{jk}^{\beta}b_{ik}^{\alpha*}R\right)+\right.\nonumber \\
 & \qquad+m_{i}m_{j}\left(2C_{11}-C_{12}\right)\left(B_{jk}^{\beta}b_{ik}^{\alpha*}L+A_{jk}^{\beta}a_{ik}^{\alpha*}R\right)+m_{j}m_{k}\left(C_{12}+2C_{0}\right)\left(B_{jk}^{\beta}a_{ik}^{\alpha*}L+A_{jk}^{\beta}b_{ik}^{\alpha*}R\right)+\nonumber \\
 & \left.\qquad+m_{i}m_{k}\left(C_{11}-C_{12}-C_{0}\right)\left(A_{jk}^{\beta}b_{ik}^{\alpha*}L+B_{jk}^{\beta}a_{ik}^{\alpha*}R\right)\right]u_{i},\end{align}
where $C=C\left(m_{k}^{2},m_{\alpha}^{2},m_{\beta}^{2},m_{i}^{2},m_{h}^{2},m_{j}^{2}\right).$

\newpage

\chapter{\label{app:dijcij definition}Definition of the n-point integral
functions}

We present here the definitions for 1-loop scalar, vector and tensor
integrals:

\begin{align}
B_{0};B_{\mu}\left(m_{1}^{2},m_{2}^{2},p^{2}\right) & =\int\frac{d^{4}k}{i\pi^{2}}\frac{1;k_{\mu}}{\left[k^{2}-m_{1}^{2}\right]\left[\left(k+p\right)^{2}-m_{2}^{2}\right]},\end{align}

\begin{align}
C_{0};C_{\mu};C_{\mu\nu};\tilde{C}_{0}\left(m_{1}^{2},m_{2}^{2},m_{3}^{2},p_{1}^{2},p_{2}^{2}\right) & =\int\frac{d^{4}k}{i\pi^{2}}\frac{1;k_{\mu};k_{\mu\nu};k^{2}}{\left[k^{2}-m_{1}^{2}\right]\left[\left(k+p_{1}\right)^{2}-m_{2}^{2}\right]\left[\left(k+p_{1}+p_{2}\right)^{2}-m_{3}^{2}\right]},\end{align}

\begin{align}
{\rm B}_{\mu} & =p_{\mu}{\rm B}_{1},\nonumber \\
{\rm C}_{\mu} & =p_{1\mu}{\rm C}_{11}+p_{2\mu}{\rm C}_{12},\nonumber \\
{\rm C}_{\mu\nu} & =p_{1\mu}p_{1\nu}{\rm C}_{21}+p_{2\mu}p_{2\nu}{\rm C}_{22}+\{ p_{1}p_{2}\}_{\mu\nu}{\rm C}_{23}+g_{\mu\nu}{\rm C}_{24},\end{align}

where $\{ ab\}_{\mu\nu}\equiv a_{\mu}b_{\nu}+a_{\nu}b_{\mu}$.

\chapter{\label{app:higgs width formulas}Higgs width calculation}

In this section we give the formulae that we used in calculating the
Higgs width. The main contributions to the total width $\Gamma^{tot}$
are:

\begin{align}
\Gamma^{tot} & =\Gamma^{h\rightarrow\bar{q}q}+\Gamma^{h\rightarrow VV}+\Gamma^{h\rightarrow H_{i}H_{j}}+\Gamma^{h\rightarrow VH}.\end{align}

Only leading order values were used. The decay products are all taken
to be on-shell, and their secondary decay products are not taken into
account. Each contribution of $h\rightarrow x+y$ was calculated above
the threshold: $m_{h}>m_{x}+m_{y}$. The vertices are defined here
as in Fig. \ref{fig:vertices definition}.

The leading order value of $h\rightarrow\bar{q}q$ is \cite{HHG}:\begin{align}
\Gamma\left(h\rightarrow\bar{q}q\right) & =\frac{N_{c}A_{hqq}^{2}}{8\pi}m_{h}\left(1-\frac{4m_{q}^{2}}{m_{h}^{2}}\right)^{\frac{3}{2}},\end{align}

where $A_{hqq}=-\frac{gm_{q}}{2m_{W}}\frac{\cos\alpha}{\cos\beta}$;
$\frac{gm_{q}}{2m_{W}}\frac{\sin\alpha}{\cos\beta}$ is the quarks
coupling to $H^{0};h^{0}$, respectively, and $N_{c}=3$ is the color
factor, as mentioned above. We also give explicitly the width for
the process $h^{0}\rightarrow\bar{b}b$ in the T2HDM for $\alpha=\beta$:\begin{align}
\Gamma\left(h^{0}\rightarrow\bar{b}b\right) & =\frac{3g^{2}m_{b}^{2}}{32\pi m_{W}^{2}}m_{h^{0}}\tan^{2}\beta\left(1-\frac{4m_{b}^{2}}{m_{h^{0}}^{2}}\right)^{\frac{3}{2}}.\end{align}

The leading order value of $h\rightarrow W^{+}W^{-}$ is \cite{HHG}:\begin{align}
\Gamma\left(h\rightarrow W^{+}W^{-}\right) & =\frac{g_{hWW}^{2}m_{h}^{3}}{64\pi m_{W}^{4}}\left(1-x\right)^{\frac{1}{2}}\left(1-x+\frac{3}{4}x^{2}\right),\end{align}

where $g_{hWW}=$$gm_{W}\cos\left(\beta-\alpha\right)$; $gm_{W}\sin\left(\beta-\alpha\right)$
is the $W^{+}W^{-}$ coupling to $H^{0};h^{0}$, respectively, and
$x=\frac{4m_{W}^{2}}{m_{h}^{2}}$.

The leading order value of $h\rightarrow Z^{0}Z^{0}$ is \cite{HHG}:\begin{align}
\Gamma\left(h\rightarrow Z^{0}Z^{0}\right) & =\frac{g_{hZZ}^{2}m_{h}^{3}\cos^{4}\theta_{W}}{32\pi m_{W}^{4}}\left(1-x\right)^{\frac{1}{2}}\left(1-x+\frac{3}{4}x^{2}\right),\end{align}

where $g_{hZZ}=\frac{gm_{Z}}{\cos\theta_{W}}\cos\left(\beta-\alpha\right)$;
$\frac{gm_{Z}}{\cos\theta_{W}}\sin\left(\beta-\alpha\right)$ is the
$Z^{0}Z^{0}$ coupling to $H^{0};h^{0}$, respectively, and $x=\frac{4m_{Z}^{2}}{m_{h}^{2}}$.

Since the couplings $W^{+}W^{-}h^{0}$ and $Z^{0}Z^{0}h^{0}$ are
both $\propto\sin\left(\beta-\alpha\right)$ (as we have shown in
table \ref{tab:vvh feyn rules}), then by choosing $\alpha=\beta$,
the widths of the processes $h^{0}\rightarrow W^{+}W^{-}$ and $h^{0}\rightarrow Z^{0}Z^{0}$
are both reduced to zero.

The leading order value of $h\rightarrow H_{i}H_{j}$ (where $H_{i}$
and $H_{j}$ are any two scalars) is:\begin{align}
\Gamma\left(h\rightarrow H_{i}H_{j}\right) & =\frac{g_{hH_{i}H_{j}}^{2}}{16\pi m_{h}}\lambda^{\frac{1}{2}}\left(1,\frac{m_{H_{i}}^{2}}{m_{h}^{2}},\frac{m_{H_{j}}^{2}}{m_{h}^{2}}\right),\end{align}

where $g_{hH_{i}H_{j}}$ is the triple Higgs coupling $hH_{i}H_{j}$.

The leading order value of $h\rightarrow VH$ (where $VH=W^{+}H^{-}$
or $Z^{0}+\mbox{neutral scalar}$) is:\begin{align}
\Gamma\left(h\rightarrow VH\right) & =\frac{g_{VHh}^{2}m_{V}^{2}}{16\pi m_{h}}\lambda^{\frac{1}{2}}\left(1,\frac{m_{V}^{2}}{m_{h}^{2}},\frac{m_{H}^{2}}{m_{h}^{2}}\right)\lambda\left(1,\frac{m_{h}^{2}}{m_{V}^{2}},\frac{m_{H}^{2}}{m_{V}^{2}}\right),\end{align}

where $g_{hVH}$ is the vector-scalar-scalar coupling $VHh$. This
contribution can be important, as stated in Sec. \ref{sub:Higgs-decay-BR}.

\begin{figure}
\begin{centering}
\includegraphics[width=408pt,keepaspectratio]{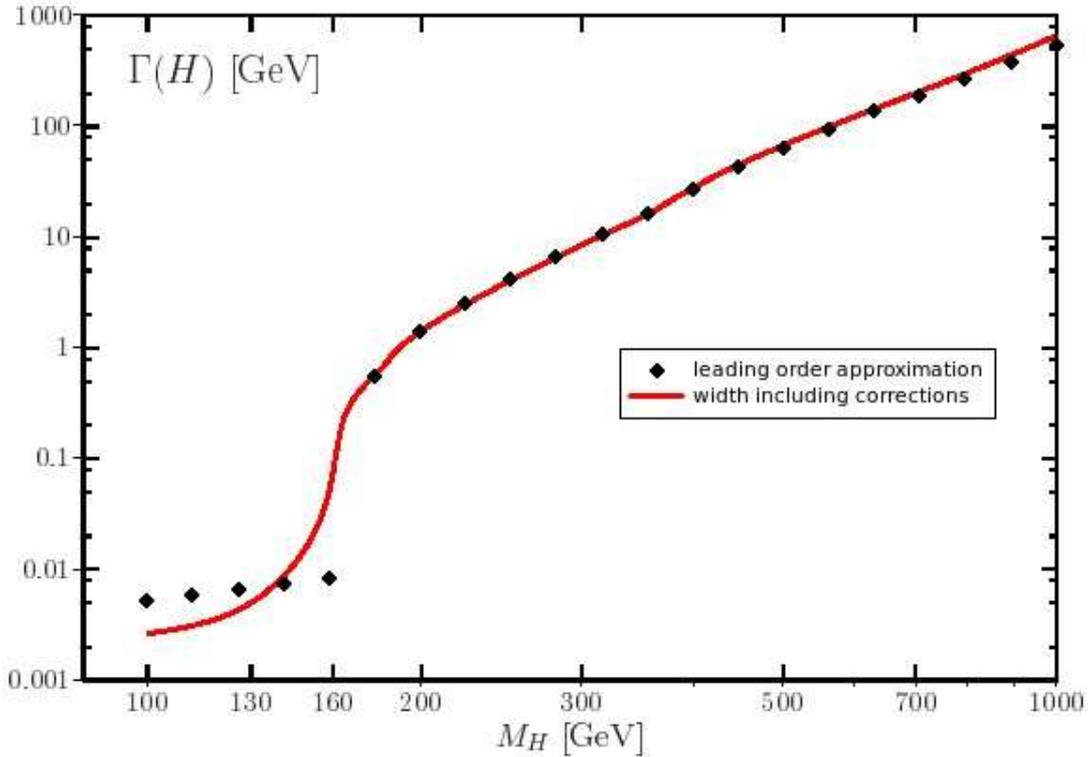}
\par\end{centering}

\caption{\label{fig:higgs width SM}The total width of the SM Higgs: leading
order approximation compared to corrected width of \cite{djouadi I}.}
\end{figure}

In order to demonstrate the error introduced by including only leading
terms in the Higgs width calculation, we give in Fig. \ref{fig:higgs width SM}
the total SM Higgs width as calculated in this work compared to the
width of \cite{djouadi I} which includes higher-order corrections.

As can be seen, below the WW threshold (at about 160 GeV) the values
are different. In this mass range $b\bar{b}$ decay dominates, and
corrections have a large impact. However, this mass range is also
below the $h\rightarrow\bar{t}c$ threshold. On the other hand, above
the WW threshold the values are very similar, and corrections have
a small impact, and so the use of leading order approximation can
be justified.

\chapter{\label{app:cancel divergences}Cancellation of divergences in the
1-loop amplitude}

Some of the 1-loop Feynman diagrams have a divergent part. These divergent
parts cancel, since this is a leading-order calculation and, therefore,
there is no renormalization.

This cancellation is also important as a means of checking the self
consistency of the calculation.

We define $\varepsilon=4-d$, where $d\rightarrow4$ is the number
of dimensions. As $\varepsilon\rightarrow0$, some n-point integrals
will have a term proportional to $\frac{1}{\varepsilon}$. These are
summarized below:

\begin{eqnarray}
B_{0}\sim-2\frac{1}{\varepsilon} & ,\qquad & C_{24}\sim-\frac{1}{2}\frac{1}{\varepsilon},\nonumber \\
B_{1}\sim1\frac{1}{\varepsilon} & ,\qquad & \tilde{C_{0}}\sim-2\frac{1}{\varepsilon}.\end{eqnarray}

The parts of the Feynman diagrams proportional to $\frac{1}{\varepsilon}$
were collected below. Only the parts of the diagrams with a left projection
operator are given, while the right-handed parts are subject to a
similar cancellation. The Feynman diagrams were shown in Fig. \ref{fig: 1-loop diags}.

\begin{align*}
M_{1L}^{\infty l=j} & =\frac{-1}{m_{i}^{2}-m_{l}^{2}}\left[-2m_{k}\left(m_{l}B_{lj}^{H^{0}*}A_{lk}^{\alpha}B_{ik}^{\alpha*}+m_{i}B_{lj}^{H^{0}*}B_{lk}^{\alpha}A_{ik}^{\alpha*}\right)-\right.\\
 & \left.\hphantom{=\frac{-1}{m_{i}^{2}-m_{l}^{2}}2}-m_{i}\left(m_{l}B_{lj}^{H^{0}*}A_{lk}^{\alpha}A_{ik}^{\alpha*}+m_{i}B_{lj}^{H^{0}*}B_{lk}^{\alpha}B_{ik}^{\alpha*}\right)\right],\\
M_{2L}^{\infty l=i} & =\frac{-1}{m_{j}^{2}-m_{l}^{2}}\left[-2m_{k}\left(m_{l}A_{jk}^{\alpha}B_{lk}^{\alpha*}B_{il}^{H^{0}*}+m_{j}B_{jk}^{\alpha}A_{lk}^{\alpha*}B_{il}^{H^{0}*}\right)-\right.\\
 & \left.\hphantom{=\frac{-1}{m_{i}^{2}-m_{l}^{2}}2}-m_{j}\left(m_{l}B_{jk}^{\alpha}B_{lk}^{\alpha*}B_{il}^{H^{0}*}+m_{j}A_{jk}^{\alpha}A_{lk}^{\alpha*}B_{il}^{H^{0}*}\right)\right],\\
M_{3L}^{\infty l=j} & =\frac{1}{m_{i}^{2}-m_{l}^{2}}\left[-8m_{k}\left(m_{l}B_{lj}^{H^{0}*}b_{lk}^{\alpha}a_{ik}^{\alpha*}+m_{i}B_{lj}^{H^{0}*}a_{lk}^{\alpha}b_{ik}^{\alpha*}\right)-\right.\\
 & \left.\hphantom{=\frac{-1}{m_{i}^{2}-m_{l}^{2}}}-2m_{i}\left(m_{l}B_{lj}^{H^{0}*}b_{lk}^{\alpha}b_{ik}^{\alpha*}+m_{i}B_{lj}^{H^{0}*}a_{lk}^{\alpha}a_{ik}^{\alpha*}\right)\right],\\
M_{4L}^{\infty l=i} & =\frac{1}{m_{j}^{2}-m_{l}^{2}}\left[-8m_{k}\left(m_{l}b_{jk}^{\alpha}a_{lk}^{\alpha*}B_{il}^{H^{0}*}+m_{j}a_{jk}^{\alpha}b_{lk}^{\alpha*}B_{il}^{H^{0}*}\right)-\right.\\
 & \left.\hphantom{=\frac{-1}{m_{i}^{2}-m_{l}^{2}}}-2m_{j}\left(m_{l}a_{jk}^{\alpha}a_{lk}^{\alpha*}B_{il}^{H^{0}*}+m_{j}b_{jk}^{\alpha}b_{lk}^{\alpha*}B_{il}^{H^{0}*}\right)\right],\end{align*}

\begin{eqnarray}
M_{5L}^{\infty q=k} & = & 2A_{jq}^{\alpha}A_{kq}^{H^{0}*}B_{ik}^{\alpha*},\nonumber \\
M_{6L}^{\infty q=k} & = & -8b_{jq}^{\alpha}B_{kq}^{H^{0}*}a_{ik}^{\alpha*},\nonumber \\
M_{7L}^{\infty} & = & 0,\nonumber \\
M_{8L}^{\infty} & = & 0,\nonumber \\
M_{9L}^{\infty} & = & -2g_{\beta\alpha H^{0}}^{vhh}b_{jk}^{\beta}B_{ik}^{\alpha*},\nonumber \\
M_{10L}^{\infty} & = & 2g_{\alpha\beta H^{0}}^{vhh}A_{jk}^{\beta}a_{ik}^{\alpha*}.\end{eqnarray}

After inserting the Feynman rules of the T2HDM, we were able to show
that the terms proportional to $\frac{1}{\varepsilon}$ cancel as
shown below:

\begin{align}
 & M_{1}^{\infty}+M_{2}^{\infty}+M_{5}^{\infty}=0,\nonumber \\
 & M_{3}^{\infty}+M_{4}^{\infty}=0,\nonumber \\
 & M_{6}^{\infty}=0,\nonumber \\
 & M_{7}^{\infty}=0,\nonumber \\
 & M_{8}^{\infty}=0,\nonumber \\
 & M_{9}^{\infty}=0,\nonumber \\
 & M_{10}^{\infty}=0.\end{align}

This cancellation was also verified numerically in the FORTRAN code.

\end{document}